\documentclass[aps,groupedaddress]{revtex4}
\usepackage{epsfig}
\usepackage{graphicx}
\usepackage[T1]{fontenc}
\usepackage{ae}
\usepackage{color}
\usepackage[latin1]{inputenc}
\usepackage{amssymb,amsbsy,amsmath}
\usepackage{bbm}

\newtheorem{defi}{Definition}
\newtheorem{prop}{Proposition}

\begin{document}



\title[CST]
      {The classical spin triangle as an integrable system}
\author{Heinz-J\"urgen Schmidt$^1$
}
\address{$^1$  Universit\"at Osnabr\"uck,
Fachbereich Physik,
 D - 49069 Osnabr\"uck, Germany}


\begin{abstract}
The classical spin system consisting of three spins with Heisenberg interaction is an example
of a completely integrable mechanical system. In this paper we explicitly calculate its time evolution
and the corresponding action-angle  variables. This calculation is facilitated by splitting the six
degrees of freedom into three internal and three external variables, such that the internal variables
evolve autonomously. Their oscillations can be explicitly calculated in terms of the Weierstrass elliptic function.
We test our results by means of an example and comparison with direct numerical integration. A couple of special
cases is analyzed where the general theory does not apply, including the aperiodic limit case for special
initial conditions. The extension to systems with a time-depending
magnetic field in a constant direction is straightforward.
\end{abstract}

\maketitle

\section{Introduction}\label{sec:Intro}

\textit{Spin} is a genuine quantum concept.
Nevertheless, electronic spin produces macroscopic magnetic effects described in classical terms.
A key to understanding this theoretically lies in the classical limit of quantum spin systems, see \cite{L73,FKL07}.
Many single spins with spin quantum number $s=1/2$ can be combined to yield systems with larger $s$, integer- or half-integer-valued.
In the limit $s\to\infty$ the spin vector operator, after re-scaling, can be replaced by a classical spin vector,
i.~e., a unit vector ${\mathbf s}\in {\mathcal S}^2$.
This limit can also be extended to systems of $N$ classical spins including their interaction.
The advantage of considering the classical limit is at least threefold:
\begin{itemize}
  \item The  theory of classical spin systems which is simpler than its quantum analogue can be used to approximate the behaviour of real systems
  with localized relatively large spins, e.g., of Gadolinium atoms \cite{P15} with $s=7/2$ embedded into a magnetic molecule \cite{GOB14,Qetal17,Qetal21}.
  \item The classical limit $s\to\infty$ can be used as a test for theoretical calculations that are made for arbitrary $s$. For example,
  the high temperature expansion of the specific heat or susceptibility leads to certain polynomials in $s$ the leading coefficient
  of which can be calculated by classical theory, see, e.~g., \cite{SLR11}.
  \item A classical spin system can be understood as a system with $2N$-dimensional phase space
  ${\mathcal S}^2\times \ldots \times {\mathcal S}^2$ and analyzed with the methods of classical mechanics.
  In this way one can extend the realm of classical mechanics by examples of quantum origin.
\end{itemize}
The focus of the present work lies on the last item. We consider a classical ``spin triangle", i.~e., a spin system of $N=3$ spins with Heisenberg interaction.
This is a phenomenological ansatz to describe the exchange interaction of spins by an isotropic Hamiltonian which is bi-linear
in the spin observables and has a straightforward classical analogue, see, e.~g., \cite[Eq.~(1.40)]{W15}. Every quantum spin system
with Heisenberg Hamiltonian has three commuting observables that are constants of motion: The Hamiltonian itself, the square of the total
spin and its $3$-component. Since the corresponding classical functions on the six-dimensional phase space Poisson-commute we obtain a
\textit{completely integrable} classical system in the sense of the Arnol'd-Liouville theorem, see \cite{A78}.
Hence the time evolution can be explicitly calculated up to integrations, in contrast to the situation for the quantum spin triangle \cite{S13}.
Although the integrability of the classical spin triangle can thus be taken as given,
it will be nevertheless instructive to consider the details of the corresponding calculations,
taking into account the specific properties of the problem at hand.
\\

Since we fear that the reader might lose the thread due to the extensive
material that has arisen on the general spin triangle, we have divided the text into two parts.
The first part containing the Sections \ref{sec:CM} to \ref{sec:SO} gives the basic definitions and
calculations without going into the mathematical details and justifications.
This part should be suitable for those readers who want to get a first overview.
The second part, see Appendices \ref{sec:MF} to \ref{sec:FT}, then deals with extensions and additions,
as well as a more geometrically oriented approach to the time evolution of the system under study.

In section \ref{sec:CM} we recapitulate the basic features of classical mechanics of
spin systems with Heisenberg-Hamiltonian. The time evolution of the spin triangle
is discussed in Section \ref{sec:TE}. The six degree of freedom of the spin triangle can be split into
three internal degrees describing the ``form" of the spin configuration and three external degrees describing its ``position"
in spin space. The time evolution of the internal degrees turns out to be autonomous, see subsection \ref{sec:TID},
and the external degrees evolve in dependence of a given solution for the internal degrees, see subsection \ref{sec:TED}.
Section \ref{sec:TID} also contains the definition of the ``generic case'' assumed throughout the first part of the paper.
Interestingly, the integrals resulting for the internal degrees are of elliptic type and hence their time
evolution can be explicitly given in terms of the Weierstrass elliptic function. In contrast,
the integrals obtained for the external degrees could only be numerically calculated.
We close the first part with a Summary and Outlook in Section \ref{sec:SO}.

The second part starts with Appendix \ref{sec:MF} that shows how to reduce the case of an additional  time-dependent
magnetic field with constant direction to the pure Heisenberg case.
This reduction is known and is only recapitulated here for the sake of completeness.
The Appendix \ref{sec:G} is devoted to
the geometric approach to the problem at hand and contains further mathematical elaborations.
Subsection \ref{sec:SM} deals with the symplectic structure of phase space and contains a proof
that the Landau-Lifshitz equations result from the Heisenberg Hamiltonian.
In the subsection \ref{sec:PG} it is explained how the internal degrees of freedom are represented
by the Gram matrix $G(s)$ of the spin configuration $s$ and that the Gram matrices $G(s)$ can be viewed as the points of a
three-dimensional convex ``Gram set" ${\mathcal G}$ with tetrahedral symmetry, already defined in \cite{SL03}.
The conserved quantities give rise to linear constraints of the internal degrees of freedom.
Correspondingly, the time evolution can be visualized as an oscillation on
a line $L$ intersecting ${\mathcal G}$, see subsection \ref{sec:TG}.
Moreover, we prove that in the generic case the Weierstrass polynomial has three real simple roots
and hence the analysis of Section \ref{sec:TID} completely covers this case.
Further, it turns out to be mathematically more convenient to represent
the internal degrees of freedom by the ``double" ${\mathcal G}'$ of ${\mathcal G}$ analogously to the introduction
of Riemannian surfaces for the domain of definition of otherwise many-valued complex functions.
Subsection \ref{sec:TP} contains more details of the time evolution of the position of the spin configuration and
subsection \ref{sec:AR} deals with some subtle points in the definition of the averaged rotation of the system in spin space.
We proceed with an example in Subsection \ref{sec:EX} where we compare the semi-analytical solution with a numerical
integration of the equations of motion for a particular choice of the values of the conserved quantities.

In Appendix \ref{sec:AA} we analyze in more detail the definition of action-angle  variables for the spin system under consideration.
An interesting byproduct is the finding that the partial energy derivative of the total area swept by the three spin vectors
with averaged rotation in spin space is proportional to the period ${\sf T}$ of the internal oscillation,
analogous to the well-known result for one-dimensional mechanical problems.
Next, several special cases that have to be excluded in the main part are reconsidered in Appendix \ref{sec:SC}.
These are the isosceles spin triangle case, subsections \ref{sec:IST} and \ref{sec:SCIST}, and the case
of stationary Gram matrices in subsection \ref{sec:SN}.
Additionally, the aperiodic limit case occurring for special initial conditions is treated in subsection \ref{sec:TSG}.
An enumeration of all stationary states corresponding to critical values of the Hamiltonian is given in subsection  \ref{sec:SS}.
The evolution of the external degrees of freedom can also be described in terms of Floquet theory,
see some related remarks in the Appendix \ref{sec:FT}.

\section{Classical mechanics of spin systems}\label{sec:CM}
As mentioned in the Introduction, classical spin systems are examples of systems that can be described by classical mechanics.
Although this is well-known, see, e.~g., \cite{SSHL15}, we will recapitulate the essential facts.
In this Section we will consider general systems of $N$ classical spins since there is no advantage in restricting ourselves to $N=3$,
except for the last part.

The $N$ spins are represented by unit vectors ${\mathbf s}_\mu,\; \mu=1,\ldots,N$,
with components ${\mathbf s}_\mu^{(i)},\;i=1,2,3$. Hence the phase space of the spin system
can be taken as the $N$-fold Cartesian product of unit spheres (sometimes called ``Bloch spheres")
\begin{equation}\label{defphase}
 {\mathcal P}={\mathcal S}^2 \times \ldots \times {\mathcal S}^2= \left({\mathcal S}^2\right)^N
 \;,
\end{equation}
and hence is $2\,N$-dimensional and compact.
Canonical coordinates $\left(\phi_\mu,\,z_\mu\right)_{\mu=1,\ldots,N}$ are defined
via the representation
\begin{equation}\label{defcan}
  {\mathbf s}_\mu =
  \left(
\begin{array}{c}
 \sqrt{1-z_{\mu }^2} \cos\,\phi _{\mu } \\
 \sqrt{1-z_{\mu }^2} \sin \,\phi _{\mu } \\
 z_{\mu } \\
\end{array}
\right)
\;,
\end{equation}
It is clear that these coordinates are not defined on a cut from the north pole to the south pole of the unit spheres
and that we would need at least one more coordinate chart to cover ${\mathcal P}$ completely, but we will not dwell into these details.
Note that the canonical coordinates yield the (up to a factor) unique surface element
$d\phi_\mu\wedge d z_\mu$ on the $\mu$-th unit spheres that is invariant under rotations.

As the standard example we consider the bilinear, isotropic \textit{Heisenberg Hamiltonian}
\begin{equation}\label{defH}
 H=\sum_{\mu<\nu}J_{\mu\nu} {\mathbf s}_\mu\cdot {\mathbf s}_\nu=\sum_{\mu<\nu}J_{\mu\nu}
 \left( \sqrt{1-z_\mu^2}\,\sqrt{1-z_\nu^2}\, \cos\left(\phi_\mu-\phi_\nu \right)+z_\mu\,z_\nu\right)
 \;,
\end{equation}
where the $J_{\mu\nu}$ are ${N\choose 2}$ real coupling coefficients. The special case where all $J_{\mu\nu} \equiv 1$
will be denoted by
\begin{equation}\label{defH1}
 H_1:=\sum_{\mu<\nu}{\mathbf s}_\mu\cdot {\mathbf s}_\nu
 \;.
\end{equation}

Then the following holds:
\begin{prop}\label{P1}
 For the Heisenberg Hamiltonian (\ref{defH}) the Hamiltonian  equations of motion
 \begin{eqnarray}
 \label{Ham1}
   \dot{z}_\mu&=& \frac{\partial H}{\partial \phi_\mu},\quad \mu=1,\ldots,N\\
   \label{Ham2}
   \dot{\phi}_\mu&=& -\frac{\partial H}{\partial z_\mu},\quad \mu=1,\ldots,N
   \;,
 \end{eqnarray}
 are equivalent to the \textit{Landau-Lifshitz} equations of motion
\cite{LL35,F1}
 \begin{equation}\label{LL}
    \dot{\mathbf s}_\mu=\left(\sum_\nu J_{\mu\nu} {\mathbf s}_\nu\right)\times {\mathbf s}_\mu
    \;.
 \end{equation}
\end{prop}
In Eq.~\ref{LL} we have tacitly assumed that the array of coupling constants $J_{\mu\nu}$ has been extended to a symmetric
$N\times N$-matrix with vanishing diagonal. Moreover, $t$  denotes the (dimensionless) time and $\;\dot{}\equiv \frac{d}{dt}$.
The {\bf proof} of Proposition \ref{P1} will be omitted since it can be reduced to a straightforward calculation.
For an alternative formulation of Proposition \ref{P1} without using canonical coordinates see Proposition \ref{PropHam} in Appendix \ref{sec:SM}.\\

As usual, we define the \textit{total spin vector} ${\mathbf S}$ with components
\begin{equation}\label{defS}
  {\mathbf S}^{(i)}:= \sum_\mu  {\mathbf s}_\mu^{(i)} \quad \mbox{for } i=1,2,3
  \;.
\end{equation}
Its squared length can be written as
\begin{equation}\label{defS2}
  S^2:= \sum_{i=1}^{3} {\mathbf S}^{(i)2}= \sum_{\mu\nu i}  {\mathbf s}_\mu^{(i)}\, {\mathbf s}_\nu^{(i)}
  =N + 2  \sum_{\mu<\nu}  {\mathbf s}_\mu\cdot {\mathbf s}_\nu \stackrel{(\ref{defH1})}{=}N+2\, H_1
  \;.
\end{equation}

As $H$ does not explicitly depend on $t$ it will be a constant of motion. Further conserved quantities are
${\mathbf S}^{(i)},\;i=1,2,3$ according to
\begin{eqnarray}
\label{con1}
  \frac{d}{dt}{\mathbf S}^{(i)} &\stackrel{(\ref{defS})}{=}& \sum_\mu  \dot{\mathbf s}_\mu^{(i)}  \\
\label{con2}
   &\stackrel{(\ref{LL})}{=}&  \sum_{\mu\nu} J_{\mu\nu}\, {\mathbf s}_\nu\times {\mathbf s}_\mu=0
   \;,
\end{eqnarray}
since the terms of the double sum in (\ref{con2}) are anti-symmetric in $\mu,\nu$. In view of
Noether's theorem the invariance of ${\mathbf S}^{(i)},\;i=1,2,3$ is equivalent to the symmetry
of the Hamiltonian (\ref{defH}) under global rotations $R$. This symmetry also implies that
${\mathbf s}_\mu(t),\,\mu=1,\ldots,N$ is a solution of (\ref{LL}) iff $R\,{\mathbf s}_\mu(t),\,\mu=1,\ldots,N$ is a solution.

The canonical coordinates $\left(\phi_\mu,\,z_\mu\right)$ for the phase space ${\mathcal P}$ give rise to the
anti-symmetric  \textit{Poisson brackets} defined
for any two smooth functions $f,g:{\mathcal P}\rightarrow {\mathbbm R}$:
\begin{equation}\label{defPoisson}
  \{ f,g\}:= \sum_\mu \left(\frac{\partial f}{\partial z_\mu}\,\frac{\partial g}{\partial \phi_\mu}-
  \frac{\partial f}{\partial \phi_\mu}\,\frac{\partial g}{\partial z_\mu}\right)
  \;,
\end{equation}
such that the total time derivative of $f$ can be written in the form
\begin{eqnarray}\label{total1}
  \frac{d}{dt} f &=& \sum_{\mu} \left(\frac{\partial f}{\partial z_\mu}\,\frac{d z_\mu}{d t}+
  \frac{\partial f}{\partial \phi_\mu}\,\frac{d \phi_\mu}{d t}\right)\\
  \label{total2}
  &\stackrel{(\ref{Ham1},\ref{Ham2})}{=}&
  \sum_{\mu} \left(\frac{\partial f}{\partial z_\mu}\,\frac{\partial H}{\partial \phi_\mu}-
  \frac{\partial f}{\partial \phi_\mu}\,\frac{\partial H}{\partial z_\mu}\right)\\
  \label{total3}
  &\stackrel{(\ref{defPoisson})}{=}&  \{
  f,H\}
  \;.
\end{eqnarray}

Starting with the four constants of motion $H,  {\mathbf S}^{(i)},\;i=1,2,3,$ we may define three
constants of motion \textit{in involution}, i.~e., with mutually vanishing Poisson brackets,
namely $H, S^2$ and ${\mathbf S}^{(3)}$. The vanishing of $\{ H,S^2\}$ and $\{ H,{\mathbf S}^{(3)}\}$
follows by (\ref{total3}) since $S^2$ and ${\mathbf S}^{(3)}$ are constants of motion. Moreover,
$\{{\mathbf S}^{(3)},S^2\}=0$  since $\{{\mathbf S}^{(3)},H_1\}=0$ as special case of $\{{\mathbf S}^{(3)},H\}=0$
and further using (\ref{defS2}).

Thus we found three constants of motion $H, S^2$ and ${\mathbf S}^{(3)}$ in involution. In the special case of
$N=3$, which we consider from now on, this implies that the spin system with Hamiltonian (\ref{defH}) is
\textit{completely integrable} in the sense of the Arnol'd-Liouville theorem \cite{A78}, and its solutions
can be implicitly expressed in terms of integrals. Moreover, the system will move on an invariant $3$-torus
that can be parametrized by special canonical coordinates $\left({\sf I}_i,\psi_i\right),\; i=1,2,3$ called
\textit{action-angle  variables} such that the equations of motion assume the simple form
\begin{equation}\label{aaa}
  \dot{\sf I}_i =0,\quad \dot{\psi}_i=\Omega_i\left({\sf I}_1,{\sf I}_2,{\sf I}_3\right)),\quad i=1,2,3
  \;.
\end{equation}
The rotations according to the angles $\psi_i$ are uniquely composed of phase flows generated by the
conserved quantities. The details of the solution of the equation of motion for the integrable $N=3$ system will be presented in the
remainder of this paper.\\

Throughout this work we will denote by ${\mathcal R}({\mathbf n},\alpha)$ the rotation about the axis given by the
(not necessarily normalized) vector ${\mathbf n}$ with an angle $\alpha$.
The transposition of a matrix $A$ will be denoted by $A^\top$. As usual, $O(3)$ will denote the group of $3\times 3$ rotation/reflection matrices,
i.~e., of invertible real matrices $R$ satisfying $R^{-1} = R^\top$ and $SO(3)$ the subgroup of proper rotations defined by $\det R=1$.
Moreover, the three unit vectors forming the standard
basis in ${\mathbbm R}^3$ will be denoted by
\begin{equation}\label{defei}
 {\mathbf E}_1:= \left( \begin{array}{c}
                          1 \\
                          0 \\
                          0
                        \end{array}\right)\;,
                        {\mathbf E}_2:= \left( \begin{array}{c}
                          0 \\
                          1 \\
                          0
                        \end{array}\right)\;,
                        {\mathbf E}_3:= \left( \begin{array}{c}
                          0 \\
                          0 \\
                          1
                        \end{array}\right)\;.
\end{equation}

\section{Time evolution}\label{sec:TE}
We recapitulate the special form of the Hamiltonian (\ref{defH}) for $N=3$:
\begin{equation}\label{H3}
 H= J_1\,{\mathbf s}_2\cdot{\mathbf s}_3+ J_2\,{\mathbf s}_3\cdot{\mathbf s}_1+J_3\,{\mathbf s}_1\cdot{\mathbf s}_2
 \;,
\end{equation}
and the resulting equations of motion
\begin{eqnarray}\label{eom1}
  \dot{\mathbf s}_1&=& \left( J_2 \,{\mathbf s}_3+ J_3 \,{\mathbf s}_2\right)\times {\mathbf s}_1,\\
  \label{eom2}
  \dot{\mathbf s}_2&=& \left( J_3 \,{\mathbf s}_1+ J_1 \,{\mathbf s}_3\right)\times {\mathbf s}_2,\\
  \label{eom3}
  \dot{\mathbf s}_3&=& \left( J_1 \,{\mathbf s}_2+ J_2 \,{\mathbf s}_1\right)\times {\mathbf s}_3
  \;,
\end{eqnarray}
where we have renamed the three coupling constants.
We take the spin configuration $s=\left({\mathbf s}_1,{\mathbf s}_2,{\mathbf s}_3\right)\in{\mathcal P}$
to be a $3\times 3$-matrix with entries $s_{i,\mu}={\mathbf s}_\mu^{(i)}$ for $i,\mu=1,2,3$. The equations of motion
(\ref{eom1} - \ref{eom3}) will be written in the compact form
\begin{equation}\label{eom4}
   \dot{s} = {\mathcal J}(s)
   \;,
\end{equation}
using the bilinear matrix-valued function ${\mathcal J}(s)$ with entries
\begin{equation}\label{defopJ}
 {\mathcal J}(s)_{i \mu}:=  \left( \sum_\kappa J_{\mu\kappa} {\mathbf s}_\kappa\times {\mathbf s}_\mu\right)^{(i)}
 \;,
\end{equation}
for $i,\mu=1,2,3$. Since the vector product transforms in a natural way under rotations $R\in SO(3)$ we have
\begin{equation}\label{RopJ}
  {\mathcal J}(R\,s)= R\, {\mathcal J}(s)
  \;,
\end{equation}
for all $R\in SO(3)$.\\

If all coupling constants are shifted by a constant value, $J_i\mapsto J_i+\delta J,\; i=1,2,3,$
the time evolution of the spin configuration according to (\ref{eom1} - \ref{eom3}) would be modified
by ${\mathcal R}({\mathbf S},\delta J\,S)$,
a uniform rotation about the constant total spin ${\mathbf S}$ with angular velocity $\delta J \,S$.
Moreover, multiplication of all coupling constants by a non-zero real number $\lambda$ will lead to
a time evolution with a suitable scaled time variable $\lambda^{-1}\,t$, accompanied by a time reflection if $\lambda<0$.
These transformations could be used to restrict the coupling constants to the special case of, say,
$J_1=J, J_2=1,J_3=0$, but we will not make use of this simplification in the present paper, except
for Section \ref{sec:TSG}.

Our strategy to solve the equations of motion (\ref{eom1} - \ref{eom3}) will be to split the six degrees of freedom
of the spin triangle into three \textit{internal degrees} and three  \textit{external degrees} such that
the internal degrees evolve autonomously. As the variables corresponding to the internal degrees
we define the three scalar products between the spin vectors:
\begin{equation}\label{defuvw}
 u:= {\mathbf s}_2\cdot  {\mathbf s}_3,\quad  v:= {\mathbf s}_3\cdot  {\mathbf s}_1,\quad  w:= {\mathbf s}_1\cdot  {\mathbf s}_2
 \;.
\end{equation}
These internal variables determine the form of the spin configuration. It can be rotated or reflected in spin space:
If $r(u,v,w)$ denotes any spin configuration realizing the internal
variables $(u,v,w)$ and $R\in O(3)$, then $R\,r(u,v,w)$
will be another spin configuration realizing $(u,v,w)$. It can be shown, see Appendix \ref{sec:PG}, that all spin configurations
realizing $(u,v,w)$ can be obtained in this way, and hence it appears sensible to identify the external degrees of freedom
with the three parameters specifying the rotation/reflection matrix  $R\in O(3)$.

For later use we consider the scalar triple product
\begin{equation}\label{defdelta}
  \delta:= \left({\mathbf s}_1 \times {\mathbf s}_2\right)\cdot{\mathbf s}_3
  \;,
\end{equation}
that can be expressed by the internal variables as
\begin{equation}\label{deltasqrt}
 \delta= \pm \sqrt{ 1-u^2-v^2-w^2+2 u v w}
 \;,
\end{equation}
see (\ref{detS}).

\subsection{Time evolution of the internal degrees of freedom}\label{sec:TID}

First consider the conserved quantities ${\mathbf S}^{(3)}$, $H(s)$ and $H_1(s)={\scriptsize{\frac{1}{2}}}\left(S^2-3\right)$.
assuming the values
\begin{equation}\label{S3const}
  {\mathbf S}^{(3)} = \sigma_3
  \;,
\end{equation}
\begin{equation}\label{Huvw}
 H(s)= J_1\,{\mathbf s}_2\cdot{\mathbf s}_3+ J_2\,{\mathbf s}_3\cdot{\mathbf s}_1+J_3\,{\mathbf s}_1\cdot{\mathbf s}_2
 \stackrel{(\ref{defuvw})}{=} \left(
\begin{array}{c}
 J_1\\
 J_2 \\
 J_3 \\
\end{array}
\right)
\cdot\left(
\begin{array}{c}
 u \\
 v \\
 w \\
\end{array}
\right)=\varepsilon
\;,
\end{equation}
and
\begin{equation}\label{H1uvw}
 H_1(s)= {\mathbf s}_2\cdot{\mathbf s}_3+{\mathbf s}_3\cdot{\mathbf s}_1+{\mathbf s}_1\cdot{\mathbf s}_2
 \stackrel{(\ref{defuvw})}{=}\left(
\begin{array}{c}
 1\\
 1 \\
 1 \\
\end{array}
\right)
\cdot\left(
\begin{array}{c}
 u \\
 v \\
 w \\
\end{array}
\right)=\sigma
\;,
\end{equation}
where the constants  $\sigma_3,\varepsilon$ and $\sigma$ depend on the initial value of $s$.
We will also use the abbreviation $S=\sqrt{3+2\,\sigma}$.

The calculations of this Section will be restricted to the \textit{generic case}:
\begin{defi}\label{DefGeneric}
The {\rm generic case} is defined by the following conditions:
\begin{itemize}
  \item The coupling constants are pairwise different, i.~e.,
  \begin{equation}\label{generic1}
    J_1\neq J_2, \;  J_2\neq J_3,\;\mbox{ and }J_3\neq J_1
    \;,
  \end{equation}
  \item the constants of motion $\sigma, \sigma_3, \varepsilon$ do not assume their extremal values, i.~e.,
  \begin{eqnarray}
  \label{generic2}
    0 &<& S<3, \\
    \label{generic3}
    -S &<& \sigma_3 < S, \\
    \label{generic4}
    E_{\scriptstyle min}(\sigma) &<& \varepsilon<  E_{\scriptstyle max}(\sigma)
    \;,
  \end{eqnarray}
   \item  and the following cases are excluded:
  \begin{equation}\label{generic5}
    \mbox{if } \sigma=-1 \mbox{ then }\left(\varepsilon\neq J_1-J_2-J_3 \mbox{ and }
    \varepsilon\neq J_2-J_3-J_1 \mbox{ and } \varepsilon\neq J_3-J_1-J_2 \right)
    \;.
  \end{equation}
\end{itemize}
\end{defi}
The condition (\ref{generic1}) excludes the ``isosceles spin triangle" which will be separately treated in the Appendices
\ref{sec:IST} and \ref{sec:SCIST}. Condition (\ref{generic4}) has to be understood in the following sense:
If (\ref{generic1} - \ref{generic3}) is satisfied, then it can be shown that the possible energies $H(s)=\varepsilon$
under the constraint $H_1(s)=\sigma$ lie in a closed interval
$[  E_{\scriptstyle min}(\sigma),  E_{\scriptstyle max}(\sigma)]$ such that
$  E_{\scriptstyle min}(\sigma)<  E_{\scriptstyle max}(\sigma)$ and  (\ref{generic4}) excludes the endpoints of this interval.
Finally, condition (\ref{generic5}) excludes the cases where an aperiodic time evolution of the internal degrees of freedom occurs,
see Appendix \ref{sec:TSG}.\\

For the time derivative of the internal variables we obtain
\begin{eqnarray}
\label{ud1}
 \dot{u} &=&\frac{d}{dt}\left({\mathbf s}_2\cdot{\mathbf s}_3\right)= \dot{\mathbf s}_2  \cdot{\mathbf s}_3+{\mathbf s}_2\cdot\dot{\mathbf s}_3
  \\
 \label{ud2}
   &\stackrel{(\ref{eom2},\ref{eom3})}{=}&
   \left( \left(J_3 {\mathbf s}_1+J_1 {\mathbf s}_3 \right)\times {\mathbf s}_2\right) \cdot{\mathbf s}_3
   +{\mathbf s}_2\cdot\left( \left(J_1 {\mathbf s}_2+J_2 {\mathbf s}_1 \right)\times {\mathbf s}_3\right) \\
   \label{ud3}
  &=& \left( J_3-J_2\right)\left({\mathbf s}_1 \times {\mathbf s}_2\right)\cdot{\mathbf s}_3\\
    \label{ud4}
  &\stackrel{(\ref{deltasqrt},\ref{defdelta})}{=}&\pm\left( J_3-J_2\right)\, \sqrt{ 1-u^2-v^2-w^2+2 u v w}\\
  \label{ud5}
  &=&\left( J_3-J_2\right)\,\delta
    \;,
\end{eqnarray}
and, analogously,
\begin{eqnarray}
\label{vdot}
  \dot{v} &=& \left( J_1-J_3\right)\,\delta, \\
  \label{wdot}
   \dot{w} &=& \left( J_2-J_1\right)\,\delta
   \;.
\end{eqnarray}
In order to obtain a complete autonomous system of differential equations it is advisable to add the variable $\delta$
to the internal degrees of freedom. Since $\delta$ is, up to the sign of the square root, already a function of $u,v,w$, see (\ref{deltasqrt}),
this does not increase the number of internal degrees of freedom but only removes the sign ambiguity in, say,  (\ref{ud4}).
The geometric significance of this extension will be more closely analyzed in Appendix \ref{sec:TG}.
The time derivative of $\delta$ can be obtained by differentiating $\delta^2$ and using (\ref{deltasqrt}) as well as (\ref{ud5} - \ref{wdot}):
\begin{equation}\label{deltadot}
\dot{\delta}= J_1\,(u+1)(w-v)+J_2\,(v+1)(u-w)+J_3\,(w+1)(v-u)
\;.
\end{equation}
The system of differential equations (\ref{ud5} - \ref{deltadot}) is simplified by elimination of $v$ and $w$,
using that (\ref{Huvw}) and (\ref{H1uvw}) leads to :
\begin{eqnarray}
\label{u2v}
 v &=& \frac{J_3 (u-\sigma )+\varepsilon-J_1 u }{J_2-J_3}, \\
 \label{u2w}
 w &=& \frac{-J_2 (u-\sigma )-\varepsilon +J_1 u}{J_2-J_3}
 \;.
\end{eqnarray}
Note that $J_2\neq J_3$ due to the restriction to the generic case.
Instead of $u$ we will use the variable $x$ given by
\begin{equation}\label{defx}
x=x_0+g\,u
 \;,
\end{equation}
where the constants $x_0$ and $g$ will be determined later such that the
\textit{Weierstrass differential equation} \cite[23.3.10]{NIST21} is obtained.
Also $v$ and $w$ can be linearly expressed in terms of the variable $x$ in the form:
\begin{eqnarray}\label{x2v}
 v&=& \frac{J_3-J_1}{J_2-J_3}\frac{x}{g}+v_0,\\
 \label{x2w}
 w&=& \frac{J_1-J_2}{J_2-J_3}\frac{x}{g}+w_0
 \;.
\end{eqnarray}

Next we consider the time derivative of $x$:
\begin{eqnarray}
\label{xd1}
 \dot{x} &\stackrel{(\ref{defx})}{=}& g\,\dot{u}\\
 \label{xd2}
   &\stackrel{(\ref{ud4})}{=}&\pm g\,\left( J_3-J_2\right)\, \sqrt{ 1-u^2-v^2-w^2+2 u v w}\stackrel{(\ref{ud5})}{=}g\,\left( J_3-J_2\right)\,\delta
  \;.
\end{eqnarray}

By substituting (\ref{u2v}), (\ref{u2w}) and (\ref{defx}), the square of (\ref{xd2})
can be written as a $3^{rd}$ order polynomial $\Pi(x)$.
$g$ and $x_0$ will be chosen such that the cubic term
of $\Pi(x)$ reads $4\,x^3$ and the quadratic
term of $\Pi(x)$  vanishes and hence
\begin{equation}\label{sdPi}
\left( \frac{dx}{dt}\right)^2=  \dot{x}^2=g^2\,\left( J_3-J_2\right)^2\,\delta^2 =\Pi(x) = 4 x^3-g_2 x-g_3
\;.
\end{equation}
This is achieved by setting
\begin{equation}\label{solg}
  g=-\frac{1}{2} \left(J_1-J_2\right) \left(J_1-J_3\right)
  \;,
\end{equation}
and
\begin{equation}\label{solx0}
  x_0=\frac{1}{6} \left(-J_1^2-J_2^2-J_3^2+J_1 J_3+J_2 J_3+J_1 J_2+\left(2
   J_1-J_2-J_3\right) \epsilon +\left(2 J_2 J_3-J_1 \left(J_2+J_3\right)\right)
   \sigma \right)
  \;.
\end{equation}

The explicit form of the coefficients $g_2$ and $g_3$, let alone the roots of $\Pi(x)$,
is  more complicated and will only be given for an example, see Appendix \ref{sec:EX}.

\begin{figure}[th]
\centering
\includegraphics[width=0.7\linewidth]{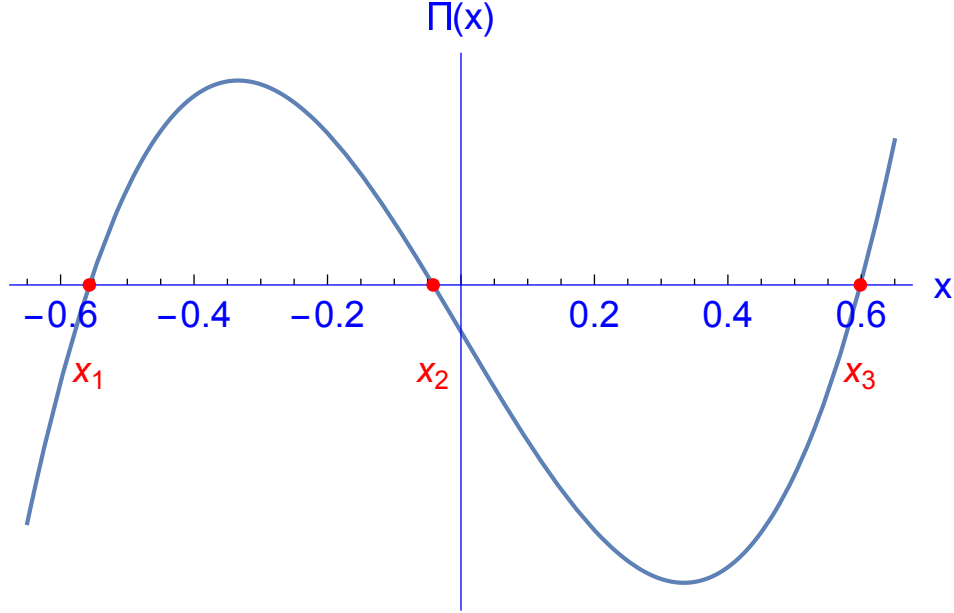}
\caption{Plot of the polynomial $\Pi(x)=4\,x^3-g_2\,x-g_3$ for the example (\ref{polyspec}) with three real roots $x_1,x_2$ and $x_3$
according to (\ref{zeroesPi1}-\ref{zeroesPi3}).
}
\label{FIGPP}
\end{figure}

\begin{figure}[th]
\centering
\includegraphics[width=0.7\linewidth]{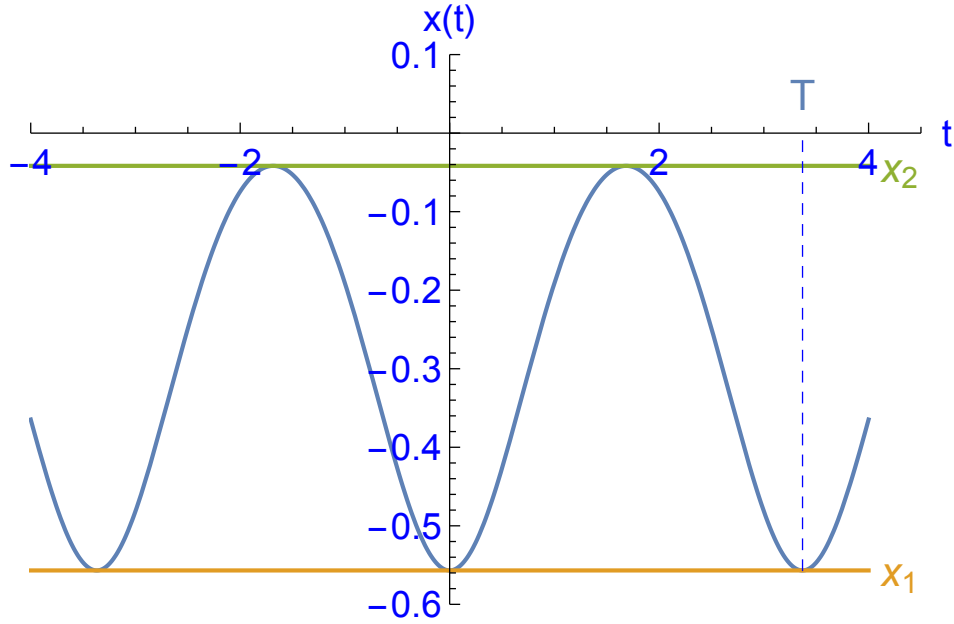}
\caption{Plot of the periodic solution $x(t)$ according to (\ref{Wei}) and corresponding to the polynomial $\Pi(x)$ shown in Figure \ref{FIGPP}.
It oscillates between the zeroes $x_1$ and $x_2$ of $\Pi(x)$.
The value of the period ${\sf T}\approx 3.3693$ follows from (\ref{period1}) and (\ref{Tspec}).
}
\label{FIGXT}
\end{figure}

It can be shown that in the generic case the polynomial $\Pi(x)$ has three real simple
roots $x_1< x_2<x_3$ such that $\Pi(x)>0$ for $x_1<x<x_2$,
see Proposition \ref{PropGeneric} in Appendix \ref{sec:TG} and Figure \ref{FIGPP} for an example.
Let us first consider the case $\dot{x}>0$, then (\ref{sdPi}) can be solved for $dt$ and integrated:
\begin{equation}\label{int1}
  t= \int dt = \int_{x_1}^x \frac{dx'}{\sqrt{ 4 x'^3-g_2 x'-g_3}}
  \;,
\end{equation}
for $x_1\le x\le x_2$. The choice of the lower limit of the integral (\ref{int1}) implies $x=x_1$ for $t=0$.
Solutions obtained for these special initial conditions can be generalized by using the time translation symmetry of (\ref{eom1} - \ref{eom3})
due to the Hamiltonian (\ref{defH}) not being explicitly time-dependent.
The inverse $x(t)$ of the elliptical integral (\ref{int1}) can be expressed in terms of the
\textit{Weierstrass elliptic function}, see, e.~g., \cite[Ch.23]{NIST21}.
Recall that the  Weierstrass elliptic function $\wp(z;g_2,g_3)$ of complex arguments $z$ has two periods $2\omega_1$ and $2\omega_3$.
As mentioned above, in the generic case the polynomial $\Pi(x)=4 x^3-g_2 x-g_3$ has three real simple roots (``rectangular case" \cite[\S 23.5(ii)]{NIST21})
and hence $\omega_1$ will be real and $\omega_3$ imaginary. In this case (\ref{int1}) is equivalent to
\begin{equation}\label{Wei}
 x(t)= \wp(t+\omega_3; g_2,g_3)
  \;,
\end{equation}
see \cite[23.6.32]{NIST21}. Moreover, the two periods of the Weierstrass elliptic function mentioned above are given by the elliptic
integrals
\begin{equation}\label{period1}
\frac{\sf T}{2}:=\omega_1=\int_{x_1}^{x_2}\frac{dx'}{\sqrt{ 4 x'^3-g_2 x'-g_3}}=\int_{x_3}^{\infty}\frac{dx'}{\sqrt{ 4 x'^3-g_2 x'-g_3}}
=\frac{1}{\sqrt{x_3-x_1}}\,K\left(\frac{x_2-x_1}{x_3-x_1}\right)
\;,
\end{equation}
and
\begin{equation}\label{period2}
t_0:=\omega_3={\sf i}\,\int_{x_2}^{x_3}\frac{dx'}{\sqrt{ \left|4 x'^3-g_2 x'-g_3\right|}}
={\sf i}\,\int_{-\infty}^{x_1}\frac{dx'}{\sqrt{\left| 4 x'^3-g_2 x'-g_3\right|}}
=\frac{\sf i}{\sqrt{x_3-x_1}}\,K\left(\frac{x_3-x_2}{x_3-x_1}\right)
\;,
\end{equation}
see  \cite[23.6.34-35]{NIST21} and \cite[17.4.61 ff]{AS72}.

To calculate the oscillation of $x(t)$ in the limit $|x_2- x_1|\to 0$ the polynomial $\Pi(x)$ can be approximated by a quadratic one,
\begin{equation}\label{limitdoubleroot}
 \Pi(x) =\Pi(x_a)+\frac{1}{2} \left.\frac{\partial^2 \Pi(x)}{\partial x^2}\right|_{x=x_a}\left(x-x_a\right)^2+ O(\left(x-x_a\right)^3)=
 \Pi(x_a)+12 x_a\left(x-x_a\right)^2+ O(\left(x-x_a\right)^3)
 \;,
\end{equation}
where $x_a= -\sqrt{\frac{g_2}{12}}$ is the $x$-coordinate of the local maximum of $\Pi(x)$ satisfying $x_1<x_a<x_2$
and $\Pi'(x_a)=0$.
From the analogy with the harmonic oscillator problem, namely
\begin{equation}\label{harmosc}
\dot{x}^2=\Pi(x) \approx \Pi(x_a)+ 12 x_a\left(x-x_a\right)^2=2\left( E-V(x)\right)=2\left( E-\frac{\omega^2}{2}\left(x-x_a\right)^2\right)
\;,
\end{equation}
we obtain in the limit case a harmonic oscillation of $x(t)$ with the period
\begin{equation}\label{limperiod}
 \lim_{x_1\to x_2}{\sf T} = \frac{2\pi}{\omega} =  \frac{2\pi}{\sqrt{12 |x_a|}}
 \;.
\end{equation}

Although we have derived the solution of the form (\ref{Wei}) under the assumption $\dot{x}>0$
it turns out that (\ref{Wei}) can be extended to all real values of $t$ and will be a periodic function
with period ${\sf T}$, see Figure \ref{FIGXT} for an example.

Upon re-substituting (\ref{u2v}),  (\ref{u2w}), and  (\ref{defx}) we obtain for the time dependence of the scalar products
of the spin vectors:
\begin{eqnarray}
\label{uvwt1}
  u(t) &=& \frac{1}{g}\left({\wp \left(t+t_0;g_2,g_3\right)-x_0} \right),\\
  \label{uvwt2}
  v(t) &=& \frac{1}{g \left(J_2-J_3\right)}
 \left( \left(J_3-J_1\right) \wp \left(t+t_0;g_2,g_3\right)-J_3 \left(g \sigma +x_0\right)+g \varepsilon +J_1   x_0
 \right),\\
 \label{uvwt3}
 w(t)&=&\frac{1}{g \left(J_2-J_3\right)}
 \left(\left(J_1-J_2\right) \wp \left(t+t_0;g_2,g_3\right)+J_2\left(g \sigma + x_0\right)-g \varepsilon -J_1 x_0
 \right)
 \;.
\end{eqnarray}

\begin{figure}[htp]
\centering
\includegraphics[width=0.7\linewidth]{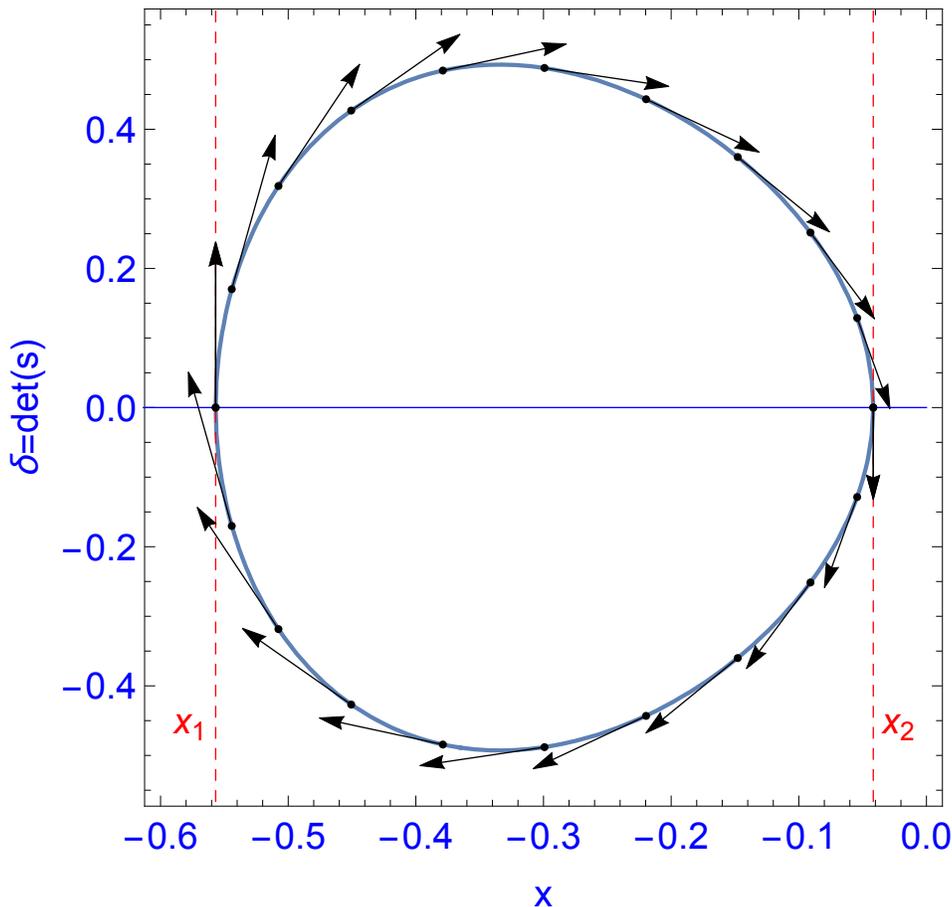}
\caption{Plot of the time evolution of the spin configuration $s(t)$  projected onto the plane with coordinates
$(x, \delta=\det(s))$ for the example of Appendix \ref{sec:EX}. This orbit is traversed in a clockwise direction,
since $\dot{x}>0$ for positive $\det(s)$. The arrows correspond to velocities $(\frac{d}{dt}x,\frac{d}{dt}\det(s))$
calculated according to (\ref{xd2}) and (\ref{deltadot}). The shown orbit is the closed part of the elliptic curve
$\delta^2=\Pi(x)$. }
\label{FIGAA1}
\end{figure}

By inserting these functions into $\delta(t)=\pm\sqrt{1-u^2(t)-v^2(t)-w^2(t)+2\,u(t)\,v(t)\,w(t)}$
and choosing the sign $\pm$ such that $t\mapsto \delta(t)$ will be a smooth function satisfying (\ref{deltadot})
we can lift the solution (\ref{Wei}) to a solution of (\ref{ud5} - \ref{deltadot}) for all $t\in {\mathbbm R}$,
see Figure \ref{FIGAA1} for a two-dimensional projection.

\subsection{Time evolution of the external degrees of freedom}\label{sec:TED}

Recall that the total spin vector ${\mathbf S}(t)$ is a constant of motion
with length $S=\sqrt{ 3+2 \sigma }$.
To simplify the presentation we will, only for this subsection, choose the coordinate system in spin space in such a way that
\begin{equation}\label{choice1}
 {\mathbf S}(t) ={\mathbf S}_0 :=
 \left(
\begin{array}{c}
 0\\
 0 \\
\sqrt{ 3+2 \sigma }\\
\end{array}
\right),
\quad \mbox{for all }t\in{\mathbbm R}
\;.
\end{equation}

We will define a ``standard  configuration"
$r(u,v,w,\delta)=\left({\mathbf r}_1, {\mathbf r}_2,{\mathbf r}_3\right)\in{\mathcal P}$
that realizes the internal degrees of freedom $(u,v,w,\delta)$
such that ${\mathbf R}:={\mathbf r}_1 + {\mathbf r}_2+{\mathbf r}_3={\mathbf S}_0$, as can be verified by a straightforward calculation.
\begin{equation}\label{defr1}
 {\mathbf r}_1=\left(
\begin{array}{c}
 \sqrt{\frac{2 (u+1)-(v+w)^2}{3+2( u+ v+ w)}} \\
 0 \\
 \frac{v+w+1}{\sqrt{3+2( u+ v+ w)}} \\
\end{array}
\right)\;,
\end{equation}

\begin{equation}\label{defr2}
 {\mathbf r}_2=\left(
\begin{array}{c}
 \frac{w (u+v+1)-(u+1) (v+1)+w^2}{ \sqrt{(2 (u+1)-(v+w)^2)(3+2( u+ v+ w)})} \\
 \frac{\delta}{\sqrt{2(u+1)-(v+w)^2}} \\
 \frac{u+w+1}{\sqrt{3+2 (u+v+w)}} \\
\end{array}
\right)\;,
\end{equation}

\begin{equation}\label{defr3}
 {\mathbf r}_3=\left(
\begin{array}{c}
 \frac{v (w+u+1)-(w+1) (u+1)+v^2}{ \sqrt{(2 (u+1)-(v+w)^2)(3+2( u+ v+ w)})} \\
 - \frac{\delta}{\sqrt{2(u+1)-(v+w)^2}} \\
 \frac{u+v+1}{\sqrt{3+2 (u+v+w)}} \\
\end{array}
\right)\;.
\end{equation}
For the details of the domain of definition of the ${\mathbf r}_\mu,\;\mu=1,2,3,$ see Appendix \ref{sec:TP}.

The time-dependent spin configuration $r(t)$ is obtained by substituting
$(u(t),v(t),w(t),\delta(t)$ according to (\ref{uvwt1}-\ref{uvwt3}) for $(u,v,w,\delta)$.
In general, $r(t)$ will not solve the equations of motion (\ref{eom1} - \ref{eom3}),
but $r(t)$ is not the only spin configuration compatible with $(u(t),v(t),w(t),\delta(t))$.
Note that
\begin{equation}\label{ansatz}
 s(t) = Z(t)\, r(t)
 \;,
\end{equation}
where $Z(t)$ is a smooth family of rotations, realizes the same internal degrees of freedom as $r(t)$.
Hence we will seek for a differential equation for $Z(t)$ which implies that
$s(t)$ solves (\ref{eom1} - \ref{eom3}). First, note that $\det s(t)=\delta(t)=\det r(t)$ implies $\det Z(t)=1$,
and hence $Z(t)\in SO(3)$. Second, since ${\mathbf R}(t)= {\mathbf S}(t)={\mathbf S}_0$ the rotation $Z(t)$ must leave ${\mathbf S}_0$
invariant and hence $Z(t)\in {\mathcal R}\left({\mathbf S}_0 ,\alpha(t) \right)$ for some smooth function $t\mapsto \alpha(t)$,
i.~e.,
 \begin{equation}\label{formZ}
 Z(t)= \left(
\begin{array}{ccc}
 \cos \alpha(t)&-\sin \alpha(t) & 0 \\
 \sin \alpha(t) & \cos \alpha(t) & 0 \\
 0& 0 & 1 \\
\end{array}
\right)
\;.
 \end{equation}

The choice of the coordinate system leading to (\ref{choice1}) leaves the freedom of a fixed rotation about the $3$-axis. This
freedom can be used to achieve the inial value
\begin{equation}\label{choice2}
r(0)=s(0)
\;,
\end{equation}
and hence  $Z(0)={\mathbbm 1}$ .
(\ref{formZ}) implies
 \begin{equation}\label{Zt}
   \frac{d}{dt}Z(t)=\Omega(t)\, Z(t)= Z(t)\,  \Omega(t)
   \;,
 \end{equation}
 where $\Omega(t)$ is the anti-symmetric ``angular velocity  matrix"
 \begin{equation}\label{defOmega}
  \Omega(t)= \dot{\alpha}(t)\,\left(
\begin{array}{ccc}
0&-1 & 0 \\
1 & 0 & 0 \\
 0& 0 &0 \\
\end{array}
\right)
\;.
 \end{equation}
 Hence
 \begin{equation}\label{sdot1}
   \dot{s}\stackrel{(\ref{ansatz})}{=}\dot{Z}r+Z\dot{r}\stackrel{(\ref{Zt})}{=}Z\left( \Omega+\dot{r}\right)
   \;,
 \end{equation}
and further
\begin{equation}\label{sdotminusJ}
   \dot{s}-{\mathcal J}(s)\stackrel{(\ref{sdot1},\ref{ansatz})}{=}Z\left( \Omega+\dot{r}\right)-{\mathcal J}(Z r)\stackrel{(\ref{RopJ})}{=}Z\left( \Omega+\dot{r}-{\mathcal J}(r)\right)
   \;.
\end{equation}
Thus the equation of motion $0= \dot{s}-{\mathcal J}(s)$ is equivalent to
\begin{equation}\label{eomext}
 0= \Omega\,r +\dot{r}-{\mathcal J}(r)
 \;,
\end{equation}
which is the differential equation for $Z(t)$ we were looking for.
In the case of an invertible $r$ the solution of (\ref{eomext}) is given by
\begin{equation}\label{soleomext}
 \Omega(t)= \left( {\mathcal J}(r(t))-\dot{r}(t)\right) r^{-1}(t)
 \;,
\end{equation}
and can be extended to $t$ being an integer multiple of $\frac{\sf T}{2}$,
where $r(t)$ is co-planar and hence not invertible, by means of continuity,
see Appendix \ref{sec:AR} for the details.
The rotation matrix $Z(t)$ is then given by an integral over $t$:
Taking into account the form of $\Omega(t)$ according to (\ref{defOmega}) we obtain
\begin{equation}\label{intalpha}
 \alpha(t)=\int_{0}^{t}\dot{\alpha}(t')\,dt'=\int_{0}^{t}\left(  \left( {\mathcal J}(r(t'))  -\dot{r}(t')\right) \, r(t')^{-1}\right)_{2,1}\,dt'
 \;,
\end{equation}
where the initial value $\alpha(0)=0$ follows from (\ref{choice2}).
Eq.~(\ref{intalpha}) yields $Z(t)$ and, finally, $s(t)=Z(t)\, r(t)$ satisfying  (\ref{eom1} - \ref{eom3}) for all $t\in{\mathbbm R}$.\\

The dependence on the special coordinate system leading to (\ref{choice1}) can be removed
by replacing $r(t)$ by  $\widetilde{r}(t):=R\,r(t)$, where $R$ is an arbitrary fixed rotation $R\in SO(3)$.
Then the corresponding solution for the spin configuration $s(t)$
can be written as
\begin{equation}\label{solext}
 s(t)={\mathcal R}\left( {\mathbf S},\alpha(t)\right)\,\widetilde{r}(t)
 \;,
\end{equation}
without using the auxiliary coordinate system introduced in this Section.
\\

For later use in Appendix \ref{sec:AA} we note the following additivity property:
\begin{equation}\label{alphadd}
  \alpha(t+{\sf T})=\alpha(t)+\alpha({\sf T})\quad \mbox{ for all } t\in{\mathbbm R}
  \;.
\end{equation}
The {\bf proof} is straightforward since $\dot{\alpha}(t)$ is a ${\sf T}$-periodic function according to (\ref{intalpha}).\\

This concludes the solution of the equation of motion for the spin triangle.
An example comparing the semi-analytical solution with a numerical integration of (\ref{eom1} - \ref{eom3})
will be given in Appendix \ref{sec:EX}.

\section{Summary and Outlook}\label{sec:SO}

We have identified the classical spin triangle as a completely integrable mechanical system due to
its $6$-dimensional phase space ${\mathcal P}$ and three first integrals $H, S^2, {\mathbf S}^{(3)}$ in involution. Moreover,
we have explicitly calculated the time evolution that can be related to the equations of motion for the action-angle variables,
see Appendix \ref{sec:AA}.

Among all classical spin systems this is a rare case; mostly one has to resort to numerical integrations, see, e.~g., \cite{KBL98}.
However, this special example gives us the possibility to understand the motion of three spins in more detail, although
our physical intuition is rather focused on systems of particles with position and momentum.
In addition to being completely integrable, the spin triangle has the pleasant property that its interior degrees of freedom $(u,v,w,\delta)$
evolve autonomously, i.~e., independently of the remaining three exterior degrees.
This fortunate circumstance is closely related to the fact that the Heisenberg Hamiltonian is bilinear and isotropic
and implies that $\dot{u}, \dot{v},\dot{w}$ will be linear functions of the triple
product of the three spins, the latter being the square root of a cubic polynomial in $u,v,w$. Moreover, since the two
constants of motion $H$ and $S^2$ are \textit{linear} in $u,v,w$ all three internal variables can be written as linear function
of a single variable $x$ satisfying the Weierstrass differential equation $\dot{x}^2=4x^3-g_2 x-g_3$. Thus the internal degrees
of freedom perform a periodic oscillation analogous to the motion of a one-dimensional
particle governed by $\dot{x}^2= \frac{2}{m}\left(E-V(x) \right)$. At the turning points of this oscillation, i.~e., for $\dot{x}=0$,
the three-dimensional spin configuration degenerates into a co-planar one. This part of the time evolution
can be visualized by the closed curve, see Figure \ref{FIGAA1}.

The time evolution of the remaining three external degrees of freedom can be facilitated by considering
the time-dependent standard configuration $r(t)$,  see (\ref{defr1} - \ref{defr3}),
that correctly reproduces the total spin ${\mathbf S}$. Hence the solution of the equation of motion $s(t)$ differs from
$r(t)$ only by a rotation $Z(t)$ about  ${\mathbf S}$ and can be calculated in terms of an integral.

So far the exact time evolution has been obtained for a vanishing external magnetic field ${\mathbf B}={\mathbf 0}$.
The latter can be included without further problems for the case of a time-varying field ${\mathbf B}(t)=B(t)\,{\mathbf e}$
with constant direction ${\mathbf e}$, see Appendix \ref{sec:MF}.
In the special case of periodically varying $B(t)$ there occur interesting overlaps
with Floquet theory that can be already applied in the case of  ${\mathbf B}={\mathbf 0}$, see Appendix \ref{sec:FT},
and will be further analyzed in the future.
Another open problem is to extend the present work, which is confined entirely to classical mechanics,
to classical statistical mechanics and to calculate quantities such as the specific heat,
the susceptibility, and the autocorrelation function of spin for the general spin triangle,
analogous to the calculations for the equilateral spin triangle in \cite{CLAL99}.

\section*{Acknowledgment}
I am deeply indebted to Hans Werner Sch\"urmann for intensive discussions on earlier versions of the paper
and especially for valuable hints on the theory of Weierstrass elliptic functions.

\appendix

\section{Time evolution with an exterior magnetic field}\label{sec:MF}
For completeness we mention the well-known fact that the time evolution with a time-dependent
exterior magnetic field ${\mathbf B}(t) =B(t)\,{\mathbf e}$ into a constant direction ${\mathbf e}$ can be reduced
to the equations  of motion (\ref{eom1} - \ref{eom3}) in a straightforward way.
The modified Hamiltonian will be (\ref{defH}) plus a \textit{Zeeman term}
\begin{equation}\label{HamZeeman}
 H_B=\sum_{\mu<\nu}J_{\mu\nu} {\mathbf s}_\mu\cdot {\mathbf s}_\nu+ {\mathbf B}(t)\cdot {\mathbf S}
 \;,
\end{equation}
which entails the modified equations of motion
\begin{eqnarray}\label{meom1}
  \dot{\mathbf s}_1&=& \left( J_2 \,{\mathbf s}_3+ J_3 \,{\mathbf s}_2+{\mathbf B}(t)\right)\times {\mathbf s}_1,\\
  \label{meom2}
  \dot{\mathbf s}_2&=& \left( J_3 \,{\mathbf s}_1+ J_1 \,{\mathbf s}_3+{\mathbf B}(t)\right)\times {\mathbf s}_2,\\
  \label{meom3}
  \dot{\mathbf s}_3&=& \left( J_1 \,{\mathbf s}_2+ J_2 \,{\mathbf s}_1+{\mathbf B}(t)\right)\times {\mathbf s}_3
  \;.
\end{eqnarray}
Let ${\sf B}(t)=B(t)\,{\sf E}$ be the anti-symmetric $3\times 3$-matrix satisfying
\begin{equation}\label{Banti}
{\sf B}(t)\,{\mathbf a} = {\mathbf B}(t)\times {\mathbf a}\quad \mbox{ for all } {\mathbf a}\in{\mathbbm R}^3
\;,
\end{equation}
and $R(t)\in SO(3)$ a time-dependent rotation satisfying  the differential equation
\begin{equation}\label{defRot}
 \dot{R}(t)=-{\sf B}(t)\,R(t)
 \;,
\end{equation}
with initial condition $R(0)={\mathbbm 1}$.
$R(t)$ will be a rotation about the axis ${\mathbf e}$ with an angle $\beta(t)$ given by
\begin{equation}\label{alphaint}
 \beta(t)=-\int_0^t B(t')\,dt'
 \;.
\end{equation}
Hence $R(t)\,{\mathbf B}={\mathbf B}$. Then the spin vectors ${\mathbf s}_\mu$, transformed into the rotating frame, i.~e.,
\begin{equation}\label{defsmu}
{\mathbf s}_\mu':= R(t)\,{\mathbf s}_\mu,\quad \mbox{ for } \mu=1,2,3,
\end{equation}
satisfy the original equations of motion
\begin{eqnarray}
\label{meomu}
 \dot{\mathbf s}_\mu' &=&R\,\dot{\mathbf s}_\mu +\dot{R}\,{\mathbf s}_\mu \\
   &\stackrel{(\ref{meom1}-\ref{meom3},\ref{defRot})}{=}& \sum_{\nu}J_{\mu\nu}R\left({\mathbf s}_\nu\times {\mathbf s}_\mu\right)+
   R\,\left({\mathbf B}\times {\mathbf s}_\mu\right)- {\sf B}\,R\,{\mathbf s}_\mu\\
  &\stackrel{(\ref{defsmu})}{=}& \sum_{\nu}J_{\mu\nu}{\mathbf s}_\nu'\times {\mathbf s}_\mu' +{\mathbf B}\times {\mathbf s}_\mu'- {\sf B}\,{\mathbf s}_\mu'\\
   &\stackrel{(\ref{Banti})}{=}& \sum_{\nu}J_{\mu\nu}{\mathbf s}_\nu'\times {\mathbf s}_\mu'
   \;,
\end{eqnarray}
without any magnetic field. In other words, the influence of an external magnetic field into the constant direction ${\mathbf e}$
can be taken into account by a rotation about the axis ${\mathbf e}$ with time-dependent angular velocity.
Hence we will neglect this modification in what follows except for some remarks in Appendix \ref{sec:AA}.

\section{Geometry of spin configurations and time evolution}\label{sec:G}
An important problem was left open in the first part of this paper, namely, whether the analysis in terms of Weierstrass elliptic function
in section \ref{sec:TID} completely covers the generic case.
This problem is solved in this Appendix using a geometric approach.
A second issue to be addressed is a more detailed analysis of the definition
of the ``standard configuration'' $r(u,v,w,\delta)$ and of the mean rotation $\alpha({\sf T})$ in section \ref{sec:TED}.

\subsection{${\mathcal P}$ as a symplectic manifold }\label{sec:SM}
Mathematically, a ``phase space" of classical mechanics can be characterized as a ``symplectic manifold", see, e.~g., \cite[Ch.~8]{A78}.
This means that it is a smooth manifold equipped with a symplectic form $\omega$, i.~e., a non-degenerate closed $2$-form.
The most frequent case is the phase space as the cotangent bundle
$T^\ast {\mathcal V}$ which carries a natural symplectic form and  ${\mathcal V}$ being a ``configuration manifold".
Our case of a compact phase space ${\mathcal P}={\mathcal S}^2\times\dots\times{\mathcal S}^2$ describing $N$ classical spins is different.
To define $\omega$ we may use the embedding ${\mathcal S}^2\subset{\mathbbm R}^3$ and
identify a vector field $U$ at the point $s\in{\mathcal P}$ with an N-tuple of $3$-dimensional tangent vectors
$U(s)=\left({\mathbf U}_1, \ldots,{\mathbf U}_N\right)$ satisfying ${\mathbf U}_\mu\cdot {\mathbf s}_\mu=0$ for $\mu=1,\ldots,N$.
Recall that a $2$-form  $\omega$ maps a pair of smooth vector fields $U,V$ onto a function $\omega(U,V):{\mathcal P}\rightarrow{\mathbbm R}$.
Then the symplectic form $\omega$ can be defined by means of the usual scalar triple product as
\begin{equation}\label{defomega}
 \omega\left( U,V\right)(s):=\sum_\mu{\mathbf s}_\mu\cdot\left( {\mathbf U}_\mu\times {\mathbf V}_\mu\right)
 \;.
\end{equation}
Obviously, $\omega$ is invariant under global rotations, i.~e., $\omega(R\,U,R\,V)(R\,s)=\omega(U,V)(s)$ for all $R\in SO(3)$.
In terms of the chart $\left(z_\mu, \phi_\mu\right)_{\mu=1,\ldots,N}$, defined by (\ref{defcan}), the symplectic form $\omega$
has the Darboux representation of $\sum_\mu d\phi_\mu\wedge d z_\mu$  (a sum of surface elements)
and hence this chart yields an example of ``canonical coordinates".

Recall that to every smooth ``Hamilton" function ${\sf H}:{\mathcal P}\rightarrow{\mathbbm R}$ there belongs a ``Hamiltonian vector field" $X$
satisfying
\begin{equation}\label{hameq}
\omega(X,U)=d{\sf H}(U)
\end{equation}
for all vector fields $U$, see \cite{A78} or \cite[Sect.~3.2]{AM78}. $X$ is uniquely determined by (\ref{hameq}) due to $\omega$ being non-degenerate.
For the special case of the Heisenberg Hamiltonian (\ref{defH}) it will be shown that (\ref{hameq}) is satisfied
with the Hamiltonian vector field
\begin{equation}\label{hamvec}
  X_\mu(s)=\sum_\nu J_{\mu\nu}\,{\mathbf s}_\nu \times {\mathbf s}_\mu,\quad\mbox{ for }\mu=1,\ldots,N
  \;,
\end{equation}
corresponding to the equations of motion (\ref{LL}). We state this result as
\begin{prop}\label{PropHam}
 The Hamiltonian $H(s)= \sum_{\mu<\nu}J_{\mu\nu} {\mathbf s}_\nu \cdot {\mathbf s}_\mu $ gives rise to the Hamiltonian vector field (\ref{hamvec}).
\end{prop}
{\bf Proof}:
The corresponding calculation can be performed in the larger space
$\left({\mathbbm R}^3\right)^N\supset{\mathcal P}$ if we only take into account that all considered vector fields are tangent to ${\mathcal P}$.
First we obtain
\begin{equation}\label{dham}
dH(U)(s)= d\left( \sum_{\mu<\nu} J_{\mu\nu} {\mathbf s}_\nu \cdot {\mathbf s}_\mu \right)(U)(s)
= \sum_{\mu\nu} J_{\mu\nu}{\mathbf s}_\nu \cdot d({\mathbf s}_\mu)({\mathbf U}_\mu) =\sum_{\mu\nu} J_{\mu\nu}{\mathbf s}_\nu \cdot {\mathbf U}_\mu
\;.
\end{equation}
Here we have used
\begin{equation}\label{dsU}
   d({\mathbf s}_\mu^i)({\mathbf U}_\mu)=  d({\mathbf s}_\mu^i)\left(\sum_j {\mathbf U}_\mu^j \frac{\partial}{\partial {\mathbf s}_\mu^j} \right)
   = \sum_j  {\mathbf U}_\mu^j \delta_{ij}={\mathbf U}_\mu^i
   \;,
\end{equation}
for $i=1,2,3$ since $\left(\frac{\partial}{\partial {\mathbf s}_\mu^j} \right)_{j=1,2,3}$ is a basis in the tangent space $T_{{\mathbf s}_\mu}{\mathbbm R}^3$
with dual basis $ \left( d({\mathbf s}_\mu^i)\right)_{i=1,2,3}$ in the dual tangent space $T_{{\mathbf s}_\mu}^\ast{\mathbbm R}^3$ for $\mu=1,\ldots,N$.
Further,
\begin{eqnarray}
\label{omux1}
  \omega(X,U)(s) &\stackrel{(\ref{defomega},\ref{hamvec})}{=}&
  \sum_\mu {\mathbf s}_\mu\cdot\left( \sum_\nu J_{\mu\nu}\left({\mathbf s}_\nu \times {\mathbf s}_\mu \right)\times{\mathbf U}_\mu \right)\\
  \label{omux2}
  &=& \sum_{\mu\nu}J_{{\mu\nu}}{\mathbf s}_\mu \cdot
  \left({\mathbf s}_\mu {\mathbf U}_\mu\cdot {\mathbf s}_\nu- {\mathbf s}_\nu
  \underbrace{{\mathbf U}_\mu\cdot {\mathbf s}_\mu }_{=0}\right)\\
  \label{omux3}
  &=& \sum_{\mu\nu}J_{{\mu\nu}}{\mathbf s}_\nu \cdot {\mathbf U}_\mu
  \;.
\end{eqnarray}
Comparison of (\ref{omux3}) and (\ref{dham})
proves that (\ref{hamvec}) is the correct Hamiltonian vector field corresponding to the Heisenberg Hamiltonian (\ref{defH}).
 \hfill$\Box$\\

\subsection{Polar decomposition and Gram set}\label{sec:PG}

\begin{figure}[th]
\centering
\includegraphics[width=0.7\linewidth]{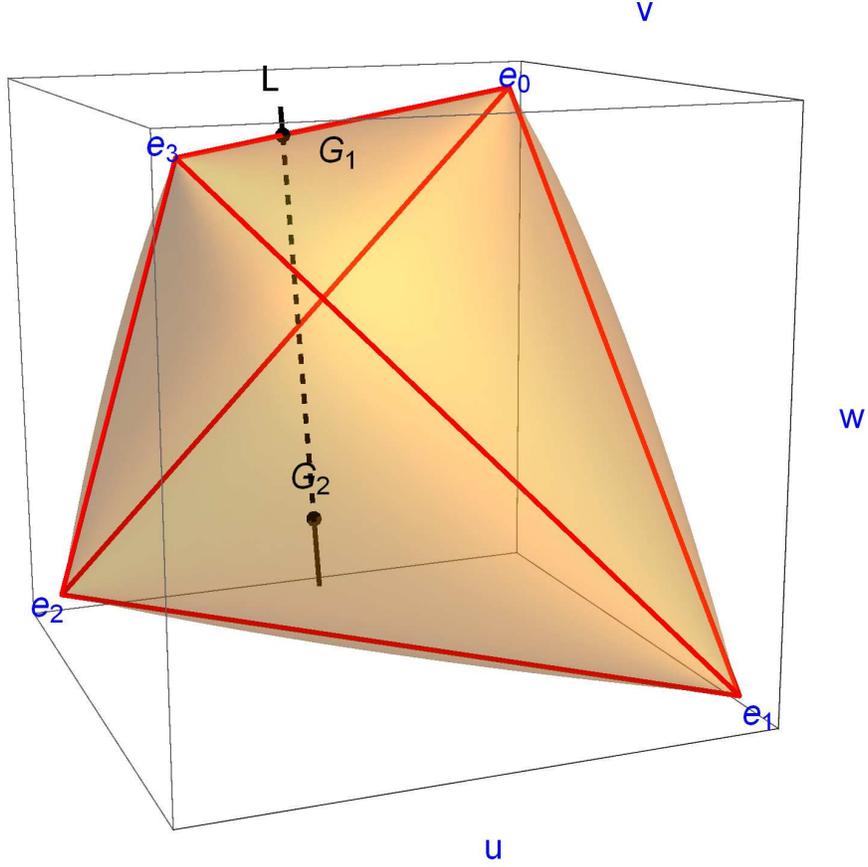}
\caption{Plot of the convex Gram set ${\mathcal G}$ defined in (\ref{defGS}) with four singular extremal points
${\mathbf e}_0,{\mathbf e}_1,{\mathbf e}_2,{\mathbf e}_3$.
Its six one-dimensional faces forming a tetrahedron are displayed as red lines.
The black (dashed) line $L$, representing spin configurations where the two conserved quantities
(\ref{Huvw}) and (\ref{H1uvw}) assume definite values, intersects the boundary of ${\mathcal G}$ at two points $G_1$ and $G_2$,
see also Figure \ref{FIGCP1} for a two-dimensional section.
}
\label{FIGGRAM}
\end{figure}

A spin configuration $s=\left({\mathbf s}_1,{\mathbf s}_2,{\mathbf s}_3\right)\in{\mathcal P}$ can be visualized as a spherical triangle.
It is characterized by its \textit{form} and \textit{position}. We define two configurations $s$ and $s'$ as \textit{congruent}
iff $s$ can be transformed into $s'$ by means of a $3$-dimensional rotation/reflection, i,~e., iff there exists an $R\in O(3)$ such
that $s'=R\,s$. Clearly, congruence is an equivalence relation and the \textit{form} of $s$ is defined as its corresponding equivalence class.

To define the \textit{position} of $s$ more precisely we consider the (right) polar decomposition of the real matrix $s$ of the form:
\begin{equation}\label{polar}
 s= R\,p
 \;,
\end{equation}
where $R\in O(3)$ and $p$ is \textit{positively semi-definite}, i.~e.,
a symmetric $3\times 3$-matrix with only non-negative eigenvalues, in symbols $p\ge 0$.
$p$ is uniquely determined as the positively semi-definite square root of
\begin{equation}\label{defG}
G:=s^\top\,s= \left( p\,R^\top \right)\left( R \,p\right)=p\,\underbrace{R^\top\,R}_{\mathbbm 1} p=p^2
\;.
\end{equation}
In our case the three columns of $s$ are unit vectors and hence also the three columns of $p$ are so.
This means that $p$ can be considered as another spin configuration that is congruent to $s$,
i.~e., as a uniquely defined representative of the equivalence class of $s$.
Therefore, it is meaningful to define the rotation/reflection $R$ of the polar decomposition
(\ref{polar}) as the \textit{position} of $s$. In general $R$ is not unique,
but if $s$ is an invertible matrix, it is.

The matrix $G\ge 0$ defined in (\ref{defG}) can be considered as the \textit{Gram matrix} of the spin configuration $s$,
sometimes also denoted as $G(s)$, see also \cite{SL03,S17a,S17b}.
It has the entries
\begin{equation}\label{entG}
  G_{\mu\nu}= {\mathbf s}_\mu\cdot {\mathbf s}_\nu \quad \mbox{for } \mu,\nu=1,2,3,
\end{equation}
and hence $G_{\mu\mu}=1$ for $\mu=1,2,3$. Conversely, every real, symmetric, positively semi-definite $3\times 3$-matrix $G$ with
a unit diagonal is the Gram matrix of some spin configuration $s$ (choose, for example, $s=\sqrt{G}$).
Since the Gram matrix $G$ is in $1:1$ relation to its square root $p$, it can also be taken as a representative of the form of $s$.
The set ${\mathcal P}$ of spin configurations $s$ can be divided into three disjoint parts
${\mathcal P}={\mathcal P}_1 \uplus{\mathcal P}_2 \uplus {\mathcal P}_3$
according to the matrix rank of $G$:
There are three-dimensional configurations corresponding to $\mbox{rank } G=3$, forming ${\mathcal P}_3$,
co-planar configurations corresponding to $\mbox{rank } G=2$, forming ${\mathcal P}_2$, and, finally,
collinear configurations corresponding to $\mbox{rank } G=1$, forming ${\mathcal P}_1$.
These three cases correspond to $G$ having three positive eigenvalues, two positive and one zero eigenvalues, and
one positive and two zero eigenvalues, resp.~.
For later use, we define $\dot{\mathcal P}:={\mathcal P}_2 \uplus {\mathcal P}_3$
as the phase space without collinear spin configurations. It is an open subset of ${\mathcal P}$.
The neglect of the collinear configurations is all the more reasonable since we will show in Appendix \ref{sec:TSG}
that they cannot be reached in finite times from the other points of ${\mathcal P}$,
but, if at all, asymptotically in the limit $t\to\pm \infty$.

Let the  \textit{Gram set} ${\mathcal G}$  be the set of Gram matrices of three-dimensional spin configurations, i.~e.,
\begin{equation}\label{defGS}
{\mathcal G}:=\left \{ \left.G=  \left(
\begin{array}{ccc}
 1 & w & v \\
 w & 1 & u \\
 v & u & 1 \\
\end{array}
\right)
\right|\;-1\le u,v,w\le 1 \mbox{ and } \det(G)= 1-u^2-v^2-w^2+2 u v w\ge 0\right\}
\;.
\end{equation}
Note that all matrices $G\in{\mathcal G}$ are positively semi-definite by virtue of Sylvester's criterion, namely that
all principal minors of $G$ have to be non-negative: $\det G \ge 0$ by definition, and $G_{\mu\mu}=1\ge 0$ for $\mu=1,2,3$.
The determinant of the upper left $2\times 2$-submatrix of $G$ is $1-w^2\ge0$ and analogous for the two remaining
principal minors.

Sometimes it will be convenient to view ${\mathcal G}$ as a subset of ${\mathbbm R}^3$ in which case we will write
$G\widehat{=}(u,v,w)$.
The map
\begin{equation}\label{defprojection}
  \pi: {\mathcal P} \rightarrow {\mathcal G}
\end{equation}
will denote the \textit{natural projection} defined by $\pi(s):=s^\top s=G(s)$. Note that
\begin{equation}\label{detS}
 \det s = \pm \sqrt{1-u^2-v^2-w^2+2\,u\,v\,w}
 \;,
\end{equation}
since
\begin{equation}\label{detdelta}
  1-u^2-v^2-w^2+2 u v w= \det G(s) = \det\left( s^\top \,s\right) = \left( \det s\right)^2
 \;.
\end{equation}

${\mathcal G}$ is a convex set, see Figure \ref{FIGGRAM}, that can be first divided into the interior $\stackrel{\circ}{\mathcal G}$,
the open subset of  ${\mathcal G}$ consisting of Gram matrices $G$ with $\det G>0$ (or, equivalently, $\mbox{rank } G=3$)
and its boundary $\partial{\mathcal G}$  consisting of Gram matrices $G$ with $\det G=0$. The boundary, in turn, consists
of regular points forming a smooth surface and four singular points $\mathbf{e}_n,\,n=0,1,2,3$, see Figure \ref{FIGGRAM}.
The smooth surface satisfies the condition that its normal vector
\begin{equation}\label{graddet}
 \nabla \det G=2\,
 \left(
\begin{array}{c}
v\, w -u\\
u\,w -v \\
u\,v-w\\
\end{array}
\right)
\end{equation}
is non-zero.
Conversely, the equations $\nabla \det G={\mathbf 0}$ and $\det G=0$ are only satisfied
by the four Gram matrices with entries $(u,v,w)^\top$ of the form
\begin{equation}\label{singular}
 \mathbf{e}_0\widehat{=}\left(
\begin{array}{r}
1\\
1\\
1\\
\end{array}
\right),\;
 \mathbf{e}_1\widehat{=}\left(
\begin{array}{r}
1\\
-1\\
-1\\
\end{array}
\right),\;
 \mathbf{e}_2\widehat{=}\left(
\begin{array}{r}
-1\\
1\\
-1\\
\end{array}
\right),\;
 \mathbf{e}_3\widehat{=}\left(
\begin{array}{r}
-1\\
-1\\
1\\
\end{array}
\right).
\end{equation}
These singular extremal points
correspond to the collinear configurations symbolized by
\begin{equation}\label{collinear}
\mathbf{e}_0\simeq \uparrow\uparrow\uparrow,\quad
\mathbf{e}_1\simeq \downarrow\uparrow\uparrow,\quad
\mathbf{e}_2\simeq \uparrow\downarrow\uparrow,\quad
\mathbf{e}_3\simeq \uparrow\uparrow\downarrow
\;.
\end{equation}
The other four possible configurations
($\downarrow\downarrow\downarrow,\; \uparrow\downarrow\downarrow,\;\downarrow\uparrow\downarrow,\; \downarrow\downarrow\uparrow$)
can be obtained from those in (\ref{collinear})
by rotations/reflections and hence have the same Gram matrix.

$\partial{\mathcal G}$ can alternatively be divided into one-dimensional faces and zero-dimensional ones, i.~e., extremal points.
Consider first the extremal points.
They correspond to either coplanar spin configurations with $\mbox{rank } G=2$ lying in the smooth part of $\partial{\mathcal G}$,
but not on the one-dimensional faces, or to the four collinear spin configurations (\ref{collinear}) with $\mbox{rank } G=1$.
There are exactly six one-dimensional faces joining
the four extremal points $\mathbf{e}_n,\,n=0,1,2,3$ that form a tetrahedron, see Figure \ref{FIGGRAM}.
The points in the interior
of these one-dimensional faces correspond to those co-planar states and Gram matrices in the smooth
part of $\partial{\mathcal G}$ that are characterized by possessing exactly two collinear spin vectors.
If the third spin vector is chosen  parallel or anti-parallel to the two collinear ones we obtain the two endpoints
$\mathbf{e}_n,\;\mathbf{e}_m,\,n\neq m,$  of the one-dimensional face.

Let us denote by $\dot{\mathcal G}$ the Gram set ${\mathcal G}$ with the four singular extremal points removed.
Then $\dot{\mathcal G}$ will become a \textit{manifold with boundary}, consisting of a three-dimensional manifold $\stackrel{\circ}{\mathcal G}$
and its boundary $\partial\dot{\mathcal G}$ forming a two-dimensional manifold.\\

The symmetry group of ${\mathcal G}$ consists of linear transformations
of the form $G\mapsto R_i\,G\,R_i^\top,\,R_i\in O(3)$,
that generate all permutations of $\{{\mathbf e}_n\left| n=0,1,2,3\right.\}$
and is hence isomorphic to the symmetric group $S_4$ consisting of $4!=24$ permutations or, equivalently, to the tetrahedral group.
The tetrahedral symmetry is also evident from Figure \ref{FIGGRAM}.
On the level of spin configurations this symmetry group
is generated by permutations of three spins (corresponding to the above $R_i$ chosen as permutation matrices)
and local inversions ${\mathbf s}_\mu \mapsto -{\mathbf s}_\mu$ (corresponding to the above $R_i$ chosen as diagonal matrices with entries $\pm 1$) .
This would give $3!\times 2^3=48$ symmetry transformations, but due to the above-mentioned invariance of any Gram matrix under global inversions
only $24$ symmetries are left, see also \cite{S17b}.

\subsection{Time evolution of Gram matrices}\label{sec:TG}

\begin{figure}[htp]
\centering
\includegraphics[width=0.7\linewidth]{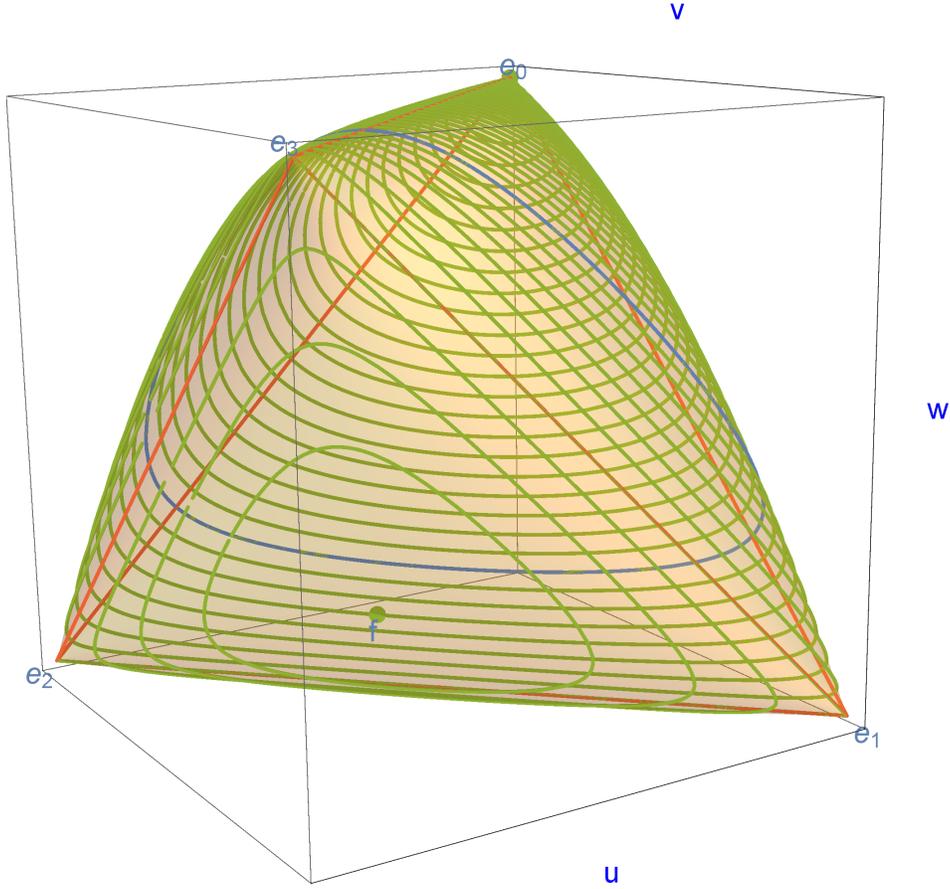}
\caption{Plot of the intersections $P_\sigma\cap \partial {\mathcal G}$, where $\sigma= -3/2,-11/8,\ldots, 23/8,3$.
These are closed smooth (green) curves, except for the values $\sigma=-3/2$ (green dot marked by ${\mathbf f}$),
$\sigma=3$ (green dot marked by ${\mathbf e}_0$)
and $\sigma=-1$ (red triangle joining ${\mathbf e}_1$, ${\mathbf e}_2$, ${\mathbf e}_3$).
The planes $P_\sigma$ represent sets of Gram matrices of spin configurations with the same conserved total spin length $S=\sqrt{3+2\sigma}$.
A two-dimensional section for the value $\sigma=0$ (blue curve) is also shown in Figure \ref{FIGCP1}.
}
\label{FIGGSL}
\end{figure}

We will reconsider the time evolution of the internal variables treated in Section \ref{sec:TID} in connection with the geometry of the Gram set.
Geometrically, the equations (\ref{Huvw}) and (\ref{H1uvw}) represent two planes $\widetilde{P}_\varepsilon$ and $P_\sigma$, resp.~,  in ${\mathbbm R}^3$.
The intersection $P_\sigma\cap {\mathcal G}$ will be a two-dimensional convex set with interior points,
except for the extremes $\sigma=3$ and $\sigma=-{\scriptsize \frac{3}{2}}$, which have been excluded for the generic case by (\ref{generic3}).
In the latter two cases
$P_\sigma\cap {\mathcal G}$ will consist of a single point. In the cases $-{\scriptsize \frac{3}{2}}<\sigma <-1$ and $-1<\sigma <3$ the
intersection $P_\sigma\cap {\mathcal G}$ will be a strictly convex set with a smooth boundary, see Figure \ref{FIGGSL} and Figure \ref{FIGCP1}
for the value $\sigma=0$,
whereas for $\sigma=-1$ the intersection $P_\sigma\cap {\mathcal G}$ will be the equilateral triangle spanned by the three singular extremal
points ${\mathbf e}_n,\;n=1,2,3$.

The intersection  $\widetilde{P}_\varepsilon\cap P_\sigma$ will be a line $L$ parallel to the vector
\begin{equation}\label{line}
{\mathbf n}:=  \left(
\begin{array}{c}
 1\\
 1 \\
 1 \\
\end{array}
\right)
\times
\left(
\begin{array}{c}
 J_1\\
 J_2\\
 J_3\\
\end{array}
\right)
=
\left(
\begin{array}{c}
 J_3-J_2\\
 J_1-J_3\\
 J_2-J_1\\
\end{array}
\right)
\;.
\end{equation}
Due to the exclusion of the case $J_1=J_2=J_3$ implied by (\ref{generic1}) the case of parallel planes
$\widetilde{P}_\varepsilon,\;P_\sigma$ can be excluded and the line $L$ is well-defined.
This line intersects the convex Gram set ${\mathcal G}$ and thus the motion of $G(s)$ is confined to the set $L\cap{\mathcal G}$.
In the generic case, $L\cap{\mathcal G}$ will be an interval bounded by two distinct
points $G_1,G_2$ given by $L\cap\partial{\mathcal G}=[G_1,G_2]$, see Figure \ref{FIGGRAM}, but such that neither $G_1$ nor $G_2$ will be a singular
extremal point. The latter case will be separately treated in Appendix \ref{sec:TSG}.
Note that condition (\ref{generic5}) defining the the generic case excludes the cases where $\sigma$ and $\varepsilon$ are the constants of motion assumed by
the three singular extremal points ${\mathbf e}_n,\;n=1,2,3$ and ${\mathbf e}_0$ is already excluded by (\ref{generic2}).

It follows that the time evolution of Gram matrices is confined to a closed interval $[x_1,x_2]$ of the variable $x$ defined in (\ref{defx}),
such that the endpoints $x_1<x_2$ correspond to the Gram matrices $G_1$ and $G_2$ (in suitable order) where $\delta=\det s$
vanishes and hence $\Pi(x_1)=\Pi(x_2)=0$, see (\ref{sdPi}). Moreover, the case of a double zero $x_1$ or $x_2$ of $\Pi(x)$ is excluded
for the generic case. This can be seen as follows.

In the case $L\cap\partial{\mathcal G}=[G_1,G_2]$ where $G_1$ and $G_2$ are regular points of
$\partial {\mathcal G}$ the directional derivative
\begin{equation}\label{dirder}
\gamma:= {\mathbf n}\cdot \nabla \,\det G
\end{equation}
does not vanish at $G_1$ or $G_2$.
This would only happen if $L$ were tangent to $\partial {\mathcal G}$ and $L\cap\partial{\mathcal G}$ would consist of a
single point or of a one-dimensional face which is excluded here. The following calculation shows that $\gamma$ is proportional to
$\frac{\partial \Pi(x)}{\partial x}$:
\begin{eqnarray}
\label{gamma1}
 \gamma= {\mathbf n}\cdot \nabla \,\det G &=& \left( n_1\frac{\partial}{\partial u}+\ n_2\frac{\partial}{\partial v}+ n_3\frac{\partial}{\partial w}\right)\,\det G \\
   &=&  \left( n_1\frac{\partial x}{\partial u}+\ n_2\frac{\partial x}{\partial v}+ n_3\frac{\partial x}{\partial w}\right)\,\frac{\partial}{\partial x}\det G \\
   &\stackrel{(\ref{line},\ref{defx},\ref{x2v},\ref{x2w})}{=}& \left(
   \left( J_3-J_2\right)\,g + \left( J_1-J_3\right)\frac{J_2-J_3}{J_3-J_1}\,g+\left( J_2-J_1\right)\frac{J_2-J_3}{J_1-J_2}\,g
   \right)\,\frac{\partial}{\partial x}\det G \\
   &\stackrel{(\ref{sdPi})}{=}& 3 g \left( J_3-J_2\right) \frac{\partial }{\partial x}\frac{\Pi(x)}{g^2\,\left(J_3-J_2 \right)^2}\\
   &=& \frac{3}{g\left(J_3-J_2 \right)}\frac{\partial \Pi(x)}{\partial x}
   \;.
\end{eqnarray}
Therefore, $\frac{\partial \Pi(x)}{\partial x}$ cannot vanish for $x=x_1$ or $x=x_2$ as it would for a double root.
We will formulate this result as

\begin{prop}\label{PropGeneric}
 In the generic case defined by Definition (\ref{DefGeneric}) the polynomial $\Pi(x)$ has three real simple roots.
\end{prop}
Thus the analysis in terms of the Weierstrass elliptic function in Section \ref{sec:TID} completely covers the generic case.\\

In other possible ``non-generic" cases $L\cap{\mathcal G}$ would consist of a single point, necessarily lying in ${\partial\mathcal G}$,
or of a one-dimensional face of ${\mathcal G}$ of the form $[{\mathbf e}_n,{\mathbf e}_m],\;0\le n<m\le 3$.
In this Section we will only consider the generic case and defer the special cases to  Appendix \ref{sec:SC}.

We have remarked at various places that the three internal degrees of freedom  $(u,v,w)$ represented by Gram matrices $G \in {\mathcal G}$
evolve autonomously. As already mentioned in Section \ref{sec:TID}, it would be more adequate to represent the internal degrees of freedom by
the variables $(u,v,w,\delta)$ subject to the condition (\ref{deltasqrt}).
This amounts to replacing the dotted Gram set $\dot{\mathcal G}$ by it ``double" ${\mathcal G}'$ defined as:
\begin{equation}\label{defGprime}
 {\mathcal G}':=\{(u,v,w,\delta)\in{\mathbbm R}^4\left| \right. \delta^2=1-(u^2+v^2+w^2)+2\,u\,v\,w\;\text{ and } -1\le u,v,w\le 1,
 \mbox{ but not }\left| u\right| =\left| v\right| =\left| w\right| =1\}
 \;.
\end{equation}
Here we have excluded the singular extremal points by the condition that $\left| u\right| =\left| v\right| =\left| w\right| =1$
must not hold.  ${\mathcal G}'$ is an open subset of a real
algebraic variety characterized by the vanishing of the polynomial $P(u,v,w,\delta)=\delta^2-1+u^2+v^2+w^2-2\,u\,v\,w$.
It can be shown that $\nabla P\neq {\mathbf 0}$ and hence  ${\mathcal G}'$ consists only of regular points and is a three-dimensional manifold.
In the mathematical literature the above construction of ${\mathcal G}'$  is known under the name ``the double $D(M)$
of a manifold $M$ with boundary" and is proven to be a smooth manifold, see, e.~g., \cite[Example 9.32]{L13}.
It consists of the union of two copies of $M$ glued together at their boundary $\partial M$.
In our case the manifold with boundary is $M=\dot{\mathcal G}$.

Recall that, in the generic case, the time evolution is an oscillation between the endpoints
$G_1, G_2\in {\mathcal G}$ along an interval $L\cap {\mathcal G}=[G_1,G_2]$.
Therefore, each point in the interior of $[G_1,G_2]$ is traversed in two directions, ``forward" and ``backward",
and the corresponding spin configurations differ by their orientation.
Therefore, it appears more appropriate to represent the time evolution of the internal degrees of freedom by a cyclic
motion in $L' \cap {\mathcal G}'$, where $L'$ represents the constraint due to the conservation laws $H(s)=\varepsilon$ and $H_1(s)=\sigma$.
Figure \ref{FIGAA1} can be viewed as a projection of this cyclic motion and shows that there is no discontinuous reflection
at $\partial{\mathcal G}$ but a smooth transition.

\subsection{Time evolution of the position of $s$}\label{sec:TP}
We have used the (left) polar decomposition $s=R\,p$ with $R\in O(3)$ and $p\ge 0$ in order to represent
the form of a spin configuration by $p$ and its position by $R$. It follows that $\det s=\det R \,\det p=\pm \det p$ depending
on whether $R$ is a reflection or a proper rotation.
It will be convenient at this point to consider another, modified decomposition $s= R'\,p'$ defined such that $R'=R$ and $p'=p$ in the
case $\det s\ge 0$ and $R'=-R$ and $p'=-p$ in the case $\det s<0$. This entails that $R'$ is always a proper rotation, $R'\in SO(3)$,
but $p'$ has either only non-negative eigenvalues or only non-positive ones. $p'$ will be said to represent the \textit{oriented form}
of $s$ and the equivalence relation ``there exists an $R\in SO(3)$ such that $s'=R \,s$ " will be called the
\textit{oriented congruence} of $s$ and $s'$. In this case $s$ and $s'$ will be said to be \textit{congruent and equally oriented}.
Each Gram matrix $G(s)$ in the interior of the above-mentioned interval $[G_1,G_2]$ corresponds to two different oriented forms
$\pm p$; for $\dot{x}>0$ we have $\det p'>0$ and for $\dot{x}<0$ we have $\det p'<0$ with vanishing $\det p'$ at the points $G_1$ and $G_2$.
This means that we can describe the motion of $s(t)$ as following a closed curve $\widetilde{\mathcal C}_1$ if projected onto the plane with coordinates
$(x, \det(s))$, see Figure \ref{FIGAA1}. \\

Let us formalize these considerations to some extent. Recall that $\dot{\mathcal P}$ denotes the
phase space with collinear spin configurations removed and consider
\begin{equation}\label{projection}
 \pi': \dot{\mathcal P}\rightarrow  \dot{\mathcal P}/SO(3) \cong {\mathcal G}'
\end{equation}
the projection of the phase space onto the set of equivalence classes w.~r.~t.~oriented congruence, or, what is the same,
onto the set of orbits of the left action of $SO(3)$ on spin configurations. These orbits can be represented
by symmetric spin configuration $p'=\pm\sqrt{G},\,G\in{\mathcal G}$, that have to be identified if $G\in\partial {\mathcal G}$.
The latter follows from the fact that for coplanar spin configurations, corresponding to $\mbox{rank } G=2$
($\mbox{rank } G=1$ has been excluded),
the two configurations $+\sqrt{G }$ and  $-\sqrt{G }$ can be mapped onto each other by a proper rotation $R\in SO(3)$ and are
hence congruent and equally oriented. Equivalently, the equivalence classes can be represented by points of ${\mathcal G}'$,
the double of ${\mathcal G}$,  see (\ref{defGprime}). In fact, each point $(u,v,w,\delta)\in{\mathcal G}'$ determines
a class of congruent spin configurations via $(u,v,w)$, and, in the case of three-dimensional configurations $s$, the sign
of $\delta$ also determines the subclass of congruent spin configurations with the same orientation.

Moreover, it follows that the fibers of the projection (\ref{projection}) are $1:1$ copies of $SO(3)$.
In the case of three-dimensional spin configurations this is obvious. But also for two congruent coplanar configurations
$s$ and $s'=R\,s$ the rotation $R\in SO(3)$ will be uniquely determined. (This would no longer be true if we consider reflections $R$
or collinear configurations $s$.) Consequently, it is plausible that the left action $s\mapsto R\,s$ of $SO(3)$ on $\dot{\mathcal P}$
makes $\dot{\mathcal P}$ a {\it principal fiber bundle} with projection (\ref{projection}) onto the base space ${\mathcal G}'$.
However, we will not go into the details of this concept,  see, e.~g., \cite[Chapter III.10]{KMS93}.

Next recall  the standard spin configuration $r(u,v,w,\delta)=\left({\mathbf r}_1, {\mathbf r}_2,{\mathbf r}_3\right)$
with Gram matrix $G(r)\widehat{=}(u,v,w)$ defined in (\ref{defr1} - \ref{defr3})
such that ${\mathbf R}:={\mathbf r}_1 + {\mathbf r}_2+{\mathbf r}_3={\mathbf S}_0$.

\begin{figure}[th]
\centering
\includegraphics[width=0.7\linewidth]{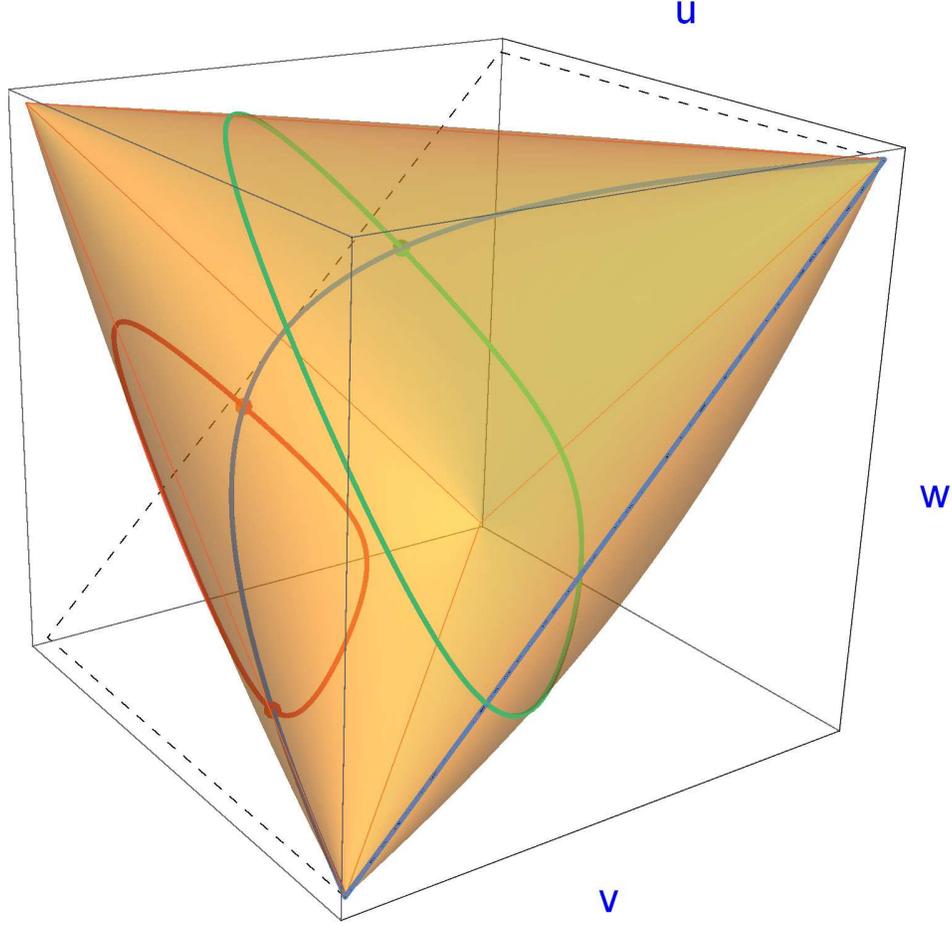}
\caption{Plot of the intersection of the plane given by $v=w$ (dashed black lines) and the Gram set ${\mathcal G}$. The boundary of this
intersection  is given by the (blue) line $u=1, v=w$ and the (blue) parabola $u=2v^2-1$, the latter defining a set of critical points
for the definition of (\ref{defr1} - \ref{defr3}) and (\ref{soleomext}). We have further displayed the cases $\sigma=0$ (green curve)
with one critical point and $\sigma=-4/3$ (red curve) with two critical points. These two cases also appear in Figure \ref{FIGOM}.
}
\label{FIGCP}
\end{figure}

We will determine the domain ${\mathcal G}''$ where (\ref{defr1} - \ref{defr3}) is well-defined. This includes the cases
where all square roots have non-negative arguments or positive arguments if they occur in the denominator and possibly an extension
by means of continuity.
The first condition is $S^2=3+2(u+v+w)>0$ and hence the point ${\mathbf f}$ with coordinates $u=v=w=-\frac{1}{2}$ has to be excluded.
At this point some matrix entries of $r(u,v,w,\delta)$ are of the form $\frac{0}{0}$, but such that their limit depends
on the direction in which one approaches  ${\mathbf f}$.

Next consider the argument  $2(u+1) -(v+w)^2$. Here we have
the preliminary result:
\begin{prop}\label{proppos}
If $\delta>0$ or $v\neq w$ then
  \begin{equation}\label{posterm}
   2(1+u) -(v+w)^2 > 0
   \;.
  \end{equation}
\end{prop}
{\bf Proof}:
The value $u=-1$ is only assumed by the singular extremal point ${\mathbf e}_2$ and hence can be excluded.
So we can assume $u+1>0$. Then, according to the assumptions,
\begin{eqnarray}
\label{post1}
  0 &<& 2\,\delta^2+(v-w)^2(u+1) \\
   &=& 2-2(u^2+v^2+w^2)+4uvw+(v^2+w^2-2vw)(u+1)\\
   &=& 2-v^2-w^2-2u^2-2vw+uv^2+uw^2+2uvw\\
   &=& 2(1+u)(1-u)-(v+w)^2+u(v+w)^2\\
   &=& \left( 2(1+u)-(v+w)^2\right)(1-u)
   \:.
\end{eqnarray}
This implies $u<1$ and $ 2(1+u)-(v+w)^2>0$.   \hfill$\Box$\\

Hence $r(u,v,w,\delta)$ is well-defined in the case of Proposition \ref{proppos} and we are left with
the extension of $r$ to the ``critical points" with $\left(\delta=0 \mbox{ and } v= w\right)$.
The plane given by $v=w$ intersects the Gram set ${\mathcal G}$ at a two-dimensional convex set
bounded by the interval $[{\mathbf e}_0,{\mathbf e}_1]$ and satisfying $u=1$ and the parabola $u=2 v^2-1$,
see Figure \ref{FIGCP}.
If $G$ is a point in the interior of $[{\mathbf e}_0,{\mathbf e}_1]$, satisfying $-1<v=w<1$,  then $2(1+u) -(v+w)^2=4(1-v^2)>0$
and $r$ is well-defined at the corresponding point $(u=1,v,w=v,\delta=0)$.

In the remaining case $\delta=0$, $G_1\widehat{=}\left(u=2 v^2-1,v,v\right)$ and $-1<v<1$ we obtain terms of the form $\frac{0}{0}$ in
$r(u,v,w,\delta)_{2,2}=-r(u,v,w,\delta)_{2,3}$ and will determine their limit by means of l'Hospitale's rule when
approaching $G_1$ from the interior of the interval $[G_1,G_2]$.
We write $x=x_2+\xi$ (assuming without loss of generality that $G_1$ corresponds to $x_2$),
express $\delta$ as well as $\sqrt{2(u+1)-(v+w)^2}$ in terms of $x$, and expand numerator and denominator w.~r.~t.~$\xi$.
 After some calculations we obtain for this case
\begin{equation}\label{limitr22}
\lim_{\xi\to 0} \frac{\delta }{\sqrt{2(u+1)-(v+w)^2}}=
\sqrt{1-v^2} >0
\;.
\end{equation}
The same limit is assumed it we approach $G_1$ from both sides of the curve $P_\sigma \cap \partial{\mathcal G}$.

Summarizing, the domain of definition ${\mathcal G}''$ of the map $r$ can be chosen as ${\mathcal G}'$, but with the
point ${\mathbf f}\widehat{=}(u,v,w,\delta)=\left( -\frac{1}{2}, -\frac{1}{2}, -\frac{1}{2},0\right)$ removed.
We thus obtain a smooth local section $r:{\mathcal G}''\rightarrow \dot{\mathcal P}$ of the projection (\ref{projection}),
i.~e., satisfying $\pi'\, r= \mbox{ id }_{{\mathcal G}''}$, that cannot be extended to the whole space ${\mathcal G}'$.
This can be understood by means of the following
\begin{prop}\label{Pglobalsection}
 There do not exist global sections   $\widehat{r}:{\mathcal G}'\rightarrow \dot{\mathcal P}$ of the projection (\ref{projection}),
 i.~e., the principal fiber bundle  $\dot{\mathcal P}$ is not globally trivial.
\end{prop}
{\bf Proof}: A globally trivial principal fiber bundle $\dot{\mathcal P}\cong {\mathcal G}'\times{SO(3)}$ would possess a non-trivial
fundamental group $\pi_1\left(\dot{\mathcal P}\right)\cong \pi_1\left({\mathcal G}'\right)\times\pi_1\left({SO(3)}\right)$
since $\pi_1\left({SO(3)}\right)={\mathbbm Z}_2$, see, e.~g., \cite[Prop.~13.10]{H15}.
This would contradict the finding of $\pi_1\left( \dot{\mathcal P}\right)$ being trivial, i.~e.,
that every loop in $\dot{\mathcal P}$ can be continuously deformed to a point. The latter can be seen as follows.
For every loop $\lambda\mapsto s(\lambda)\in\dot{\mathcal P}$ there exists a $\lambda_0$ and two spin vectors, say,
${\mathbf s}_1$ and ${\mathbf s}_2$, such that $\left({\mathbf s}_1(\lambda_0),{\mathbf s}_2(\lambda_0)\right)$ will
be coplanar. By continuity, this also holds for a small interval $I=(\lambda_0-\epsilon,\lambda_0+\epsilon)$. For
$\lambda\in I$ the loop described by ${\mathbf s}_3(\lambda)$ can be continuously deformed to a point, such that the total
deformed configuration $\widetilde{s}(\lambda)$ will never be collinear. By repeating the procedure the analogous deformation
of $\widetilde{\mathbf s}_2(\lambda)$ and $\widetilde{\mathbf s}_1(\lambda)$ to a point can be achieved.
 \hfill$\Box$\\

Finally, we reconsider (\ref{intalpha}).
Another way to write the solution (\ref{intalpha}) starts from a Fourier series representation of the ${\sf T}$-periodic function
$\dot{\alpha}(t)$ of the form
\begin{equation}\label{fourier}
 \dot{\alpha}(t)= \sum_{n\in{\mathbbm Z}} a_n\, \exp\left(\frac{2\pi{\sf i}n t}{\sf T} \right)
 \;,
\end{equation}
which yields, by termwise integration, and taking into account the initial value $\alpha(0)=0$:
\begin{equation}\label{fourierint}
 \alpha(t)= a_0\,t +\sum_{n\neq 0} a_n\,\frac{\sf T}{2\pi {\sf i} n}\, \left(\exp\left(\frac{2\pi{\sf i} n t}{\sf T} \right)-1\right)
 \;.
\end{equation}
For $t={\sf T}$ the terms of the series in (\ref{fourierint}) vanish and we obtain
\begin{equation}\label{a0}
 \alpha({\sf T})= a_0\,{\sf T}
 \;.
\end{equation}

Eq.~(\ref{eomext}), written in the form $\dot{r}={\mathcal J}(r) -\Omega\,r$, can be interpreted as an equation of motion
for the spin configuration $r(t)$ under the additional influence of a ${\sf  T}$-periodic magnetic field ${\mathbf B}(t)$ of the form
\begin{equation}\label{Bt}
{\mathbf B}(t)= \left(
\begin{array}{c}
  0 \\
  0 \\
  -\dot{\alpha}(t)
\end{array}
\right)
\;.
\end{equation}
The configuration $r(t)$ obtained by substituting $(u(t),v(t)$ into (\ref{defr1} -\ref{defr3}) is a special ${\sf  T}$-periodic
solution of this extended equation of motion. Hence the transformation $r(t)\mapsto s(t)=Z(t)\, r(t)$ can be
understood as a time-dependent rotation compensating for the influence  of the magnetic field (\ref{Bt}),
analogous to the considerations in Appendix \ref{sec:MF}.

\subsection{Detailed definition of the averaged rotation}\label{sec:AR}

\begin{figure}[htp]
\centering
\includegraphics[width=1.0\linewidth]{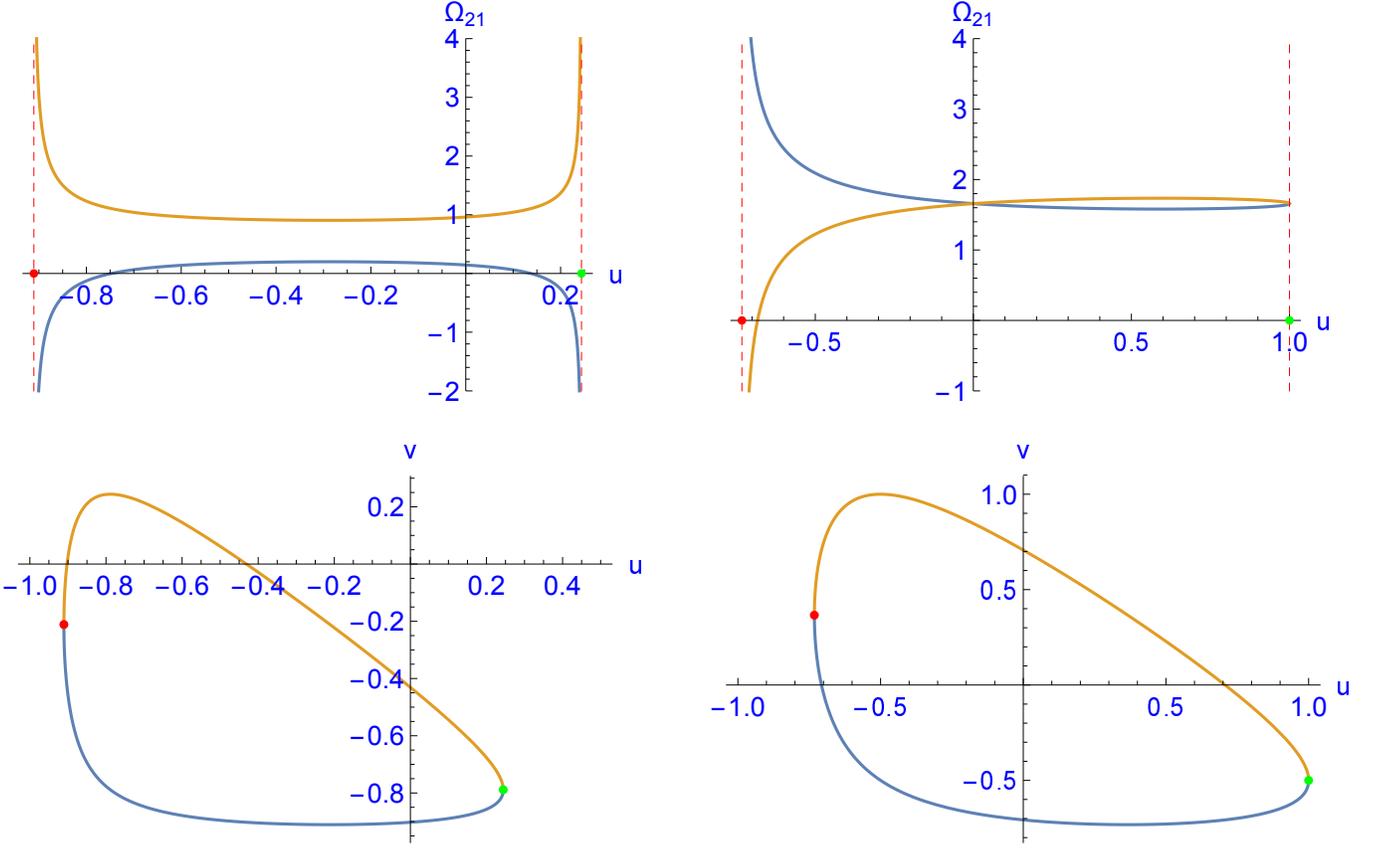}
\caption{Plot of the limiting values of $\Omega_{2,1}$ as a function of $u$ (upper panels) evaluated along the curves $P_\sigma \cap\partial{\mathcal G}$
(lower panels).
The left panels corresponds to the value of $\sigma=-{\scriptstyle \frac{4}{3}}$ where two critical points occur for minimal and maximal $u$.
The right panels corresponds to the value of $\sigma=0$ where only one critical point occurs for minimal $u$.
$\Omega_{2,1}$ diverges to $\pm\infty$ at the critical points with different signs depending on whether the critical points are approached
from above (dark yellow curves) or from below (blue curves). For the position of the selected critical points on the Gram set see also Figure \ref{FIGCP}.
}
\label{FIGOM}
\end{figure}

For the definition of the action-angle variables in Appendix \ref{sec:AA} it will be necessary to consider some details of the ``averaged rotation" of the
spin vectors ${\mathbf s}_\mu$ about the total spin ${\mathbf S}$ with angular velocity $\frac{\alpha({\sf T})}{{\sf T}}$.
Recall that $\alpha(t)$ was obtained as the $t$-integral over $\dot{\alpha}(t)$.
In the  definition of $ \dot{\alpha}(t)=\Omega_{2,1}(t)$,  see (\ref{intalpha}),
it is required that $r=r(u,v,w,\delta)$ is invertible and hence this definition
fails for $\delta = \det r =0$ if taken at face value. Thus we will try to extend the definition (\ref{soleomext}) of $\Omega$ by means
of continuity. This can be achieved by cancelling the factor $\delta$ before taking the limit $\delta\to 0$.
In fact, it turns out that, with some exceptions to be considered below, only the second column of the matrix $r^{-1}$ diverges for  $\delta\to 0$.
Since we are looking for
\begin{equation}\label{Omega21}
 \dot{\alpha}(t)=\Omega_{2,1}(t)\stackrel{(\ref{soleomext})}{=} \sum_{i}\left( {\mathcal J}(r)-\dot{r}\right)_{2,i}\,r^{-1}_{i,1}
 \;,
\end{equation}
it suffices to calculate the second row of ${\mathcal J}(r)-\dot{r}$ and the first column of $r^{-1}$ in the limit $\delta\to 0$
which is assumed for integer multiples of $t=\frac{\sf T}{2}$.

First, we split $\Omega_{2,1}(t)$ into two terms $\Omega^{(c)}_{2,1}(t)$ and $\Omega^{(d)}_{2,1}(t)$ according to
\begin{eqnarray}
\label{Omegasplitc}
  \Omega^{(c)}_{2,1}(t) &=& \sum_i {\mathcal J}(r)_{2,i} \,r^{-1}_{i,1}\;,\\
  \label{Omegasplitd}
  \Omega^{(d)}_{2,1}(t) &=& - \sum_i \dot{r}_{2,i}\,r^{-1}_{i,1}\;,
\end{eqnarray}
and write $r_{2,i}$ as a function of $u$ and $\delta$ along the line $L$ defined by $H(s)=\varepsilon$ and $H_1(s)=\sigma$.
Hence the $t$-derivative of $r_{2,i}$ leads to the sum of two terms proportional to
$\dot{u}\stackrel{(\ref{ud5})}{=}\left(J_3-J_2 \right)\delta$ and $\dot{\delta}$, resp.~.
The first term vanishes in the limit $\delta\to 0$. The second term only occurs in $\dot{r}_{2,2}=-\dot{r}_{2,3}\sim \dot{\delta}$
and does not contribute to  $\Omega^{(d)}_{2,1}(t)$ since $r^{-1}_{2,1}(t)=r^{-1}_{3,1}(t)$ for all $t\in{\mathbbm R}$.

So we only need to consider the first term $ \Omega^{(c)}_{2,1}(t)$ and, after some calculations, will write it as a function of $u,v,\sigma$:
\begin{equation}\label{Omegauvsigma}
  \Omega^{(c)}_{2,1}=\frac{\sqrt{2 \sigma +3} \left(J_3 \left(\sigma ^2-\sigma  (2 u+v)+u
   (u+v-1)-1\right)-J_2 (u v+u-\sigma  v+1)\right)}{\sigma ^2+u^2-2 (\sigma +1) u-2}
   \;.
\end{equation}
Along the line $L$ we obtain
\begin{equation}\label{Omegausigmaeps}
  \Omega^{(c)}_{2,1}= \frac{\sqrt{2 \sigma +3} \left(\left(J_2+J_3\right) (u+1)-(u-\sigma ) \left(J_1
   u-\epsilon \right)\right)}{2 (u+1)-(u-\sigma )^2}
   \;.
\end{equation}
If $2 (u+1)-(u-\sigma )^2\neq 0$ for all $u(t)$ the limit of this expression exists for
$t\to n\frac{\sf T}{2},\;n\in{\mathbbm Z}$,  and $\dot{\alpha}(t)$
is well-defined for all $t\in{\mathbbm R}$.

Hence the case $2 (u+1)-(u-\sigma )^2=0$ for some $u=u(t)$ is left. According to Proposition \ref{proppos} this case only occurs
for $\delta=0$ and $v=w$. It leads to two one-parameter families of critical points in $\partial{\mathcal G}$ with coordinates
\begin{eqnarray}
\label{uvcrit1}
 u = u_c^{(1)}(\sigma) &=& 1+\sigma -\sqrt{3+2\sigma},\quad v=v_c^{(1)}(\sigma) =
 {\scriptsize\frac{1}{2}}\left(\sqrt{3+2\sigma} -1\right)\;, \mbox{ for } \sigma\in\left[-{\scriptsize\frac{3}{2}},3\right]\;, \\
 \label{uvcrit2}
 u = u_c^{(2)}(\sigma) &=& 1+\sigma +\sqrt{3+2\sigma},\quad v=v_c^{(2)}(\sigma) =
  \frac{1}{2}\left(-\sqrt{3+2\sigma} -1\right)\;, \mbox{ for } \sigma\in\left[-{\scriptsize\frac{3}{2}},-1\right]
 \;.
\end{eqnarray}
The corresponding values of the critical energies are
\begin{eqnarray}
\label{ecrit1}
  \varepsilon_c^{(1)}(\sigma) &=& \frac{J_2+J_3}{2}\left( \sqrt{3+2\sigma}-1\right)+J_1\left(1+\sigma -  \sqrt{3+2\sigma}\right)\;,
  \mbox{ for } \sigma\in\left[-{\scriptsize\frac{3}{2}},3\right]\;,\\
  \label{ecrit2}
   \varepsilon_c^{(2)}(\sigma) &=&- \frac{J_2+J_3}{2}\left( \sqrt{3+2\sigma}+1\right)+J_1\left(1+\sigma + \sqrt{3+2\sigma}\right)\;,
  \mbox{ for } \sigma\in\left[-{\scriptsize\frac{3}{2}},-1\right]
  \;.
\end{eqnarray}
These are the exceptions referred to at the beginning of the Section.
It can be shown that $\Omega^{(c)}_{2,1}$ diverges to $\pm\infty$ if the critical points are approached along the curve
$P_\sigma \cap\partial{\mathcal G}$, and
moreover, the sign of the diverging values depend on whether the critical points are approached from above or from below, see
Figure \ref{FIGOM}. This result is verified by a series expansion of $\Omega^{(c)}_{2,1}$ at the critical point which will not be given here but shows that the divergence is of the form $\pm \left(u-u_c\right)^{-1/2}$. Curiously, $\Omega^{(c)}_{2,1}$ has a finite limit at the critical points as one approaches the critical point along line $L$, but this does not change its singular behavior.\\

We now turn to the definition of $\alpha({\sf T})$. According to the equation $s(t)= {\mathcal R}\left({\mathbf S},\alpha(t) \right)\,r(t)$
the angle $\alpha(t)$ is only defined modulo $2\pi$. The integral representation (\ref{intalpha}) of $\alpha(t)$ yields a choice that is locally
a smooth function of $\varepsilon$ (for fixed $\sigma$). The above discussion, however, shows that the integral representation of $\alpha({\sf T})$
could lead to a jump of the values of $\alpha({\sf T})$ obtained from approaching the critical points from below or above.
Numerical examples show a jump of exactly $2\pi$ and consequently $\alpha({\sf T})$ can be chosen as a smooth function of
$\varepsilon$ after adding a suitable integer multiple of $2\pi$ if the energy $\varepsilon$ crosses the critical energy  $\varepsilon_c^{(1)}$
or $\varepsilon_c^{(2)}$ given by (\ref{ecrit1}), resp.~(\ref{ecrit2}).

In order to give a more rigorous justification of the smooth dependence of $\varepsilon \mapsto \alpha({\sf T})$ we
recall that for any two congruent coplanar configurations $s$ and $s'$ the rotation $R\in SO(3)$ such that $s'=R\,s$ is uniquely
determined. Moreover, $R$ depends smoothly on $(s,s')$. Let $\lambda\mapsto (u(\lambda),v(\lambda))$ be a smooth parametrization
of the curve $P_\sigma \cap \partial{\mathcal G}$ in the neighbourhood of a critical point $(u_c,v_c)$. For example, one could choose
$\lambda=\varepsilon$.
Let $s_\lambda=s(0)=r(0)=r(u(\lambda),v(\lambda),w=\sigma-u(\lambda)-v(\lambda),\delta=0)$ be the initial value of a time evolution
and $s_\lambda'=s({\sf T})$ the final value after one period ${\sf T}$. Consider
$s({\sf T})= {\mathcal R}\left( {\mathbf S},\alpha({\sf T},\lambda)\right)\,r({\sf T})$ and note that $r(0)=r({\sf T})$.
Hence, according to what has been said,  $\alpha({\sf T},\lambda)$ will depend smoothly on $\lambda$
(after adding a suitable integer multiple of $2\pi$ if necessary),
even it a critical point is passed.


\subsection{Example}\label{sec:EX}

\begin{figure}[th]
\centering
\includegraphics[width=0.7\linewidth]{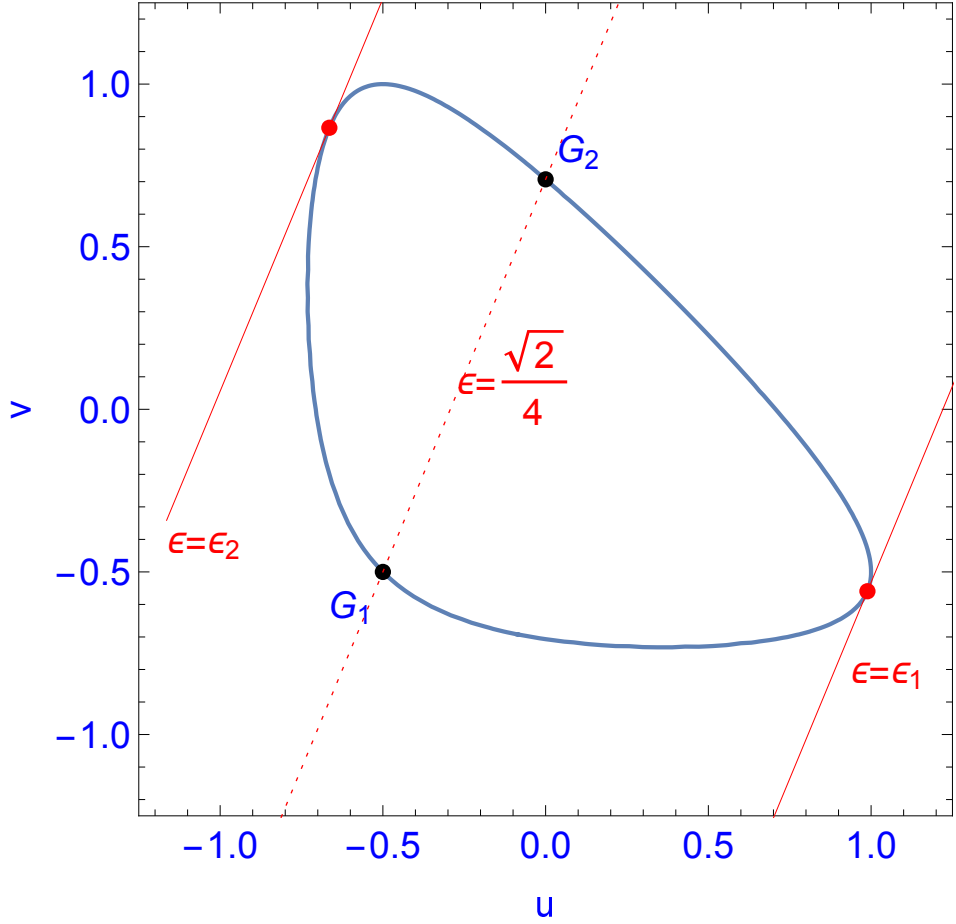}
\caption{Plot of the intersection ${\mathcal G}_0$ of the plane given by $u+v+w=\sigma=0$ with the Gram set ${\mathcal G}$.
We have chosen the two Gram matrices $G_1$ and $G_2$ according to (\ref{CG1},\ref{CG2})
at the boundary of ${\mathcal G}$ that lie on the (red, dotted) line
given by $J_1 u +J_2 v + J_3 (-u-v))=\varepsilon=\frac{\sqrt{2}}{4}$.
The corresponding coupling constants $J_1,J_2,J_3$ are given in (\ref{J123}).
The energies $\varepsilon$ of Gram matrices in ${\mathcal G}_0$ lie between the extremal values
$\varepsilon_1\approx -1.47328$ and $\varepsilon_2\approx 1.23498$ corresponding to the two tangents of ${\mathcal G}_0$
indicated by red lines.
}
\label{FIGCP1}
\end{figure}

\begin{figure}[htp]
\centering
\includegraphics[width=1.0\linewidth]{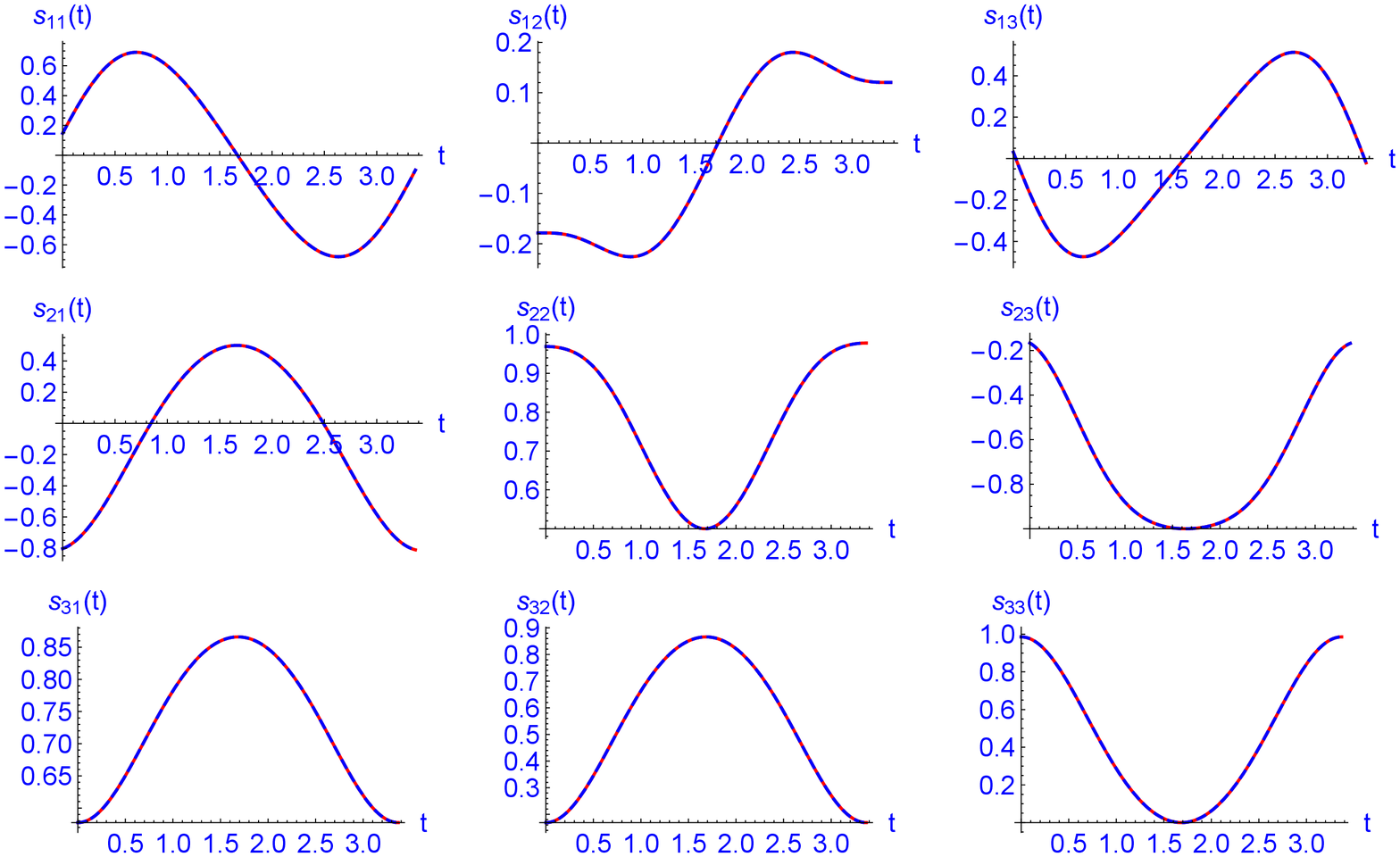}
\caption{Plot of the solution $s_{i,\mu}(t), \; i,\mu=1,2,3,$ of the equations of motion (\ref{eom1} - \ref{eom3})
for the example of this Section and for a full period $0\le t\le {\sf T}$.
The red curves are obtained by a (semi-)analytical solution and the blue, dashed
ones by numerical integration. We note a perfect agreement.
}
\label{FIGPEX}
\end{figure}

The results of the Sections \ref{sec:TID} and \ref{sec:TED} will be illustrated by an example.
In order to keep the expressions as simple as possible we will choose the parameters of the example
in the following  particular way, but without restricting the solution to a special case:
\begin{eqnarray}
\label{J123}
J_1&=&-{\scriptsize\frac{1}{2}},\quad J_2={\frac{1}{2}+\frac{\sqrt{2}}{2}},\quad J_3={\scriptsize\frac{\sqrt{2}}{2}},\\
\label{sigma0}
  \sigma &=& u+v+w=0, \\
  \label{eps1}
 \varepsilon&=&J_1 u+J_2v +J_3 w={\frac{\sqrt{2}}{4}}
 \;.
\end{eqnarray}
In the plane $P_\sigma$ given by (\ref{sigma0}) the equation (\ref{eps1}) defines a line $L$ that
intersects $\partial {\mathcal G}$ in the two points $G_1$ and $G_2$ with the coordinates $(u,v,w)$ according to
\begin{eqnarray}
\label{CG1}
  G_1 &\widehat{=}& \left(-\frac{1}{2},-\frac{1}{2},1 \right) \\
  \label{CG2}
  G_2 &\widehat{=}& \left(0,\frac{1}{\sqrt{2}} ,-\frac{1}{\sqrt{2}}\right)
  \;,
\end{eqnarray}
see Figure \ref{FIGCP1}. The parameters $g$ and $x_0$ defined in (\ref{solg}) and (\ref{solx0}) assume the special values
\begin{equation}\label{gspec}
 g=\frac{1}{8} \left(-4-3 \sqrt{2}\right)\approx -1.03033
 \;,
\end{equation}
and
\begin{equation}\label{x0spec}
x_0=\frac{1}{48} \left(-14-9 \sqrt{2}\right)\approx-0.556832
\;,
\end{equation}
resulting in the special form of the  polynomial (\ref{sdPi})
\begin{equation}\label{polyspec}
 \Pi(x)=\frac{1}{6912}(24 x+1) \left(48 x (24 x-1)-135 \sqrt{2}-193\right)=
 4 x^3+\frac{\left(-4680-3240 \sqrt{2}\right) }{6912}\,x+\frac{-193-135 \sqrt{2}}{6912}
 \;.
\end{equation}
From this the parameters $g_2$ and $g_3$ can be read off. The zeroes of $\Pi(x)$ are
\begin{eqnarray}\label{zeroesPi1}
x_1&=&\frac{1}{48} \left(-14-9 \sqrt{2}\right)\approx-0.556832, \\
\label{zeroesPi2}
x_2&=&-\frac{1}{24} \approx -0.0416666,\\
\label{zeroesPi3}
x_3&=&\frac{1}{48} \left(16+9 \sqrt{2}\right)\approx0.598498
\;,
\end{eqnarray}
see Figure \ref{FIGPP}.
This implies the following values of the period ${\sf T}$, see (\ref{period1}), and  $t_0$, see (\ref{period2}):
\begin{equation}\label{Tspec}
 {\sf T}=4 \sqrt{\frac{2}{7} \left(5-3 \sqrt{2}\right)} K\left(\frac{1}{14} \left(2+3
   \sqrt{2}\right)\right)
   \approx 3.3693
   \;,
\end{equation}
and
\begin{equation}\label{t0spec}
  t_0=2 \,{\sf i}\, \sqrt{\frac{2}{7} \left(5-3 \sqrt{2}\right)} K\left(-\frac{3}{14}
   \left(-4+\sqrt{2}\right)\right)\approx 1.77031\,{\sf i}
   \;.
\end{equation}
Using these data we can analytically  determine the angular velocity $\dot{\alpha}(t)$ according to
(\ref{defOmega}) and (\ref{soleomext}), but the integral (\ref{intalpha}) could only be calculated numerically.

For the numerical integration of the equations of motion (\ref{eom1} - \ref{eom3}) we have to use an initial
value corresponding to the spin configuration (\ref{defr1}-\ref{defr3}) evaluated at, e.~g.,  $t=\frac{\sf T}{4}$.
Initial values at $t=0$  or $t=\frac{\sf T}{2}$ would not work since the time derivative of $r(t)$ vanishes here.
Nevertheless, it is possible to compare the (semi-)analytical solution of $s(t)$ with the numerical one, see
Figure \ref{FIGPEX} and to confirm our approach for the chosen example.

\section{Action-angle  variables}\label{sec:AA}

\begin{figure}[htp]
\centering
\includegraphics[width=0.7\linewidth]{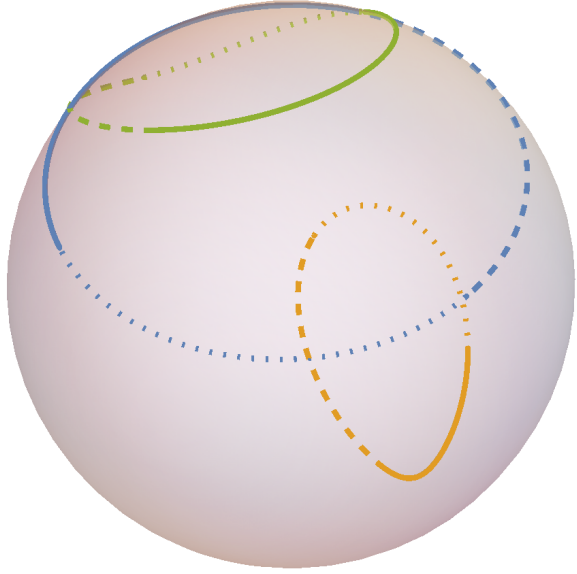}
\caption{Plot of the basic cycle ${\mathcal C}_1$ generated by $\widetilde{F}_1$ according to (\ref{defF1tilde})
for the example of Section \ref{sec:EX}. We show the three closed orbits (blue, green, yellow curves) traversed
by the spins ${\mathbf s}_i(t),\;i=1,2,3,$ resp., and have indicated the direction of the motion by choosing
different plot styles for $0\le t\le {\sf T}/3$ (solid), ${\sf T}/3\le t\le 2{\sf T}/3$ (dashed),
and $2{\sf T}/3\le t\le {\sf T}$ (dotted). The curves have been calculated by numerically solving the equations of motion
with a Hamiltonian $\widetilde{H}= H + 0.0903971\, S$.
}
\label{FIGBC}
\end{figure}

\begin{figure}[htp]
\centering
\includegraphics[width=0.7\linewidth]{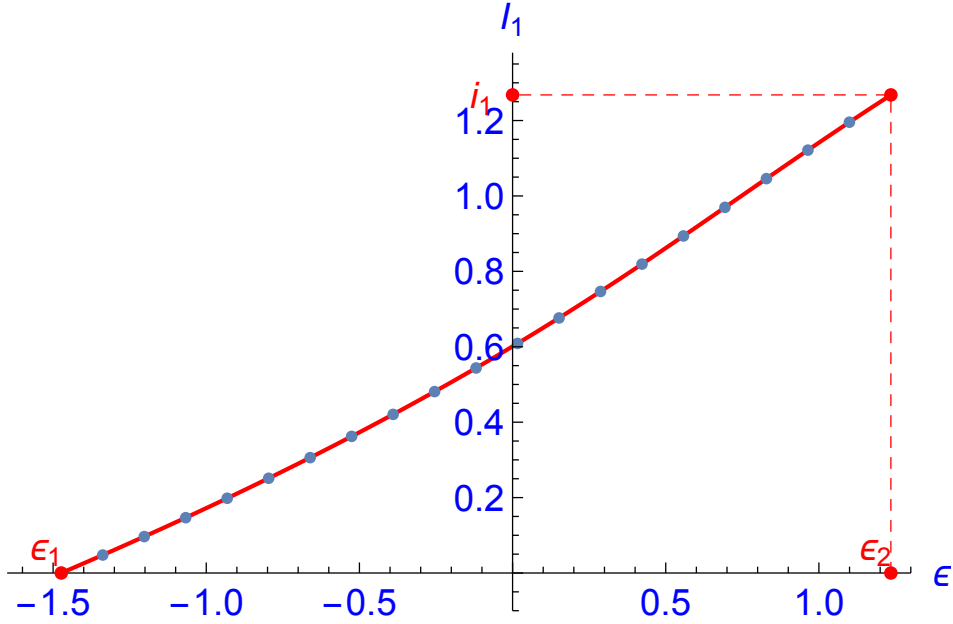}
\caption{Plot of the action variable $I_1(\varepsilon)$ for the coupling constants of the example of Section \ref{sec:EX},
constant $\sigma=0$
and varying energy $\varepsilon\in (\varepsilon_1,\varepsilon_2)$, see Figure \ref{FIGCP1}.
The calculation has been performed numerically (blue dots) and interpolated (red curve).
Exact limiting values are $I_1(\varepsilon_1)=0$ and $I_1(\varepsilon_2)=i_1=3-\sqrt{3}=1.26795\ldots$.
}
\label{FIGPINT}
\end{figure}

\begin{figure}[htp]
\centering
\includegraphics[width=0.7\linewidth]{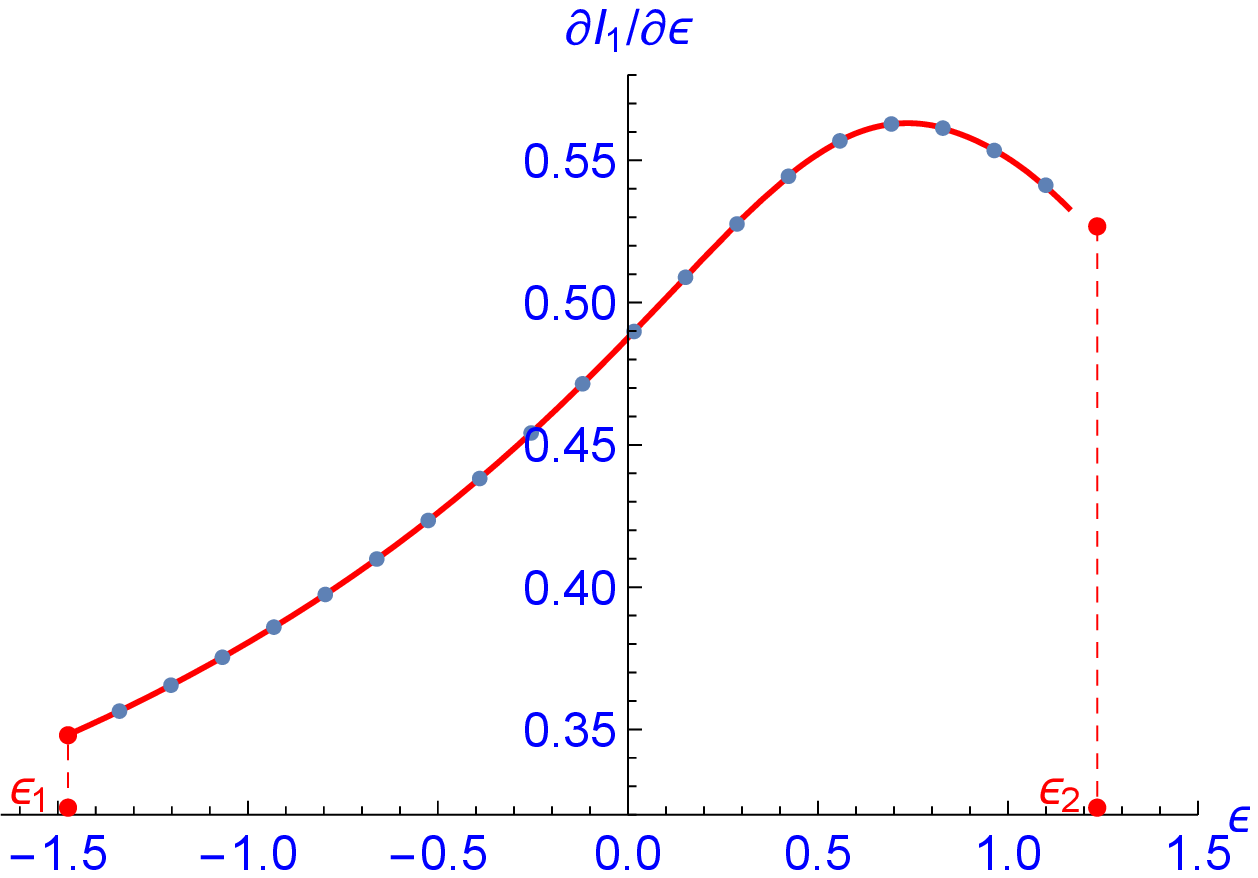}
\caption{Plot of the derivative of the action variable $\partial I_1/\partial \varepsilon$ (red curve) for the same data as in Figure \ref{FIGPINT},
calculated by numerical differentiation of  $I_1(\varepsilon)$ .
For the check of (\ref{diffaa1}) we have plotted the analytical values of $\frac{1}{2\pi}{\sf T}(\varepsilon)$ (blue dots) for $\sigma=0$.
Simple limiting values (red dots) are $\frac{1}{2\pi}{\sf T}(\varepsilon_1)= \frac{1}{\sqrt{12 |x_a|}}$ and
$\frac{1}{2\pi}{\sf T}(\varepsilon_2)= \frac{1}{\sqrt{12 |x_b|}}$, where $x_a$, resp.~$x_b$, are the coordinates of the double
zero of $\Pi(x)$ for $\varepsilon=\varepsilon_1$ and $\varepsilon=\varepsilon_2$, see (\ref{limperiod}).
}
\label{FIGLPT}
\end{figure}

In this Section we will construct the action-angle variables of the spin triangle in the generic case, see Definition \ref{DefGeneric}.
It will be sometimes helpful to check the following equations in terms of their physical dimensions and specifically
to note the identity ``action=energy $\times$ time". Here, the physical dimension of spin must be reintroduced as ``action'',
in accordance with the fact that in quantum theory eigenvalues of spin operators will be integer multiples of $\frac{\hbar}{2}$.
We do not assume any special choice of the coordinate system as, e.~g., in subsection \ref{sec:TP},
but we will have to fix an arbitrary coordinate system in what follows.
We recall the three conserved quantities, energy $H(s)$, total squared spin length $S^2(s)$ and $3$-component of the total spin
$ {\mathbf S}^{(3)}(s)$,  and their constant values
\begin{equation}\label{CQ1}
 H(s)=\varepsilon,\quad S^2(s)=3+2\sigma, \quad {\mathbf S}^{(3)}(s)=\sigma_3
 \;,
\end{equation}
and abbreviate the triple of constant values as ${\boldsymbol\sigma}:=\left(\varepsilon,\sigma,\sigma_3\right)$.
The subset of spin configurations $s$ assuming the values (\ref{CQ1}) corresponding to the triple ${\boldsymbol\sigma}$ will be denoted by
${\mathcal P}_{\boldsymbol\sigma}\subset {\mathcal P}$.

It will be in order to recapitulate the general theory of completely integrable systems, see \cite[Chapter $49$]{A78},
slightly adapted to our problem. Starting with $n=3$ first integrals $F_j,\, j=1,2,3$ in involution one considers the
corresponding phase flow
\begin{equation}\label{phaseflow}
 {\mathbf t}\in{\mathbbm R}^3 \mapsto \left( g({\mathbf t})=
 g_1^{t_1} g_2^{t_2}g_3^{t_3}:{\mathcal P}_{\boldsymbol\sigma}\rightarrow {\mathcal P}_{\boldsymbol\sigma}\right)
 \;,
\end{equation}
generated by the $F_j$. Let $s_0\in {\mathcal P}_{\boldsymbol\sigma}$ and define the corresponding \textit{stationary subgroup}
$\Gamma$ of ${\mathbbm R}^3$ by
\begin{equation}\label{defGamma}
\Gamma:=\{{\mathbf t}\in {\mathbbm R}^3\left| \right. g({\mathbf t})s_0=s_0\}
\;.
\end{equation}
It can be shown that $\Gamma$ has the form of the lattice
\begin{equation}\label{deflattice}
 \Gamma=\left \{ \sum_{i=1}^{3}n_i\,{\sf e}_i\left|  n_i\in {\mathbbm Z} \mbox{ for } i=1,2,3\right.\right\}
 \;,
\end{equation}
generated by three linearly independent generators ${\sf e}_i\in {\mathbbm R}^3,\; i=1,2,3$.
Then one obtains a diffeomorphism $\Phi:T^3 \rightarrow {\mathcal P}_{\boldsymbol\sigma}$
between the $3$-torus $T^3$ and ${\mathcal P}_{\boldsymbol\sigma}$ defined by
\begin{equation}\label{defdiffeo}
 \Phi\left(\phi_1,\phi_2,\phi_3\right) :=g\left(\frac{1}{2\pi}\left(\phi_1{\sf e}_1+ \phi_2{\sf e}_2+\phi_3{\sf e}_3\right) \right)\,s_0
 \;,
\end{equation}
see \cite[Problem $49.10$]{A78}. This diffeomorphism also yields the three basic cycles
\begin{eqnarray}
\label{defbasiccycle1}
  {\mathcal C}_1 &:& \phi_1\mapsto   \Phi\left(\phi_1,0,0\right) \\
  \label{defbasiccycle2}
  {\mathcal C}_2 &:& \phi_2\mapsto   \Phi\left(0,\phi_2,0\right) \\
  \label{defbasiccycle3}
  {\mathcal C}_3 &:& \phi_3\mapsto   \Phi\left(0,0,\phi_3\right)
 \;.
\end{eqnarray}

For our problem we may choose the first integrals  $F_1=\frac{\sf T}{2\pi}H,\;F_2=S$ and
$F_3={\mathbf S}^{(3)}$, such that $g_1^{\phi_1}$ is the scaled time evolution  with
$\phi_1 =\frac{2\pi}{\sf T} t$,
$g_3^{\phi_3}$ is a global rotation ${\mathcal R}\left({\mathbf E}_3,-\phi_3\right)$,
and, analogously, $g_2^{\phi_2}={\mathcal R}\left({\mathbf S},-\phi_2\right)$.
It follows that the flows $g_2^{\phi_2}$ and $g_3^{\phi_3}$ are already $2\pi$-periodic
and hence it is very plausible that they correspond to the generators ${\sf e}_2$ and ${\sf e}_3$
of the lattice (\ref{deflattice}), i.~e., ${\sf e}_2={\mathbf E}_2$ and ${\sf e}_3={\mathbf E}_3$.

The case of ${\sf e}_1$ is more subtle since the time evolution is, in general, not ${\sf T}$-periodic,
and hence $g_1^{\phi_1}$ would not be $2\pi$-periodic. Note, however, the result $s(t)={\mathcal R}({\mathbf S},\alpha(t))\,r(t)$,
see (\ref{solext}) in Section \ref{sec:TED}, where $r(t)$ is ${\sf T }$-periodic or its equivalent form
\begin{equation}\label{s2r}
  r(t)={\mathcal R}({\mathbf S},-\alpha(t))\,s(t)
  \;.
\end{equation}
If we define $\bar{r}(t)$ as $s(t)$ followed by a corresponding uniform ``averaged" rotation:
\begin{equation}\label{defrbar}
 \bar{r}(t):= {\mathcal R}\left({\mathbf S},-\frac{\alpha({\sf T})}{{\sf T}}t\right)\,s(t)
 \;,
\end{equation}
it can be shown that $\bar{r}(t)$ is also ${\sf T }$-periodic:
\begin{eqnarray}
\label{rbarper1}
  \bar{r}(t+{\sf T}) &\stackrel{(\ref{defrbar})}{=}&
  {\mathcal R}\left({\mathbf S},-\frac{\alpha({\sf T})}{{\sf T}}(t+{\sf T})\right)\,s(t+{\sf T}) \\
   &\stackrel{(\ref{solext})}{=}& {\mathcal R}\left({\mathbf S},-\frac{\alpha({\sf T})}{{\sf T}}(t)\right)\,
   {\mathcal R}({\mathbf S},-\alpha({\sf T}))\,
  {\mathcal R}({\mathbf S},\alpha(t+{\sf T}))\,r(t+{\sf T})\\
  &\stackrel{(\ref{alphadd})}{=}&  {\mathcal R}\left({\mathbf S},-\frac{\alpha({\sf T})}{{\sf T}}(t)\right)\,
   {\mathcal R}({\mathbf S},-\alpha({\sf T}))\,
  {\mathcal R}({\mathbf S},\alpha(t)+\alpha({\sf T}))\,r(t)\\
  &=& {\mathcal R}\left({\mathbf S},-\frac{\alpha({\sf T})}{{\sf T}}(t)\right)\,
   {\mathcal R}({\mathbf S},-\alpha({\sf T}))\, {\mathcal R}({\mathbf S},\alpha({\sf T}))\,
  {\mathcal R}({\mathbf S},\alpha(t))\,r(t)\\
    &\stackrel{(\ref{solext})}{=}& {\mathcal R}\left({\mathbf S},-\frac{\alpha({\sf T})}{{\sf T}}(t)\right)\,s(t)\\
    & \stackrel{(\ref{defrbar})}{=} &\bar{r}(t)
  \;,
\end{eqnarray}
see Figure \ref{FIGBC} for an example.
This suggests that ${\sf e}_1$ has to be chosen corresponding to the first integral
\begin{equation}\label{defF1tilde}
 \widetilde{F_1}:= \frac{\sf T}{2\pi}H-\frac{\alpha({\sf T})}{2\pi}\,S=F_1-\frac{\alpha({\sf T})}{2\pi}\,S
 \;,
\end{equation}
that generates the flow $\widetilde{g}_1^{\phi_1}$ which maps $r(0)$ onto $\bar{r}(\phi_1\frac{\sf T}{2\pi})=\bar{r}(t)$.
Hence ${\sf e}_1= {\mathbf E}_1-\frac{\alpha({\sf T})}{2\pi}\,{\mathbf E}_2$.
Figure \ref{FIGBC} can  be viewed as a plot of the basic cycle ${\mathcal C}_1$ corresponding to  ${\sf e}_1$.\\

Next we turn to the action variables ${\sf I}_i,\,i=1,2,3$. According to \cite[10.2.3]{A78},
these can be obtained by
\begin{equation}\label{AAA}
 {\sf I}_i = \frac{1}{2\pi}\oint_{{\mathcal C}_i} {\mathbf p}\, d{\mathbf q}
 \;,
\end{equation}
where the ${\mathcal C}_i,\,i=1,2,3$ are the basic cycles introduced above and $({\mathbf p},{\mathbf q})$
are canonical coordinates, which in our case can be chosen as $\left(\phi_\mu,\,z_\mu\right)_{\mu=1,2,3}$,
see Section \ref{sec:CM}. Since our phase space is simply connected each closed curve in
${\mathcal P}={\mathcal S}^2\times {\mathcal S}^2\times {\mathcal S}^2$ can be viewed as the boundary of a
surface in ${\mathcal P}$, especially ${\mathcal C}_i=\partial {\mathcal A}_i$ for $i=1,2,3$.
Further, ${\mathcal C}_i$ can be projected onto the $\mu$-th factor ${\mathcal S}^2$ of ${\mathcal P}$
and thus yields a closed curve
${\mathcal C}_{i\mu}$ that is the boundary of some surface ${\mathcal A}_{i\mu} \subset {\mathcal S}^2, \; i,\mu=1,2,3$.
In other words: If the spin system $s$ runs through a basis cycle ${\mathcal C}_i$ then each single spin ${\mathbf s}_\mu$
describes a closed curve $\partial {\mathcal A}_{i\mu}$ on its Bloch sphere. By applying Stoke's theorem, see, e.~g.,
\cite[7.5.4]{A78}, we obtain
\begin{equation}\label{aai}
 {\sf I}_i(s) = \frac{1}{2\pi}\sum_\mu \int_{{\mathcal A}_{i\mu}} \,d{\phi_\mu}\wedge d{z_\mu}=:
 \frac{1}{2\pi}\sum_\mu \left| {\mathcal A}_{i\mu}\right|
 \;,
\end{equation}
where $\left|  {\mathcal A}_{i\mu}\right|$ denotes the (oriented) area of the surface  ${\mathcal A}_{i\mu}$, the
sign depending on the orientation of the projection of the basic cycle ${\mathcal C}_i$ onto ${\mathcal S}^2$.

We will first apply these considerations to the simplest case of the basis cycle ${\mathcal C}_3$.
Here each spin vector ${\mathbf s}_\mu$ describes a clockwise rotation about the $3$-axis and hence encircles a spherical cap
${\mathcal A}_{3,\mu}$ of area $\left| {\mathcal A}_{3,\mu}\right|=2\pi \left(1-z_\mu\right)$. Due to the
clockwise direction the area has to be multiplied by a minus sign.  This implies
\begin{equation}\label{aa3}
 {\sf I}_3 (s)= - \frac{1}{2\pi}\sum_\mu \left| {\mathcal A}_{3,\mu}\right|= -\sum_\mu \left(1-z_\mu\right)=  \sigma_3-3
 \;,
\end{equation}
or, writing ${\sf I}_3$ as a function of phase space,
\begin{equation}\label{AA3}
  {\sf I}_3= {\mathbf S}^{(3)}-3
  \;.
\end{equation}
The corresponding angle variable is
\begin{equation}\label{defpsi3}
 \psi_3=\phi_3
 \;.
\end{equation}

The case of ${\mathcal C}_2$ is completely analogous and leads to the action variable
\begin{equation}\label{AA2}
  {\sf I}_2= S-3
  \;,
\end{equation}
where $S$ is understood as a function on phase space with values $\sqrt{3+2\sigma}$.
The corresponding angle variable is
\begin{equation}\label{defpsi2}
  \psi_2= \Omega_2 t =\frac{\alpha({\sf T})}{\sf T}\,t\stackrel{(\ref{a0})}{=}a_0\,t
  \;,
\end{equation}
where $a_0$ is the zeroth coefficient of the Fourier series (\ref{fourier}) of $\dot{\alpha}(t)$.
(\ref{defpsi2}) describes the averaged uniform rotation of $s(t)$ about ${\mathbf S}$ according to (\ref{defrbar}).

The case of  ${\sf I}_1$ is more complicated. Recall that the basis cycle ${\mathcal C}_1$ yields
three closed orbits generated by $\widetilde{F}_1$ in ${\mathcal S}^2$, see Figure \ref{FIGBC} for an example.
The corresponding value of ${\sf I}_1$ is hence the  sum of the signed areas
swept by the three orbits ${}$ over $2\pi$, see (\ref{aai}), and can be most conveniently calculated numerically, see
Figure \ref{FIGPINT} for an example. As a check for this procedure we consider the corresponding angle variable defined as
\begin{equation}\label{defpsi1}
  \psi_1:= \frac{2\pi}{\sf T}\,t
  \;,
\end{equation}
and the Hamiltonian equation of motion
\begin{equation}\label{Hampsi1}
  \dot{\psi_1}= \Omega_1= \frac{2\pi}{\sf T} = \frac{\partial H}{\partial {\sf I}_1}
  \;.
\end{equation}
This implies that, similar as for one-dimensional problems \cite[10.2.2]{A78}, we have
\begin{equation}\label{diffaa1}
 \frac{\partial {\sf I}_1(\sigma,\varepsilon)}{\partial \varepsilon}= \frac{1}{2\pi}{\sf T}(\sigma,\epsilon)
 \;,
\end{equation}
see Figure \ref{FIGLPT}. Here we implicitly used the considerations of Appendix \ref{sec:AR},
which lead to a smooth dependence of $\alpha({\sf T})$ and thus of $ {\sf I}_1$ on $\varepsilon$ even across critical energies.

Hence an alternative representation of ${\sf I}_1$ would be
\begin{equation}\label{altdefI1}
{\sf I}_1(\sigma,\varepsilon)={\sf I}_1(\sigma,\varepsilon_1)+\frac{1}{2\pi}\,\int_{\varepsilon_1}^{\varepsilon}{\sf T}(\sigma,\epsilon')\,d\varepsilon'
\;,
\end{equation}
where the lower limit $\varepsilon_1$ of the integral is one of the extremal values $E_{\scriptsize{min}}(\sigma)$ or $E_{\scriptsize{max}}(\sigma)$
of the energy $H$, restricted to the
subset of configurations $s\in{\mathcal P}$ satisfying $H_1(s)= \sigma$, see Figure \ref{FIGCP1}.
The values of the limits ${\sf I}_1(\sigma,\varepsilon_1)$ and ${\sf I}_1(\sigma,\varepsilon_2)$ can be calculated by using the fact that the time evolution
corresponding to such limit points is given by a uniform rotation about ${\mathbf S}$, see Appendix \ref{sec:SN}.
For $\varepsilon=\varepsilon_2$ the first spin vector ${\mathbf s}_1$ anti-clockwise
encircles a spherical cap of the area $2\pi(1-{\mathbf s}_1\cdot \frac{\mathbf S}{S})$,
and analogously for  ${\mathbf s}_2$ and  ${\mathbf s}_3$. It follows that
\begin{equation}\label{I1limit}
 {\sf I}_1(\sigma,\varepsilon_2)\stackrel{(\ref{aai})}{=} \sum_\mu \left(1-{\mathbf s}_\mu\cdot \frac{\mathbf S}{S}\right)=3-S
 =-{\sf I}_2(\sigma,\varepsilon_2)
 \;.
\end{equation}
The coincidence of two action variables, up to a sign, at these limit points does not contradict their independence in the proper domain of definition.

In the example of Figure \ref{FIGLPT} we have $S=\sqrt{3}$ and thus obtain ${\sf I}_1(\sigma=0,\varepsilon_2)= 3-\sqrt{3}\approx 1.26795$.
Further,  ${\sf I}_1(\sigma=0,\varepsilon_1)=0$ since for this limit the orbit ${\mathcal C}_1$ degenerates to a point in ${\mathcal P}$.\\

In Appendix \ref{sec:MF} we have considered the case of a time-dependent magnetic field ${\mathbf B}(t)=B(t)\,{\mathbf e}$
into a constant direction ${\mathbf e}$, that will henceforward be
identified with the  $3$-direction of the chosen coordinate system.
If we further assume that the function $t\mapsto B(t)$ is periodic with a period ${\sf P}$
we encounter an interesting generalization of the action-angle scenario that is originally confined to time-independent Hamiltonians,
but see \cite{FGS03} for a generalization to the time-dependent case.
Let
\begin{equation}\label{BFour}
 B(t)= \sum_{n\in{\mathbbm Z}} b_n\, \exp\left(\frac{2\,\pi\,{\sf i}\,n\,t}{\sf P} \right)
\end{equation}
denote the corresponding Fourier series and assume additionally that $b_0\neq 0$.
Then we re-define the angle variable $\psi_3$ as the scaled time variable
\begin{equation}\label{defpsi3}
 \psi_3:= b_0\,t
 \;,
\end{equation}
such that
\begin{equation}\label{omega3}
  \dot{\psi}_3 = b_0 =: \Omega_3
  \;.
\end{equation}
Hence the equation of motion for the angle variable ${\psi}_3$ assumes the usual form despite a time-dependent Hamiltonian
of the form (\ref{HamZeeman}).

\section{Special cases}\label{sec:SC}

We have obtained the solution of equations of motion (\ref{eom1} - \ref{eom3}) only for the ``generic case" considered in Definition
\ref{DefGeneric}, i.~e.,
for pairwise different coupling constants $J_1, J_2,J_3$ and  $L\cap \partial {\mathcal G}=[G_1,G_2]$,
$G_1$ and $G_2$ being distinct regular points of $\partial{\mathcal G}$.
In this Appendix, we will discuss the special cases in which these conditions are violated.
These cases fall into three classes,
special coupling constants and special initial conditions and singular endpoints. Further it will be in order to identify
all stationary states of the equations of motion (\ref{eom1} - \ref{eom3}).

First, since we have explicitly excluded the case where two of the three coupling constants coincide in Section \ref{sec:TG}
we will have to treat this case separately in the following Appendix \ref{sec:IST}.
Recall that in Section \ref{sec:TID} we have assumed that the two equations (\ref{Huvw}) and (\ref{H1uvw}) expressing conservation
of energy and total spin length define a line $L$ parallel to the vector ${\mathbf n}$ given by (\ref{line}).
In the case of $J_1=J_2=J_3\equiv J$ these two conservation laws are no longer independent and ${\mathbf n}$ vanishes.
However, in this case (the ``equilateral spin triangle") the time evolution is well-known and assumes a particular simple form:
All spin vectors rotate about the constant total spin vector ${\mathbf S}$ with the same angular frequency $J S$.
This will also follow as a special case of the time evolution of an isosceles spin triangle.

Second, in the case where not all $J_1, J_2,J_3$ are equal and hence the line $L$ is well-defined and, by definition, intersects the Gram set
${\mathcal G}$, there are certain exceptions from the generic case of $L\cap \partial {\mathcal G}=[G_1,G_2]$,
$G_1$ and $G_2$ being regular points of  $\partial {\mathcal G}$, that are already mentioned
in Appendix \ref{sec:TG}: $L\cap \partial {\mathcal G}$ may consist of a single  point or of a one-dimensional face of the form
$[{\mathbf e}_n,{\mathbf e}_m],\;0\le n<m\le 3$. The latter case will be considered in Appendix \ref{sec:SCIST},
and leads to a special case of the the isosceles spin triangle to be discussed in Appendix \ref{sec:IST}.
The former case of  $L\cap \partial {\mathcal G}=\{G_0\}$ is treated in Appendix \ref{sec:SN}.
Recall, that also in the general theory of completely integrable mechanical systems the foliation of the phase
space by invariant tori need not be complete and
one has to allow for special cases. There may exist certain ``separatrices" separating the domains of the phase space
filled with invariant tori, e.~g.,
the aperiodic motions of the one-dimensional pendulum separating oscillations and rotations, see \cite[10.2.3]{A78}.

Then, in Appendix \ref{sec:TSG}, we consider the case $L\cap \partial {\mathcal G}=[{\mathbf e_1},G]$, where the motion of the spin configuration
turns out to be aperiodic and, finally, in Appendix \ref{sec:SS}, enumerate all stationary states.

\subsection{Time evolution of the isosceles spin triangle}\label{sec:IST}

\begin{figure}[htp]
\centering
\includegraphics[width=1.0\linewidth]{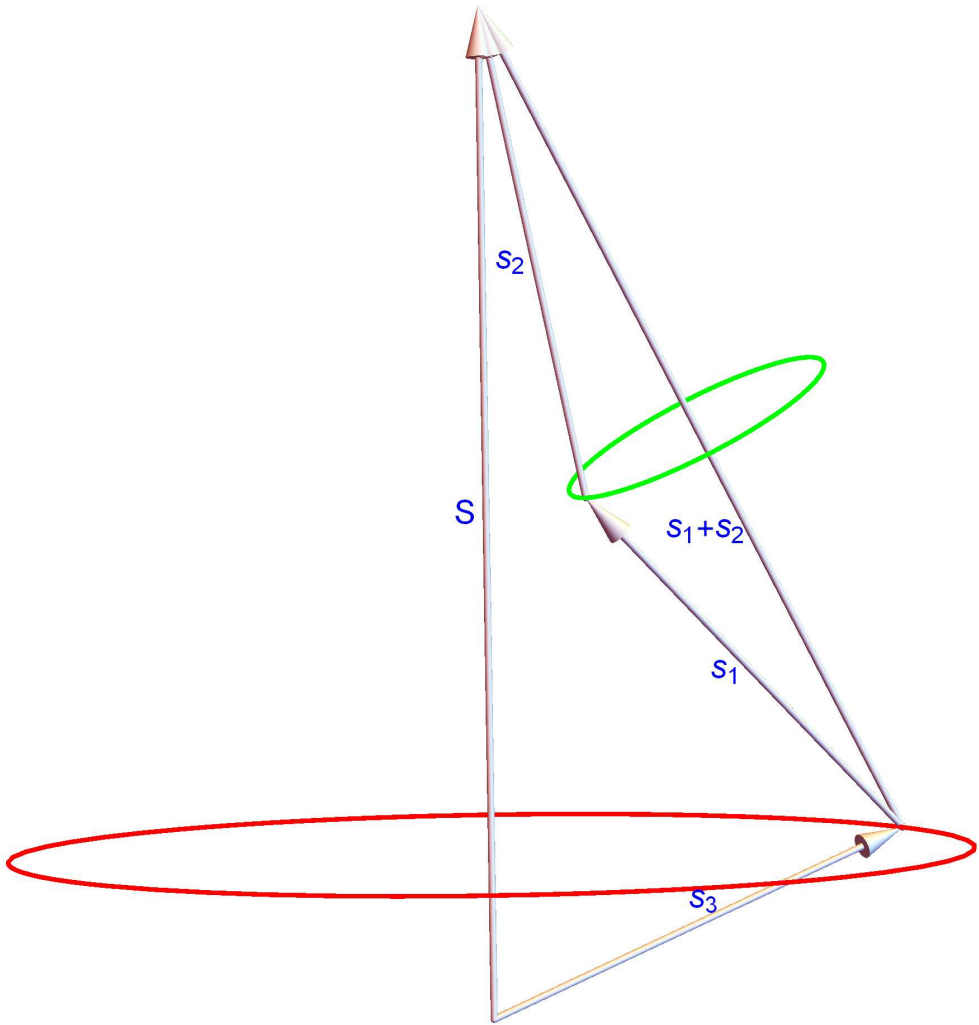}
\caption{Time evolution of the isosceles spin triangle $\left({\mathbf s}_1,{\mathbf s}_2,{\mathbf s}_3\right)$
according to (\ref{s1ti} - \ref{s3ti}). We have indicated the superposition of two uniform rotations,
a global one about the total spin vector ${\mathbf S}$ (red circle), and a local one about the momentary position of
${\mathbf s}_1+{\mathbf s}_2$ (green circle).
}
\label{FIGIST}
\end{figure}

Without loss of generality we will only consider the case $J_1=J_2\equiv J$. Then $g=0$ according to (\ref{solg}) and the transformation to
the variable $x$ is no longer possible. Instead, we will solve the equations of motion (\ref{eom1} - \ref{eom3}) directly.
The last one (\ref{eom3}) reduces to
\begin{equation}\label{eom3ist}
\dot{\mathbf s}_3= J\left({\mathbf s}_1+{\mathbf s}_2\right)\times {\mathbf s}_3 = J\,{\mathbf S}\times {\mathbf s}_3
\;,
\end{equation}
and, recalling that the total spin vector ${\mathbf S}$ is a constant of motion, leads to a uniform rotation of ${\mathbf s}_3$
about ${\mathbf S}$ with the angular frequency $J S$, where $S$ is the constant total spin length.
The remaining spin vectors will be described in a correspondingly rotating
coordinate system. This means that we pass to vectors
\begin{equation}\label{s2r}
{\mathbf r}_\mu  := R(t)\,{\mathbf s}_\mu,\; \mbox{ for } \mu=1,2,3,
\end{equation}
such that the time dependent rotation $R(t)$ satisfies
\begin{equation}\label{Rt}
  \dot{R}\,{\mathbf a} = -J {\mathbf S}\times (R {\mathbf a}),\; \mbox{ for all } {\mathbf a}\in{\mathbbm R}^3,
  \; \mbox{ and }  R\,{\mathbf S}={\mathbf S}
  \;.
\end{equation}
We obtain
\begin{eqnarray}
\label{r1dot1}
 \dot{\mathbf r}_1 &=& \dot{R} \,{\mathbf s}_1+R\, \dot{\mathbf s}_1\\
 \label{r1dot2}
  &\stackrel{(\ref{Rt},\ref{eom1})}{=}& -J\,{\mathbf S}\times {\mathbf r}_1+ \left(J\,{\mathbf r}_3+J_3\,{\mathbf r}_2 \right)\times {\mathbf r}_1\\
  \label{r1dot3}
  &=&  -J\,\left({\mathbf r}_1+{\mathbf r}_2+{\mathbf r}_3 \right)\times {\mathbf r}_1
  + \left(J\,{\mathbf r}_3+J_3\,{\mathbf r}_2 \right)\times {\mathbf r}_1\\
  \label{r1dot4}
  &=& \left( J_3 - J\right)\,{\mathbf r}_2\times {\mathbf r}_1
  \;,
\end{eqnarray}
and, analogously,
\begin{equation}\label{r2dot}
   \dot{\mathbf r}_2= \left( J_3 - J\right)\,{\mathbf r}_1\times {\mathbf r}_2
   \;.
\end{equation}
This implies $\frac{d}{dt}\left( {\mathbf r}_1+{\mathbf r}_2\right)={\mathbf 0}$ which is consistent with the
fact that $ {\mathbf r}_3$ and hence also  ${\mathbf r}_1+{\mathbf r}_2$ are constant in the rotating frame.
Moreover,
\begin{equation}\label{r1mr2d}
\frac{d}{dt}\left( {\mathbf r}_1-{\mathbf r}_2\right)\stackrel{(\ref{r1dot4},\ref{r2dot})}{=}
2\,\left(J_3 - J\right) {\mathbf r}_2\times {\mathbf r}_1
= \left(J_3 - J\right)\left( {\mathbf r}_1+{\mathbf r}_2\right) \times \left( {\mathbf r}_1-{\mathbf r}_2\right)
\;.
\end{equation}
This means that the difference vector $\left( {\mathbf r}_1-{\mathbf r}_2\right)$ uniformly rotates about the constant
vector $\left( {\mathbf r}_1+{\mathbf r}_2\right)$ with angular velocity
$\omega_{12}=\left(J_3 - J\right)\left| {\mathbf r}_1+{\mathbf r}_2\right|=:\left(J_3 - J\right)r_{12}$. We thus have a superposition of two rotations:
A ``global" one of ${\mathbf s}_3$ and  ${\mathbf s}_1+ {\mathbf s}_2$   about ${\mathbf S}$ and a ``local" one of
$\left( {\mathbf r}_1-{\mathbf r}_2\right)$  about $\left( {\mathbf r}_1+{\mathbf r}_2\right)$, see Figure \ref{FIGIST}.

We will give the explicit solution $s(t)$ with the simplification resulting from the choice of $s(0)$ being coplanar,
and assuming that $r_{12}>0$, such that
\begin{equation}\label{st0}
 {\mathbf s}_1(0)=\frac{1}{2}\left(
\begin{array}{c}
-\sqrt{1-\alpha ^2}+\frac{\beta  (S-\alpha )}{r_{12}} \\
 0 \\
 -\alpha +\frac{\sqrt{1-\alpha ^2} \beta }{r_{12}}+S \\
\end{array}
\right),\quad
 {\mathbf s}_2(0)=\frac{1}{2}\left(
\begin{array}{c}
 -\sqrt{1-\alpha ^2}-\frac{\beta  (S-\alpha )}{r_{12}} \\
 0 \\
 -\alpha -\frac{\sqrt{1-\alpha ^2} \beta }{r_{12}}+S \\
\end{array}
\right),\quad
{\mathbf s}_3(0)=   \left(
\begin{array}{c}
 \sqrt{1-\alpha ^2} \\
 0 \\
 \alpha  \\
\end{array}
\right)
\;.
\end{equation}
Recall that $S=\sqrt{3+2\sigma}$ is the constant total spin length. Further parameters are
\begin{equation}\label{defabc}
\alpha:= \frac{1}{S}(1+u(0)+v(0)),\quad \beta:= \sqrt{3-S^2+2 \alpha  S}, \mbox{ and }\quad r_{12}=\sqrt{S^2-2 \alpha  S+1}
\;.
\end{equation}

With these abbreviations the explicit solution assumes the form
\begin{equation}\label{s1ti}
 {\mathbf s}_1(t)=\frac{1}{2}\left(
\begin{array}{c}
  -\cos (J S t) \left(\sqrt{1-\alpha ^2}-\frac{\beta  (S-\alpha) \cos \left(t \omega_{12}\right)}{r_{12}}\right)
  -\beta  \sin \left(t \omega_{12}\right) \sin (J S t) \\
 -\sin (J S t) \left(\sqrt{1-\alpha ^2}-\frac{\beta  (S-\alpha ) \cos \left(t \omega_{12}\right)}{r_{12}}\right)
 +\beta  \sin \left(t \omega_{12}\right) \cos (J S t) \\
 -\alpha +\frac{\sqrt{1-\alpha ^2} \beta  \cos \left(t \omega_{12}\right)}{r_{12}}+S \\
\end{array}
\right)
\;,
\end{equation}

\begin{equation}\label{s2ti}
 {\mathbf s}_2(t)=\frac{1}{2}\left(
\begin{array}{c}
 -\cos (J S t) \left(\sqrt{1-\alpha ^2}+\frac{\beta  (S-\alpha) \cos \left(t \omega_{12}\right)}{r_{12}}\right)
 +\beta  \sin \left(t \omega_{12}\right) \sin (J S t) \\
 -\sin (J S t) \left(\sqrt{1-\alpha ^2}+\frac{\beta  (S-\alpha ) \cos \left(t \omega_{12}\right)}{r_{12}}\right)
 -\beta  \sin \left(t \omega _{12}\right) \cos (J S t)
   \\
 -\alpha -\frac{\sqrt{1-\alpha ^2} \beta  \cos \left(t \omega_{12}\right)}{r_{12}}+S \\
\end{array}
\right)
\;,
\end{equation}
and
\begin{equation}\label{s3ti}
 {\mathbf s}_3(t)=\left(
\begin{array}{c}
 \sqrt{1-\alpha ^2} \cos (J S t) \\
 \sqrt{1-\alpha ^2} \sin (J S t) \\
 \alpha  \\
\end{array}
\right)
\;.
\end{equation}

As mentioned above, the transformation of the equation of motion to the Weierstrass differential equation
is not possible in the isosceles triangle case. This can be further made plausible since $u(t)\equiv u(0)$ in this
case and hence $\det s(t)$ is the square root of a quadratic polynomial, not a cubic one, the corresponding integration
leading to trigonometric functions of $t$.

In the special case of an equilateral spin triangle, i.~e., $J_1=J_2=J_3\equiv J$ the angular frequency $\omega_{12}$ vanishes
and the solution (\ref{s1ti}.\ref{s2ti}) specializes to
\begin{equation}\label{s1te}
 {\mathbf s}_1(t)=\frac{1}{2}\left(
\begin{array}{c}
  -\cos (J S t) \left(\sqrt{1-\alpha ^2}-\frac{\beta  (S-\alpha)}{r_{12}}\right)
  \\
 -\sin (J S t) \left(\sqrt{1-\alpha ^2}-\frac{\beta  (S-\alpha ) }{r_{12}}\right)
  \\
 S-\alpha +\frac{\sqrt{1-\alpha ^2} \beta }{r_{12}} \\
\end{array}
\right)
\;,
\end{equation}
and
\begin{equation}\label{s2te}
 {\mathbf s}_2(t)=\frac{1}{2}\left(
\begin{array}{c}
 -\cos (J S t) \left(\sqrt{1-\alpha ^2}+\frac{\beta  (S-\alpha) }{r_{12}}\right)
\\
 -\sin (J S t) \left(\sqrt{1-\alpha ^2}+\frac{\beta  (S-\alpha ) }{r_{12}}\right)
\\
S -\alpha -\frac{\sqrt{1-\alpha ^2} \beta  }{r_{12}} \\
\end{array}
\right)
\;,
\end{equation}
whereas (\ref{s3ti}) is unchanged. This is the uniform rotation of all spin vectors about the constant total spin vector
mentioned above as the well-known solution of the equation of motion in the equilateral spin triangle case.

\subsection{Time evolution of the isosceles spin triangle, special case}\label{sec:SCIST}
We consider the case where $L\cap \partial {\mathcal G}$ is a one-dimensional face of ${\mathcal G}$.
Let, without loss of generality, $L\cap \partial {\mathcal G}=[{\mathbf e}_1,{\mathbf e}_2]$.
Recall that these singular extremal points correspond to the spin configurations ${\mathbf e}_1\simeq\uparrow\downarrow\downarrow$
with Gram matrix entries $u=1,\,v=w=-1$ and ${\mathbf e}_2\simeq\uparrow\downarrow\uparrow$
with Gram matrix entries $v=1,\,u=w=-1$. It follows from (\ref{line}) that $L$ is parallel to the vector
\begin{equation}\label{faceline}
{\mathbf n}= \left(
\begin{array}{c}
 J_2-J_3\\
 J_3-J_1 \\
 J_1-J_2 \\
\end{array}
\right)\sim
\left(
\begin{array}{r}
 1\\
 -1\\
 -1 \\
\end{array}
\right)-
\left(
\begin{array}{r}
 -1\\
  1\\
 -1 \\
\end{array}
\right)=
\left(
\begin{array}{r}
 2\\
 -2\\
 0\\
\end{array}
\right)
\;,
\end{equation}
and hence we encounter the case $J_1=J_2\equiv J$  of an isosceles spin triangle already treated in subsection \ref{sec:IST}.
Moreover, since ${\mathbf s}_1\cdot {\mathbf s}_2=w=-1$ for ${\mathbf e}_1$ and ${\mathbf e}_2$, this holds for all points of the one-dimensional
face spanned by these extremal points and hence ${\mathbf s}_1=-{\mathbf s}_2$ for the corresponding spin configurations $s$.
It follows that ${\mathbf s}_3={\mathbf s}_1+{\mathbf s}_2+{\mathbf s}_3={\mathbf S}$ is a constant of motion.
The ``general" solution (\ref{s1ti} - \ref{s3ti}) of the equation of motion in the isosceles case cannot be applied
since we obtain $r_{12}=\left| {\mathbf s}_1+{\mathbf s}_2\right|=0$
and (\ref{s1ti} - \ref{s3ti}) was derived under the assumption $r_{12}>0$.
Nevertheless, the equations of motion (\ref{eom1} - \ref{eom3}) can be directly solved with the result
\begin{equation}\label{solline}
  {\mathbf s}_3= \left(
\begin{array}{r}
 0\\
 0\\
 1\\
\end{array}
\right)
\quad \mbox{and}\quad
{\mathbf s}_1=-{\mathbf s}_2=
\left(
\begin{array}{c}
 \sqrt{1-\gamma^2}\,\cos J\,t\\
 \sqrt{1-\gamma^2}\,\sin J\,t\\
 \gamma\\
\end{array}
\right)
\;,
\end{equation}
with a constant parameter $-1<\gamma<1$. This implies that not only $w=-1$ but also $v=\gamma$ and $u=-\gamma$
are constant and the Gram matrix $G(s(t))$ of the spin configuration $s(t)$ is time-independent.

The latter result is also plausible if we consider the line $L$ as the limit of lines $L'$ that cut the interior of
${\mathcal G}$ and where the analysis of Section \ref{sec:TG} applies. Recall that $\dot{u}$ is proportional
to $\pm\sqrt{\det G}$, see Eq.~(\ref{ud5}), and the latter becomes smaller and smaller and hence the motion of the
Gram matrix is frozen if $L'$ approaches $L\subset \partial {\mathcal G}$.

\subsection{Time evolution corresponding to a stationary Gram matrix}\label{sec:SN}
We consider the special case where the conservation laws for $H$ and $H_1$ are only compatible with a single Gram matrix $G$, i.~e.,
$L\cap \partial {\mathcal G}=\{G\}$, and hence $G$ is stationary. If $G$ is one of the four singular extremal points ${\mathbf e}_n,\,n=0,\ldots,3$
corresponding to collinear spin configurations then the cross products in (\ref{eom1} - \ref{eom3}) vanish and these spin configurations are also stationary.

Hence we will restrict our discussion to the case where $G$ is a regular point, i.~e., belongs to the smooth part of $\partial {\mathcal G}$.
This special case can also be viewed as the limit case of the solution in terms of the Weierstrass elliptic function if the two lowest roots $x_1$ and $x_2$ of
$\Pi(x)$ merge to a double root. In this limit the oscillation of $x(t)$ becomes a harmonic one with a finite limit (\ref{limperiod}) of the period ${\sf T}$
and the amplitude $\left| x_2-x_1\right|$ of the oscillation vanishes in accordance with the stationarity of $G(t)$.

To calculate the time evolution of the spin configuration $s(t)$ we may argue as follows:
$G$ is the Gram matrix of a co-planar spin configuration $s(t)$  that, for $t=0$ and upon choosing a suitable coordinate system, assumes the form
\begin{equation}\label{coplanar}
{\mathbf s}_\mu=\left(
\begin{array}{c}
 \sin \phi_\mu\\
 0\\
\cos\phi_\mu\\
\end{array}
\right)
\quad \mbox{for }\mu=1,2,3
\;.
\end{equation}
Physically, $s(t)$ is a \textit{relative (anti-)ground state} of the spin system with Hamiltonian (\ref{defH}), i.~e.,
a spin configuration with minimal (maximal) energy among all configurations with the same total spin length.

Further, we may choose the $\phi_\mu, \mu=1,2,3$ in such a way that the total spin vector will be
\begin{equation}\label{totalS0}
 {\mathbf S}=\left(
\begin{array}{c}
 0\\
0\\
\sum_\mu \cos\phi_\mu\\
\end{array}
\right)
\;,
\end{equation}
which is equivalent to
\begin{equation}\label{sumsin}
  \sum_\mu \sin \phi_\mu=0
  \;.
\end{equation}
For $S>0$ it follows that the only possible time evolution of $s(t)$ is a uniform rotation about the $3$-axis, which is of the form
\begin{equation}\label{coplanarrot}
{\mathbf s}_\mu(t)=\left(
\begin{array}{c}
 \cos\omega t\,\sin\phi_\mu\\
 \sin\omega t\,\sin\phi_\mu\\
\cos\phi_\mu\\
\end{array}
\right)
\quad \mbox{for }\mu=1,2,3
\;,
\end{equation}
for some real parameter $\omega$. Inserting the expressions (\ref{coplanarrot}) into the equations of motion
(\ref{eom1} - \ref{eom3}) yields three solutions for $\omega$ of the form
\begin{eqnarray}
\label{omega1}
\omega&=&\csc \left(\phi _1\right) \left(J_2 \sin \left(\phi _1-\phi _3\right)+J_3 \sin \left(\phi _1-\phi _2\right)\right),\\
\label{omega2}
 \omega &=& -\csc \left(\phi _2\right) \left(J_3 \sin \left(\phi _1-\phi _2\right)-J_1 \sin \left(\phi _2-\phi _3\right)\right), \\
 \label{omega3}
 \omega &=& -\csc \left(\phi _3\right) \left(J_1 \sin \left(\phi _2-\phi _3\right)+J_2 \sin \left(\phi _1-\phi _3\right)\right),
\end{eqnarray}
that are, however, all equal by virtue of the conditions
\begin{equation}\label{cond1}
  {\mathbf n}= \left(
\begin{array}{c}
 J_2-J_3\\
 J_3-J_1 \\
 J_1-J_2 \\
\end{array}
\right)
\,
\perp
\,
 \left(
\begin{array}{c}
v\, w -u\\
u\,w -v \\
u\,v-w\\
\end{array}
\right)
=\frac{1}{2}\nabla \det G
\;,
\end{equation}
and (\ref{sumsin}). (\ref{cond1}) holds since the line $L$ with direction ${\mathbf n}$ lies in the tangent plane
of $\partial {\mathcal G}$ at the regular point $G$.

The solutions (\ref{solline}) appear as a special case of (\ref{coplanarrot}) where $\phi_3=0, \,J_1=J_2=J$ and $\phi_1-\phi_2=\pm \pi$
which together with (\ref{omega1}) yields $\omega=J$.

It remains to study the case $S=0$ of vanishing total spin ${\mathbf s}_1+{\mathbf s}_2+{\mathbf s}_3={\mathbf 0}$.
The corresponding Gram matrix will be $G\widehat{=}(-\scriptsize{\frac{1}{2}},-\scriptsize{\frac{1}{2}},-\scriptsize{\frac{1}{2}})$.
The three spin vectors form an equilateral triangle that may be rigidly rotated in spin space. We obtain
\begin{equation}\label{S0s1}
  \dot{\mathbf s}_1= \left( J_2\,{\mathbf s}_3+J_3\,{\mathbf s}_2\right)\times {\mathbf s}_1
   = \left( J_2\,\left(- {\mathbf s}_1- {\mathbf s}_2\right)+J_3\,{\mathbf s}_2\right)\times {\mathbf s}_1=
 \left(J_3-J_2\right){\mathbf s}_2\times {\mathbf s}_1
  \;,
\end{equation}
and, analogously,
\begin{equation}\label{S0s2}
\dot{\mathbf s}_2= \left(J_3-J_1\right){\mathbf s}_1\times {\mathbf s}_2
\;.
\end{equation}
Hence
\begin{equation}\label{dmdt}
 {\mathbf 0}= \frac{d}{dt}\left( \left(J_3-J_1 \right){\mathbf s}_1+\left(J_3-J_2\right){\mathbf s}_2\right)=:\frac{d}{dt}{\mathbf m}
 \;,
\end{equation}
and the vector ${\mathbf m}$ is a constant of motion. We further conclude
\begin{equation}\label{dsmudt}
  \dot{\mathbf s}_\mu = {\mathbf m}\times {\mathbf s}_\mu,\quad \mbox{ for } \mu=1,2,3
  \;,
\end{equation}
and hence the spin vectors uniformly rotate about ${\mathbf m}$:
\begin{equation}\label{smurot}
 {\mathbf s}_\mu(t) = {\mathcal R}\left({\mathbf m} ,m t\right)  {\mathbf s}_\mu(0),\quad \mbox{ for } \mu=1,2,3
 \;,
\end{equation}
with angular velocity
\begin{equation}\label{angvel}
 m=\left| {\mathbf m}\right|=\sqrt{J_1^2+J_2^2+J_3^2-J_1 J_2-J_2 J_3-J_3 J_1}
 \;.
\end{equation}
This concludes the case of stationary Gram matrices.

\subsection{Aperiodic time evolution towards a singular extremal point}\label{sec:TSG}

\begin{figure}[htp]
\centering
\includegraphics[width=0.7\linewidth]{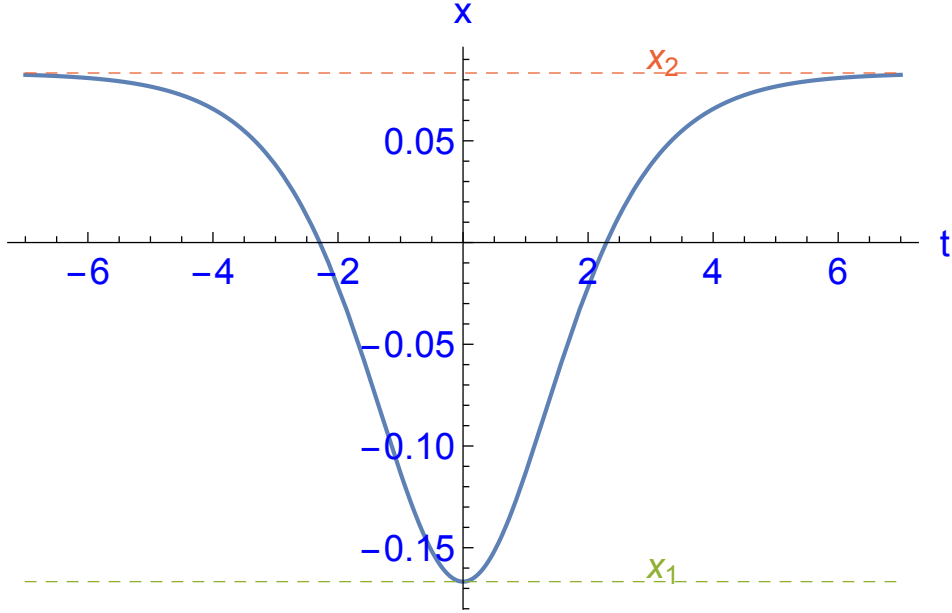}
\caption{Plot of the solution $x(t)$ for the aperiodic limit case according to (\ref{xtasy}), where the parameter $\lambda$
has been chosen as $\lambda=\frac{1}{2}$. We see that for $t\to \pm \infty$ the singular extremal point correspondig to the
value $x=x_2$ is asymptotically approached.
}
\label{FIGAL}
\end{figure}

We consider the case where one of the endpoints of the interval $L\cap \partial {\mathcal G}=[G_1,G_2]$ is a singular
extremal point of ${\mathcal G}$. The case of the singular point being ${\mathbf e}_0$ can be excluded since ${\mathbf e}_0$
is the only Gram matrix with $u+v+w=\sigma=3$. Hence we consider, without loss of generality, the case
$L\cap \partial {\mathcal G}=[{\mathbf e}_1,G]$. Recall that ${\mathbf e}_1$ has the coordinates $u=1,\,v=w=-1$, see (\ref{singular}).

In this case we have $u+v+w=\sigma=-1$ and the plane $P_{\sigma=-1}$ is spanned by the three singular extremal points ${\mathbf e}_n,\,n=1,2,3$.
Moreover, ${\mathcal G}\cap P_{\sigma=-1}$ is the equilateral triangle spanned by ${\mathbf e}_n,\,n=1,2,3$. Thus the remaining end point $G$
lies in the interior of the interval $[{\mathbf e}_2,{\mathbf e}_3]$ and has the representation
\begin{equation}\label{Grep}
  G=\lambda {\mathbf e}_2+ (1-\lambda){\mathbf e}_3\,\widehat{=}\,
  \left(
  \begin{array}{c}
    -1 \\
    2\lambda -1 \\
   1-2\lambda
  \end{array}
  \right)
 \;,
\end{equation}
with some parameter $0<\lambda<1$. According to the remarks after (\ref{defopJ}) we can,
without loss of generality, choose the coupling constants  $J_1=\lambda, J_2=1,J_3=0$
in order to satisfy (\ref{line}). In fact, this choice implies
\begin{equation}\label{npropdJ}
  G-{\mathbf e}_1 \,\widehat{=}\,2\,\left(
  \begin{array}{r}
    -1 \\
    \lambda  \\
   1-\lambda
  \end{array}
  \right)\sim
  \left(
  \begin{array}{c}
   J_3-J_2 \\
    J_1-J_3  \\
   J_2-J_1
  \end{array}
  \right)
  \;,
\end{equation}
and yields the constant value of the energy $\varepsilon=\lambda-1$.
The transformation to the function $x(t)$ according to Section \ref{sec:TG} leads to the differential equation
\begin{equation}\label{polyasy}
\dot{x}^2=\Pi(x)=-\frac{4}{27} \left(\lambda ^2-\lambda +3 x\right)^2 \left(2 \lambda ^2-2 \lambda -3
   x\right)
   \;,
\end{equation}
where the polynomial $\Pi(x)$ has a single root at
\begin{equation}\label{x1root}
  x_1=-\frac{2}{3} (1-\lambda ) \lambda
  \;,
\end{equation}
and a double one at
\begin{equation}\label{x2root}
  x_2=\frac{1}{3} (1-\lambda ) \lambda
  \;.
\end{equation}
It follows that the integral $\int dt=\int \frac{dx}{\sqrt{\Pi(x)}}$ can be evaluated in terms of elementary functions
and solved for $x(t)$ with the result:
\begin{equation}\label{xtasy}
 x(t)=x_1+\left( x_2-x_1\right) \tanh ^2\left(\gamma\, t\right)
 \;,
\end{equation}
where we have set
\begin{equation}\label{defgamma}
  \gamma:= \sqrt{(1-\lambda )\lambda}
  \;,
\end{equation}
see Figure \ref{FIGAL} for an example. For $t\to \pm\infty$ the extremal value ${\mathbf e}_1$ corresponding to the
value $x=x_2$ is asymptotically approached.
According to the prefactor $\gamma= \sqrt{(1-\lambda ) \lambda }$ of $t$ this approach is slowed down if $\lambda$ is close to $0$ or $1$,
i.~e.~, if $G$ is close to one of the singular extremal points ${\mathbf e}_2$ or ${\mathbf e}_3$.

The solution (\ref{xtasy}) has to be inserted into  (\ref{defr1} - \ref{defr3}) using the substitutions
(\ref{u2v}), (\ref{u2w}) and (\ref{defx}) to obtain the curve $r(t)\in{\mathcal P}$. After some calculations
we obtain the following simple time-independent expression for (\ref{soleomext}):
\begin{equation}\label{Omegasy}
 \Omega = \left( {\mathcal J}(r(t))  -\dot{r}(t)\right) \, r(t)^{-1}=
 \lambda \,\left(
\begin{array}{ccc}
0&-1 & 0 \\
1 & 0 & 0 \\
 0& 0 &0 \\
\end{array}
\right)
 \;.
\end{equation}
This implies that the final solution has the form  $s(t)= {\sf e}^{\Omega t}\,r(t)$ and, after some simplifications,
can be written as
\begin{eqnarray}\label{sasy1}
  {\mathbf s}_1(t)&=&
  \left(
\begin{array}{c}
 2 \tanh (\gamma \, t) \text{sech}(\gamma \, t) \cos (\lambda \, t) \\
 2 \tanh (\gamma \, t) \text{sech}(\gamma \, t) \sin (\lambda \, t) \\
 1-2 \tanh ^2(\gamma \, t) \\
\end{array}
\right),\\
\label{sasy2}
 {\mathbf s}_2(t)&=&
 \left(
\begin{array}{c}
 2\, \text{sech}(\gamma \, t) (\gamma  \sin (\lambda \, t)-\lambda  \tanh (\gamma \, t) \cos (\lambda \, t)) \\
 -2\, \text{sech}(\gamma \, t) (\gamma  \cos (\lambda\,  t)+\lambda  \tanh (\gamma \, t) \sin(\lambda \, t)) \\
 1-2 \lambda \, \text{sech}^2(\gamma\, t) \\
\end{array}
\right),\\
\label{sasy3}
 {\mathbf s}_3(t)&=&
\left(
\begin{array}{c}
 -2\, \text{sech}(\gamma \, t) (\gamma  \sin (\lambda \, t)+(1-\lambda ) \tanh (\gamma \, t)
   \cos (\lambda\,  t)) \\
 2\, \text{sech}(\gamma\,  t) (\gamma  \cos (\lambda \, t)-(1-\lambda ) \tanh (\gamma \, t)
   \sin (\lambda \, t)) \\
 1-2 (1-\lambda )\, \text{sech}^2(\gamma \, t) \\
\end{array}
\right)
\;.
\end{eqnarray}
At time $t=0$ this solution represents the coplanar state
\begin{equation}\label{st0}
{\mathbf s}_1(0)=\left(
\begin{array}{c}
 0 \\
 0 \\
 1 \\
\end{array}
\right),\;
{\mathbf s}_2(0)=
\left(
\begin{array}{c}
 0 \\
 -2\gamma \\
1-2 \lambda  \\
\end{array}
\right),\;
{\mathbf s}_3(0)=\left(
\begin{array}{c}
 0 \\
2 \gamma \\
2 \lambda -1 \\
\end{array}
\right)\;.
\end{equation}
Since $\text{sech}(\gamma t)\to 0$ and $\tanh^2(\gamma t)\to 1$ for $t\to \pm\infty$
it follows immediately that in this limit the spin configuration $s(t)$ approaches the collinear state
$\downarrow \uparrow\uparrow$ oriented into $3$-direction, the latter due to the choice of $r(t)$.

\subsection{Stationary spin configurations}\label{sec:SS}
We will determine all stationary solutions of the equations of motion (\ref{eom1} - \ref{eom3}).
As already mentioned, for collinear spin configurations corresponding to the four singular extremal points
${\mathbf e}_n,\;n=0,1,2,3,$ of the Gram set ${\mathcal G}$ all vector products in (\ref{eom1} - \ref{eom3}) vanish
and hence every collinear configuration will be stationary. Hence these configurations can be excluded from the following considerations.

We consider first the case
\begin{equation}\label{J1J2J3}
J_\mu\neq 0\quad\mbox{ for }\mu=1,2,3,
\end{equation}
and define
\begin{equation}\label{defti}
  {\mathbf t}_1:= {\mathbf s}_2\times {\mathbf s}_3,\;
   {\mathbf t}_2:= {\mathbf s}_3\times {\mathbf s}_1,\;
    {\mathbf t}_3:= {\mathbf s}_1\times {\mathbf s}_2
    \;.
\end{equation}
Then the stationarity conditions assume the form
\begin{equation}\label{stationary}
 {\mathbf 0} = \dot{\mathbf s}_1= J_2{\mathbf t}_2-J_3{\mathbf t}_3,\;
 {\mathbf 0} = \dot{\mathbf s}_2= J_3{\mathbf t}_3-J_1{\mathbf t}_1,\;
 {\mathbf 0} = \dot{\mathbf s}_3= J_1{\mathbf t}_1-J_2{\mathbf t}_2\;
 \;.
\end{equation}
It follows that all three spin vectors ${\mathbf s}_\mu$ must be orthogonal to the vector ${\mathbf t}\equiv J_\mu {\mathbf t}_\mu\neq{\mathbf 0}$
and hence every stationary spin configuration must be coplanar (remember that we have already excluded the collinear configurations).
We choose a coordinate system such that
\begin{equation}\label{stationaryconfig}
 {\mathbf s}_1=\left( \begin{array}{c}
                        x_1 \\
                        y_1 \\
                        0
                      \end{array}
 \right),\;
 {\mathbf s}_2=\left( \begin{array}{c}
                        x_2 \\
                        y_2 \\
                        0
                      \end{array}
 \right),\;
{\mathbf s}_3=\left( \begin{array}{c}
                        1 \\
                        0 \\
                        0
                      \end{array}
 \right)\;,
\end{equation}
and
\begin{equation}\label{normalization}
 x_\mu^2+y_\mu^2=1\quad \mbox{ for }\mu=1,2
 \;.
\end{equation}
Then the third stationarity condition, $J_1{\mathbf t}_1=J_2{\mathbf t}_2$, leads to
\begin{equation}\label{statcond3}
 -J_1\, y_2= J_2\,y_1
 \;,
\end{equation}
and the first one,  $J_2{\mathbf t}_2=J_3{\mathbf t}_3$, to
\begin{equation}\label{statcond1}
 J_2\,y_1= J_3(x_1\, y_2 -y_1\,x_2)
 \;.
\end{equation}
The latter condition can be written as
\begin{equation}\label{statcons1}
 (J_2+J_3\, x_2)\,y_1= J_3 \,x_1\,y_2\stackrel{(\ref{statcond3})}{=}-J_3\,x_1\,\frac{J_2}{J_1}\,y_1
 \;.
\end{equation}
If $y_1$ would vanish then, due to (\ref{statcond3}), also $y_2=0$, and the configuration would be collinear.
So we can assume $y_1\neq 0$ and when dividing by $y_1$ we conclude
\begin{equation}\label{statcons2}
 J_2+J_3\, x_2=-\frac{J_2\,J_3}{J_1}\,x_1
 \;,
\end{equation}
and, further,
\begin{equation}\label{statcons3}
  x_2=-\frac{J_2}{J_3}-\frac{J_2}{J_1}\,x_1
  \;.
\end{equation}
Normalization (\ref{normalization}) implies
\begin{eqnarray}
 \label{statcons4a}
  1 &=& x_2^2+y_2^2\stackrel{(\ref{statcons3},\ref{statcond3})}{=}\left( \frac{J_2}{J_3}+\frac{J_2}{J_1}x_1\right)^2
  +\left(\frac{J_2}{J_1}y_1 \right)^2 \\
  \label{statcons4b}
    &=& \frac{J_2^2}{J_3^2}+ \frac{J_2^2}{J_1^2}\underbrace{\left( x_1^2+y_1^2\right)}_{=1}
        +2\frac{J_2^2}{J_1 J_3}\,x_1=
        \frac{J_2^2}{J_3^2}+ \frac{J_2^2}{J_1^2}+2\frac{J_2^2}{J_1 J_3}\,x_1
    \;,
 \end{eqnarray}
which can be solved for $x_1$ with the result
\begin{equation}\label{solx1}
  x_1={\mathbf s}_1\cdot {\mathbf s}_3 = v = \frac{1}{2}\left( \frac{J_1 J_3}{J_2^2}-\frac{J_1}{J_3}-\frac{J_3}{J_1}\right)
  \;.
\end{equation}
Inserting this result into (\ref{statcons3}) yields, after some simplifications,
\begin{equation}\label{solx2}
   x_2={\mathbf s}_2\cdot {\mathbf s}_3 = u = \frac{1}{2}\left( \frac{J_2 J_3}{J_1^2}-\frac{J_2}{J_3}-\frac{J_3}{J_2}\right)
  \;,
\end{equation}
which also could be obtained from (\ref{solx1}) by cyclic permutation of indices.
Finally,
\begin{eqnarray}
\label{solwa}
 w &=& {\mathbf s}_1\cdot {\mathbf s}_2 = x_1\,x_2+y_1\,y_2\stackrel{(\ref{statcons3},\ref{statcond3})}{=}
- x_1\,\left(\frac{J_2}{J_3}+\frac{J_2}{J_1}x_1 \right)-y_1\,\left(\frac{J_2}{J_1}y_1  \right)\\
\label{solwb}
   &=& -\frac{J_2}{J_3}\,x_1 -\frac{J_2}{J_1}\underbrace{\left(x_1^2+y_1^2 \right)}_{=1}\\
   \label{solwc}
   &\stackrel{(\ref{solx1})}{=}&\frac{1}{2}\left( \frac{J_1 J_2}{J_3^2}-\frac{J_1}{J_2}-\frac{J_2}{J_1}\right)
   \;,
\end{eqnarray}
again following from  (\ref{solx1}) by cyclic permutation of indices. The corresponding energy will be
\begin{equation}\label{energystat}
  E_c=J_1\,u+J_2\,v+J_3\,w= J_1+J_2+J_3 -\frac{\left(J_1 J_2+J_3 J_2+J_1 J_3\right){}^2}{2 J_1 J_2 J_3}
  \;.
\end{equation}

Thus we have derived that every stationary state has a Gram matrix $G\widehat{=}(u,v,w)$ according to (\ref{solx1},\ref{solx2},\ref{solwc}).
Conversely, if $G\widehat{=}(u,v,w)$ is given satisfying (\ref{solx1},\ref{solx2},\ref{solwc}), it is straightforward to show
that the corresponding spin configuration
\begin{equation}\label{stationaryconfigu}
 {\mathbf s}_1=\left( \begin{array}{c}
                       v \\
                       \pm\sqrt{1-v^2} \\
                        0
                      \end{array}
 \right),\;
 {\mathbf s}_2=\left( \begin{array}{c}
                        u \\
                       \mp\sqrt{1-v^2} \\
                        0
                      \end{array}
 \right),\;
{\mathbf s}_3=\left( \begin{array}{c}
                        1 \\
                        0 \\
                        0
                      \end{array}
 \right)\;,
\end{equation}
will be stationary. It follows that every stationary, coplanar spin configuration will be congruent to (\ref{stationaryconfigu}).

Using the analogous characterization of \textit{ground states} of the classical spin triangle given in \cite{S17b} we may add
the remark that the stationary spin configurations $s$, coplanar or collinear, are \textit{critical states} of $H$, i.~e., ground states or anti-ground states,
or, more general, states where the first order variation of the energy $H$ vanishes in an infinitesimal neighbourhood of $s$.
This result can be additionally made plausible by
considering the condition that $H(s)$ has a critical value under the constraints ${\mathbf s}_\mu\cdot {\mathbf s}_\mu=1$
for $\mu=1,2,3$ which leads to
\begin{equation}\label{Hextremal}
  \sum_\nu J_{\mu\nu}{\mathbf s}_\nu = -\kappa_\mu\,{\mathbf s}_\mu,\quad \mbox{for }\mu=1,2,3,
\end{equation}
see, e.~g., \cite[Eq.~(13)]{S17a}, where the $\kappa_\mu$ are the Lagrange parameters due to the above constraints.
Note that (\ref{Hextremal}) is equivalent to the stationarity condition (\ref{stationary}).
Then there are, up to congruence, exactly five stationary states for each Hamiltonian $H$:
The coplanar state $s$ with a  Gram matrix $G\widehat{=}(u,v,w)$ according to (\ref{solx1},\ref{solx2},\ref{solwc})
and energy (\ref{energystat})
that is a regular extremal point of ${\mathcal G}$ and represents either a ground state or an anti-ground state.
Of the four collinear states  corresponding to the four singular extremal points of ${\mathcal G}$ one is again an (anti)-ground state
and the three remaining collinear states have a critical value of the energy.\\

Due to (\ref{stationary}) the next case
\begin{equation}\label{J1nullJ2J3}
J_1=0,\quad J_2\neq 0\neq J_3
\;,
\end{equation}
leads to ${\mathbf t}_2={\mathbf t}_3={\mathbf 0}$ and hence each stationary spin configuration must be collinear.\\

Hence we are left with
\begin{equation}\label{J1nullJ2nullJ3}
J_1=J_2=0,\quad  0\neq J_3\equiv J
\;.
\end{equation}
In this case the spin system effectively consists of only two spins ${\mathbf s}_1,\,{\mathbf s}_2$ and the third one
${\mathbf s}_3$ can be chosen arbitrarily and will remain constant. The time evolution of two spins with Heisenberg coupling
is well-known and consists of a uniform rotation about the constant total spin vector.
If we choose the coordinate system such that ${\mathbf s}_1(t)+{\mathbf s}_2(t)=(0,0,S)^\top$
the time development of the two spins hence assumes the form
\begin{equation}\label{twospins}
  {\mathbf s}_1(t)=\left(
  \begin{array}{c}
    \cos(J S t)\,\cos \beta \\
   \sin(J S t)\,\cos \beta  \\
    \sin \beta
  \end{array}
  \right),\quad
  {\mathbf s}_2(t)=\left(
  \begin{array}{c}
    -\cos(J S t)\,\cos \beta \\
   -\sin(J S t)\,\cos \beta  \\
    \sin \beta
  \end{array}
  \right)
  \;,
\end{equation}
where $0\le\beta<2\pi$ is some real parameter related to the total spin by means of $S=2\,\sin\beta$.
Since $J\neq 0$ the stationary case is only obtained for two cases:
Either $S=2\,\sin\beta=0$, i.~e., for $\beta=0$ or $\beta=\pi$,
symbolically $\uparrow\downarrow$, or $\beta=\frac{\pi}{2}$, symbolically $\uparrow\uparrow$.
These two configurations are also the (anti-)ground states for the ferromagnetic case $J<0$ or the anti-ferromagnetic case $J>0$.
Taking into account the arbitrary spin vector ${\mathbf s}_3$ we can localize the Gram matrices $G$ of the stationary
states in the case (\ref{J1nullJ2nullJ3}) as lying on one of the six one-dimensional faces of ${\mathcal G}$:
Either $G\in[{\mathbf e}_n,{\mathbf e}_m],\;1\le n<m\le 3$ in the case $\uparrow\downarrow$,
or $G\in[{\mathbf e}_0,{\mathbf e}_n],\;n=1,2,3,$ in the case $\uparrow\uparrow$. If ${\mathbf s}_3$ coincides
with $\pm{\mathbf s}_1$ or $\pm{\mathbf s}_2$ a collinear configuration is obtained; in all other cases $G$  lies
in the interior of a one-dimensional face and the stationary spin configuration will be coplanar.

\section{Echoes of Floquet theory}\label{sec:FT}
Floquet theory \cite{F83,YS75} applies to \textit{linear} differential equations with \textit{periodic} coefficients,
whereas our equations of motion (\ref{eom1} - \ref{eom3}) are bi-linear differential equations with constant coefficients.
Hence, at first sight, Floquet theory is not responsible for our problem.
However, note that equation (\ref{Zt}) of the form
\begin{equation}\label{Zt1}
 \dot{Z}(t)=\Omega(t)\,Z(t)
 \;,
\end{equation}
with $\Omega(t)$ given by the ${\sf T}$-periodic function (\ref{soleomext}) satisfies the requirements of Floquet theory.
Then it follows by this very theory that there exist solutions of (\ref{Zt1}) of the form
\begin{equation}\label{Flo}
  Z(t)=P(t)\,{\sf e}^{F\,t}
  \;,
\end{equation}
where $F$ is a real anti-symmetric $3\times 3$-matrix and $P(t)\in SO(3)$ a ${\sf T}$-periodic rotation matrix,
see also \cite{S20}.
From $Z(0)=P(0)=P({\sf T})={\mathbbm 1}$ it follows that ${\sf e}^{F\,{\sf T}}=Z({\sf T})$ leaves ${\mathbf S}_0=(0,0,S)^\top$
invariant and hence $F$ is of the form
\begin{equation}\label{Fform}
 F=f\,\left(
 \begin{array}{ccc}
   0 &-1 & 0 \\
   1 & 0 & 0 \\
   0 & 0 & 0
 \end{array}
 \right),\quad f\in {\mathbbm R}
 \;.
\end{equation}
The eigenvalues $1,{\sf e}^{\pm {\sf i}\,f\,{\sf T}}$ of ${\sf e}^{F\,{\sf T}}$ are called the \textit{characteristic multipliers}
of the differential equation (\ref{Zt1}) and their logarithms divided by ${\sf i\, T}$, i.~e.,
the numbers $0,\pm f$, are often referred to as the corresponding \textit{quasienergies}, see \cite{S20}.
Comparison with (\ref{fourierint}) shows that $f=a_0$, the zeroth coefficient of the Fourier series (\ref{fourier}).
It follows that also $P(t)$ leaves $(0,0,S)^\top$ invariant and hence can be written as
\begin{equation}\label{Pt}
  P(t)=\left(
 \begin{array}{ccc}
   \cos \tilde{\alpha}(t) &- \sin \tilde{\alpha}(t) & 0 \\
   \sin \tilde{\alpha}(t) &  \cos \tilde{\alpha}(t) & 0 \\
   0 & 0 & 1
 \end{array}
 \right),\quad f\in {\mathbbm R}
 \;,
\end{equation}
with the ${\sf T}$-periodic function
\begin{equation}\label{tildealpha}
  \tilde{\alpha}(t)=\sum_{n\neq 0} a_n\,\frac{\sf T}{2\pi {\sf i} n}\, \left(\exp\left(\frac{2\pi{\sf i} n t}{\sf T} \right)-1\right)
  \;,
\end{equation}
see (\ref{fourierint}).

In summary, we did not find any new solutions, but rather transferred the already known
solutions into the language of Floquet theory and found an unexpected connection between
the quasi-energy $f=a_0$ and the angular velocity $\Omega_2=a_0$ of the angular variable $\psi_2$, see (\ref{defpsi2}).
Physically, Floquet theory is often associated with periodic driving of systems due to some external forces. The present
example, however, shows that the periodic driving can also occur due to periodic variations of the internal degrees of freedom.



\begin{thebibliography}{99}


\bibitem{L73}
E.~H.~Lieb,
The Classical Limit of Quantum Spin Systems,
\textit{Commun. math. Phys.} {\bf  31}, 327 -- 340 (1973)



\bibitem{FKL07}
J.~Fr\"ohlich, A.~Knowles, and E.~Lenzmann,
Semi-Classical Dynamics in Quantum Spin Systems,
\textit{Lett. Math. Phys.} {\bf  82}, 275 -- 296 (2007)


\bibitem{P15}
P.~Pyykk\"o,
Magically magnetic gadolinium,
\textit{Nature Chem.} {\bf 7},  680  (2015).

\bibitem{GOB14}
K.~B.~Ghiassi, M.~M.~Olmstead,  and  A.~L.~Balch,
Gadolinium-containing endohedral fullerenes: structures and function as magnetic resonance imaging (MRI) agents,
\textit{Dalton Trans.} {\bf 43},  7346-7358   (2014).

\bibitem{Qetal17}
   L.~Qin, G.-J.~Zhou, H.~Nojiri, C.~Schr\"oder, R.~E.~P.~Winpenny, and Y.~Z.~Zheng,
   Topological Self-Assembly of Highly Symmetric Lanthanide Clusters: A Magnetic Study of Exchange-Coupling ``Fingerprints" in Giant Gadolinium(III) Cages,
    \textit{J. Am. Chem. Soc.} {\bf 139} (45), 16405 -- 16411  (2017).

\bibitem{Qetal21}
   L.~Qin, H.-L.~Zhang, Y.-Q.~Zhai, H.~Nojiri, C.~Schr\"oder, and Y.~Z.~Zheng,
  A giant spin molecule with ninety-six parallel unpaired electrons,
    \textit{iScience} {\bf 24} (4), 102359  (2021).


\bibitem{SLR11}
H.-J.~Schmidt, A.~Lohmann, and J.~Richter,
Eighth-order high-temperature expansion for general Heisenberg Hamiltonians,
\textit{Phys. Rev. B} {\bf 84},  104443  (2011).


\bibitem{W15}
G.~M.~Wysin,
\textit{Magnetic Excitations and Geometric Confinement},
IOP Publishing, Bristol, England,  2015.


\bibitem{A78}
V.~I.~Arnol'd,
\textit{Mathematical Methods of Classical Mechanics},
Springer, Berlin, 1978.




\bibitem{S13}
H.-J.~Schmidt,
The general spin triangle,
\textit{Int. J. Mod. Phys. B} {\bf 278} (16), 1350064 (2013).



\bibitem{SL03}
H.-J.~Schmidt and M.~Luban,
Classical ground states of symmetric Heisenberg spin systems,
\textit{J. Phys. A: Math. Gen.} {\bf 36} (23), 6351  (2003).


\bibitem{SSHL15}
H.-J.~Schmidt, C.~Schr\"oder, E.~H\"agele, and M.~Luban,
Dynamics and thermodynamics of a pair of interacting dipoles,
\textit{J. Phys. A: Math. Theor.} {\bf 48}, 185002 (2015).

\bibitem{LL35}
L.~D.~Landau and J.~M.~Lifschitz,
Theory of the dispersion of magnetic permeability in ferromagnetic bodies,
\textit{Phys. Z. Sowjetunion} {\bf 8}, 153 (1935).

\bibitem{F1}
Strictly speaking, these equations correspond to those published in \cite{LL35} but without dissipation terms.


\bibitem{NIST21}
NIST Digital Library of Mathematical Functions. http://dlmf.nist.gov/, Release 1.1.1 of 2021-03-15.
F.~W.~J.~Olver, A.~B.~Olde Daalhuis, D.~W.~Lozier, B.~I.~Schneider, R.~F.~Boisvert, C.~W.~Clark, B.~R.~Miller, B.~V.~Saunders, H.~S.~Cohl, and M.~A.~McClain, eds.



\bibitem{AS72}
M.~Abramowitz and I.~A.~Stegun (eds),
\textit{Handbook of Mathematical Functions},
Dover, New York, 1972.

\bibitem{KBL98}
M.~Krech, A.~Bunker, and D.~P.~Landau,
Fast spin dynamics algorithms for classical spin systems,
\textit{Comput. Phys. Commun.} {\bf 111} (3), 1 -- 13   (1998).


\bibitem{CLAL99}
O.~Ciftja, M.~Luban, M.~Auslender, and J.~H.~Luscombe,
Equation of state and spin-correlation functions of ultrasmall classical Heisenberg magnets,
\textit{Phys. Rev. B} {\bf 60} (14, 10122 - 10133  (1999).


\bibitem{AM78}
R.~Abraham and J.~E.~Marsden,
\textit{Foundations of Mechanics},
 Benjamin-Cummings, London, 1978.



\bibitem{S17a}
H,.-J.~Schmidt,
Theory of ground states for classical Heisenberg spin systems I,
\textit{Preprint} cond-mat:1701.02489v2 (2017).


\bibitem{S17b}
H.-J.~Schmidt,
Theory of ground states for classical Heisenberg spin systems III,
\textit{Preprint} cond-mat:1707.06512 (2017).

\bibitem{L13}
J.~M.~Lee,
\textit{Introduction to smooth manifolds}, Graduate Texts in Mathematics, $2^{nd}$ edition,
Springer, New York, 2013.


\bibitem{KMS93}
    I.~Kol\'{a}\v{r}, P.~W.~Michor, and J.~Slov\'{a}k,
    \textit{Natural operations in differential geometry},
    Springer, New York, 1993.

\bibitem{H15}
B.~C.~Hall,
\textit{ Lie Groups, Lie Algebras, and Representations: An Elementary Introduction}, Graduate Texts in Mathematics, 222 $2^{nd}$ edition,
Springer, New York, 2015.

\bibitem{F83}
	G. Floquet,
	Sur les \'equations diff\'erentielles lin\'eaires \`a coefficients
	p\'eriodiques,
	Annales de l' \'Ecole Normale Sup\'erieure {\bf 12}, 47 (1883).

\bibitem{YS75}
	V. A. Yakubovich and V. M. Starzhinskii,		
	{\em Linear differential equations with periodic coefficients\/},
	2~volumes (Wiley, New York, 1975).

\bibitem{S20}
    H.-J.~Schmidt,
    Geometry of the Rabi problem and duality of loops,
    \textit{Z. Naturforsch. A } {\bf 75},  381  (2020).


\bibitem{FGS03}
    E.~Fiorani, G.~Giachetta, and G.~Sardanashvilym,
    An extension of the Liouville-Arnold theorem for the
    non-compact case,
    \textit{Il Nuovo Cimento B} {\bf 118} (3), 307 -- 317   (2003).




\end{thebibliography}
\end{document}